\documentclass[11pt,a4paper]{article}
\usepackage[utf8]{inputenc}
\usepackage{amsmath}
\usepackage{amsfonts}
\usepackage{amssymb}
\usepackage[english]{babel}
\usepackage{graphicx}
\usepackage[left=2cm,right=2cm,top=2cm,bottom=2cm]{geometry}
\usepackage{subfloat}
\usepackage{subcaption} 
\usepackage{float} 
\usepackage{wrapfig}
\usepackage{fancyhdr}
\usepackage{indentfirst}
\usepackage{cite}
\usepackage{physics}
\usepackage{setspace}
\usepackage{authblk}
\usepackage{xcolor}
\usepackage{ulem}

\newcommand*\MC{{MoC}}

\begin{document}

\begin{center}
\begin{LARGE}

\textbf{Description of ultrastrong light-matter interaction through coupled harmonic oscillator models and their connection with cavity-QED Hamiltonians}

\end{LARGE}

\medskip

\bigskip

\begin{large}
Unai Muniain$^{*1}$, Javier Aizpurua$^{1,2,3}$, Rainer Hillenbrand$^{2,3, 4}$, Luis Martín-Moreno$^{5, 6}$ and Ruben Esteban$^{*1, 7}$
\end{large}
\begin{small}

$^1$\textit{Donostia International Physics Center, Paseo Manuel de Lardizabal 4, 20018 Donostia-San Sebastián, Spain}

$^2$\textit{IKERBASQUE, Basque Foundation for Science, María Díaz de Haro 3, 48013 Bilbao, Spain}

$^3$\textit{Department of Electricity and Electronics, University of the Basque Country (UPV/EHU), 48940 Leioa, Spain }

$^4$\textit{CIC nanoGUNE BRTA, Tolosa Hiribidea 76, 20018 Donostia-San Sebastián, Spain }

$^5$\textit{Instituto de Nanociencia y Materiales de Aragón (INMA), CSIC-Universidad de Zaragoza, 50009 Zaragoza, Spain }

$^6$\textit{Departamento de Física de la Materia Condensada, Universidad de Zaragoza, 50009 Zaragoza, Spain }

$^7$\textit{Centro de Física de Materiales (CFM-MPC), CSIC-UPV/EHU, Paseo Manuel de Lardizabal 5, 20018 Donostia-San Sebastián, Spain}

Corresponding authors: Unai Muniain (unaimuni@gmail.com); Ruben Esteban (ruben.esteban@ehu.eus)

\end{small}
\medskip

\end{center}

\textbf{Abstract:} 
Classical coupled harmonic oscillator models are capable of describing the optical and infrared response of nanophotonic systems where a cavity photon couples to dipolar matter excitations. The distinct forms of coupling adopted in these classical models leads to different results in the ultrastrong coupling regime. To clarify the specific classical model required to address particular configurations, we establish a connection between each oscillator model and the equivalent cavity Quantum Electrodynamics description. We show that the proper choice of coupled harmonic oscillator model depends on the presence or absence of the diamagnetic term in the quantum models, linked to whether transverse or longitudinal electromagnetic fields mediate the coupling. This analysis also shows how to exploit the classical oscillator models to extract measurable information of the optical response, as demonstrated in three canonical photonic systems, and to describe the opening of the Reststrahlen band in the bulk dispersion of phononic materials


\addtocontents{toc}{\setcounter{tocdepth}{0}}

\section{Introduction} \label{sec_introduction}

The optical properties of molecules, quantum dots, two-dimensional materials, or other systems supporting matter excitations are strongly modified when these excitations are coupled to the electromagnetic modes of a cavity or a resonator. The strong coupling regime is reached when the coupling strength $g$ between the cavity modes and the matter excitations exceeds their losses\cite{torma15, dovzhenko18}. In this regime, hybrid modes known as polaritons emerge, exhibiting modified frequencies and new properties as compared to the uncoupled constituents.  Strongly-coupled system can also exhibit effects beyond the classical realm, including nonlinearities due to the Jaynes-Cummings ladder \cite{fink08}, emission of strongly correlated light \cite{saezblazquez17}, and changes on the chemical reactivity \cite{thomas19} or on the conductivity \cite{orgiu15} of molecules located inside the cavity. 

After the first observations of strong coupling for a single \cite{rempe87,thompson92} and many emitters \cite{kaluzny83,brune87,raizen89}, very large coupling strengths have been successfully measured in subsequent experiments, exploiting semiconductors \cite{reithmaier04,christopoulos07}, superconducting circuits \cite{wallraff04}, plasmonic nanoparticle crystals \cite{mueller20}  or ensembles of organic molecules \cite{hakala09,vasa10,chikkaraddy16,feist18}, for instance. It is now possible to reach coupling strengths that are several times larger than the threshold that usually marks the onset of the ultrastrong coupling regime \cite{niemczyk10,kockum19,forndiaz19}, which roughly occurs when the coupling strength is $\approx 10 \%$ of the uncoupled cavity mode and matter excitation resonant frequencies. In this ultrastrong coupling regime, additional quantum effects emerge, such as a shift of the ground state energy and the appearance of virtual excitations in this state \cite{ciuti05}, which cannot be accounted for within the rotating-wave approximation (RWA).

Models based on the Cavity Quantum Electrodynamics (cavity-QED) framework offer a natural description of these effects. However, two different QED Hamiltonians have been considered when studying the ultrastrong coupling regime, with differences stemming from the presence or absence of a contribution to the energy, the so-called diamagnetic term (also known as the $A^2$ term, with $A$  the transverse vector potential of the electromagnetic mode). Introducing this term avoids a superradiant phase transition \cite{rzazewski75}, for example. However, the inclusion of the diamagnetic contribution is still under discussion  \cite{nataf10,vukics12,tufarelli15,debernardis18,schafer20} and depends on the specifics of the system \cite{galego19,feist21}. Furthermore, in the presence of a diamagnetic term, if the Hilbert space must be truncated when performing the calculations (as is often the case), care needs to be taken as the results can become dependent on the chosen gauge\cite{distefano19,salmon22}.

On the other hand, the response of nanophotonic systems in the strong and ultrastrong coupling regime is often described using phenomenological classical models based on coupled harmonic oscillators\cite{rodriguez16,novotny10,rudin99}. Such a simple description turns out to be adequate when the optical cavity couples with many quantum emitters (such as molecules, quantum dots, color centers in diamond...) or with matter excitations in an extended material. In this case, the nonlinearities behind many quantum effects are strongly attenuated compared to the single-emitter scenario. Here, we focus on nanophotonic systems for simplicity, but the discussion presented in this work is also valid for systems of micrometer dimensions unless otherwise stated.  The classical coupled harmonic oscillator models have successfully described phenomena such as the avoided crossing of the hybrid modes \cite{lockhart18}, Fano resonances \cite{joe06},  stimulated Raman scattering \cite{hemmer88}, and electromagnetically induced transparency \cite{garridoalzar02,harden11,souza15}. They are used to fit experimental data and to extract the coupling strength $g$, the frequencies of the hybrid modes, and the fraction of light and matter corresponding to each mode \cite{harder18, liu14}. However, in these phenomenological models, it is often unclear which exact physical quantity each oscillator represents, making it difficult to determine the value of a given observable in an experiment. To further complicate the situation, and similarly to the coexistence of cavity-QED Hamiltonian descriptions with and without diamagnetic term, different classical oscillator models have been used to analyze coupled systems, in both the strong and ultrastrong coupling regimes. In some models, the coupling terms are proportional to the amplitudes of the harmonic oscillators, while in others, they are proportional to the time derivatives of the amplitudes. The choice of coupling terms and the connections with the cavity-QED description are often not clearly justified. \cite{rudin99,kats07,george16,yoo21}.

In this work, we first present a cavity-QED model describing the emitter-cavity coupling and derive several classical harmonic oscillator models that reproduce the same spectral properties and expectation values of  any operator. These classical descriptions feature coupling terms that are proportional either to the amplitudes of the harmonic oscillators or to their time derivatives, accompanied by corresponding coupling-induced dressing of the oscillator frequencies. The choice of description is, in principle, a matter of preference. However, this flexibility disappears if one requires that the cavity frequencies in the phenomenological model are the (non-dressed) bare ones, which is the standard choice in nanophotonics, where bare cavity frequencies can be measured or computed.  Specifically, the presence or absence of the diamagnetic term in the original cavity-QED Hamiltonian determines the form of the coupling term in the classical model with bare cavity frequencies. We illustrate this scenario using several standard nanophotonic systems as examples. Furthermore, these examples serve to clarify how the amplitude of the oscillator modes relates to physical observables, such as the electric field within the cavity.

The paper is organized as follows:

In Sec. \ref{sec_quantum}, we analyze in detail the connection between the cavity-QED descriptions and several equivalent classical harmonic oscillator models that can be derived from them.

\begin{figure}
\centering
\includegraphics[scale=0.7]{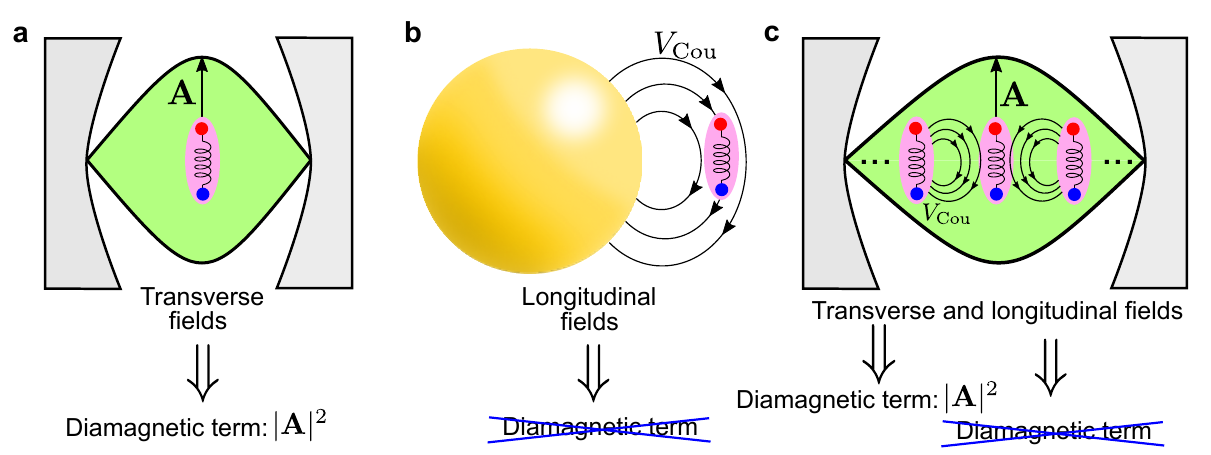}
\caption{Schematics of the interaction between matter excitations and cavity modes in the three systems considered in this work. a) A molecular emitter (as a representative quantum emitter) placed inside a dielectric (Fabry-Pérot) cavity. The transverse field of the single cavity mode considered is described with the vector potential $\mathbf{A}$, which leads to the presence of the diamagnetic term $\propto |\mathbf{A}|^2$ in the cavity-QED Hamiltonian that describes this system. b) A molecular emitter close to a metallic spherical nanoparticle and coupled to a single plasmonic mode. Within the quasistatic approximation, the molecular emitter only interacts with the longitudinal fields of the spherical nanoparticle, via the Coulomb potential $V_\text{Cou}$. Since the vector potential  $\mathbf{A}$ is not considered, the diamagnetic term is absent in the corresponding cavity-QED description. c) An ensemble of molecular emitters placed inside a Fabry-Pérot cavity. The molecular emitters behave as a homogeneous bulk material. In this system, each emitter interacts with the transverse cavity mode characterized by the vector potential $\mathbf{A}$, as well as with the longitudinal fields associated with the Coulomb potential $V_\text{Cou}$ induced by the other molecular emitters. Whereas the interaction of each emitter with cavity modes requires a diamagnetic term in the cavity-QED description, the coupling with other emitters is described without this term.} \label{figure_intro}
\end{figure}

In Sec. \ref{sec_examples}, we apply these results to three canonical situations arising in nanophotonics:  (i) a molecular emitter (or another quantum emitter) coupled to a conventional dielectric cavity (a Fabry-Pérot cavity, Fig. \ref{figure_intro}a), (ii) a molecular emitter coupled to a small metallic nanoparticle supporting plasmonic resonances (Fig. \ref{figure_intro}b), and (iii) an ensemble of molecular emitters or a homogeneous material inside a Fabry-Pérot cavity (Fig. \ref{figure_intro}c). These examples emphasize the importance of the type of coupling.  The choice of the classical coupled harmonic model (which depends on the presence of the diamagnetic term in the cavity-QED Hamiltonian) depends on whether the coupling is mediated by the transverse fields in a dielectric cavity or by the Coulomb interaction.  Additionally, we demonstrate that identifying the amplitudes of the classical harmonic oscillators with the expectation values of quantum operators allows for the calculation physical observables within the classical description. Last, we use the third canonical configuration to discuss the bulk dispersion of materials and the emergence of the Reststrahlen band within harmonic oscillator models, a point discussed in more detail in the Supplementary Material.

\section{Comparison of classical and cavity-QED models} \label{sec_quantum}

In this section, we  examine first a cavity-QED Hamiltonian that describes the interaction between a quantum emitter and a cavity optical mode. In Section \ref{sec_derivation_classical}, we derive the Heisenberg equations of motion for the displacements of the quantum operators, which take the form of classical oscillator equations. We present two equivalent descriptions, related by unitary transformations of the original quantum Hamiltonian. In one description, the coupling term between the oscillators is proportional to their amplitudes, while in the other it is proportional to their time derivatives. Both approaches yield the same results, as the coupling strength and cavity frequency are appropriately renormalized in each case.

In nanophotonics, bare cavity frequencies, which can be measured or computed, are typically used when fitting experimental and simulated spectra, without considering their potential renormalization. We therefore focus on classical models with un-renormalized cavity frequencies, referring to them as the Spring Coupling (SpC) model for amplitude-based coupling, and the Momentum Coupling (MoC) model for coupling based on time derivatives of the amplitudes.

For specific values of the diamagnetic term in the Hamiltonian, the Heisenberg equations align naturally with either the SpC (Section \ref{subsec_spring_coupling}) or MoC (Section \ref{subsec_MoC_coupling}) models, making each of them the most appropriate choice for fitting different experimental data. Section \ref{sec:comparison} illustrate the differences between these two models.

\subsection{Derivation of the classical models from the Hamiltonians} \label{sec_derivation_classical}

In this subsection, we introduce the classical harmonic oscillator models. To this purpose, we first analyze the light-matter interaction using the cavity-QED framework. The cavity modes and the matter excitations are quantized using bosonic operators. The use of bosonic operators is valid for the cavity modes, and for matter excitations such as vibrations or phonons associated with a potential with a harmonic dependence on the degrees of freedom. The correspondence with classical harmonic oscillators (and thus the use of bosonic operators) is also valid to treat the coupling with matter excitations of fermionic nature provided that the number of excitations is much smaller than the number of quantum emitters  (molecules, quantum dots...) and that any other effects induced by the saturation of the fermionic states can be discarded. Under these conditions, for example, the Quantum Rabi model (a generalization of the Jaynes-Cummings model to the ultrastrong coupling regime that includes a fermionic excitation \cite{kockum19}) becomes analogous to an appropriate bosonic Hamiltonian with a single matter excitation. Under this prescription based on bosonic operators, we can use a Hopfield-type Hamiltonian \cite{hopfield58} in the form
\begin{equation}
\hat{H}_1 = \hbar \omega_\text{cav} \left( \hat{a}^\dagger \hat{a} +\frac{1}{2} \right) + \hbar \omega_\text{mat} \left( \hat{b}^\dagger \hat{b} + \frac{1}{2} \right) + \hbar g_{\scriptscriptstyle{\text{QED}}} (\hat{a} + \hat{a}^\dagger)(\hat{b} + \hat{b}^\dagger) + \hbar D (\hat{a} + \hat{a}^\dagger)^2, \label{hopfield_hamiltonian_1}
\end{equation}
as shown in the Supplementary Material. In this Hamiltonian, the creation operator $\hat{a}^\dagger$ and the annihilation operator $\hat{a}$ act on the cavity mode, while the equivalent operators $\hat{b}^\dagger$ and $\hat{b}$ are associated to the matter excitation, obeying commutation rules $[\hat{a},\hat{a}^\dagger] =[\hat{b},\hat{b}^\dagger] = 1$. The first two terms on the right-hand side of Eq. \eqref{hopfield_hamiltonian_1} indicate the energy of the uncoupled (or bare) cavity modes and matter excitations at (angular) frequencies $\omega_\text{cav}$ and $\omega_\text{mat}$ , respectively, with $\hbar$ the reduced Planck constant. The third term describes the light-matter coupling, which is parameterized by the coupling strength  $g_{\scriptscriptstyle{\text{QED}}}$, and where we include the anti-resonant terms $\hat{a}\hat{b}$ and $\hat{a}^\dagger \hat{b}^\dagger$ required to describe the ultrastrong coupling regime correctly. $g_{\scriptscriptstyle{\text{QED}}}$  can in principle depend on $\omega_\text{cav}$ and $\omega_\text{mat}$ in specific systems (Sec. \ref{Appendix_eigenvalues-nonconstantg} in Supplementary Material). Last, we introduce the diamagnetic term, scaled by a parameter $D$ that is initially considered to have an arbitrary value (including the zero value). This diamagnetic term, which is included in many (but not all) studies of ultrastrong coupling, is negligible in the strong coupling regime, but becomes important under ultrastrong coupling. It typically originates from the $|\mathbf{A}_\perp|^2$ term of the minimal coupling Hamiltonian, where $\mathbf{A}_\perp$ is the transverse vector potential. In the main text, we work in the Coulomb gauge, where the vector potential is completely transverse ($\nabla \cdot \mathbf{A} = 0$, and thus, $\mathbf{A}_\perp = \mathbf{A}$), so that hereafter we omit the symbol $\perp$ in $\mathbf{A}$ for brevity. 

From the Hopfield Hamiltonian,  we can obtain the equations of motion of the displacements (or oscillation amplitudes) of two quantum oscillators.  With this aim, we connect the creation and annihilation operators from the Hamiltonian in Eq. \eqref{hopfield_hamiltonian_1} with the quantum operators $\hat{x}_\text{cav} = \sqrt{\frac{\hbar }{2  \omega_\text{cav}}}(\hat{a} + \hat{a}^\dagger)$, $\hat{x}_\text{mat} = \sqrt{\frac{\hbar }{2  \omega_\text{mat}}}(\hat{b} + \hat{b}^\dagger)$, $\hat{p}_\text{cav} = -i\sqrt{\frac{\hbar  \omega_\text{cav}}{2}}(\hat{a} -  \hat{a}^\dagger)$ and $\hat{p}_\text{mat} = -i \sqrt{\frac{\hbar  \omega_\text{mat} }{2 }}(\hat{b} - \hat{b}^\dagger)$. These operators correspond to the canonical position and momentum operators of harmonic oscillators (except that no mass has been included in their definitions). They fulfill the canonical commutation relations $[\hat{x}_\text{mat},\hat{x}_\text{cav}]=[\hat{p}_\text{mat},\hat{p}_\text{cav}]=[\hat{x}_\text{mat},\hat{p}_\text{cav}] = [\hat{x}_\text{cav},\hat{p}_\text{mat}] = 0$, and $[\hat{x}_\text{mat},\hat{p}_\text{mat}]=[\hat{x}_\text{cav},\hat{p}_\text{cav}]=i\hbar$. The dynamics of these operators are calculated from the general Heisenberg equation of motion of an operator $\hat{O}$, $\frac{d}{dt}\hat{O} = \frac{1}{i\hbar}[\hat{O},\hat{H}]$. We convert the four resulting first-order differential equations into two second-order equations by eliminating the momentum operators and obtain the following equations of motion for the expectation values $ \langle \hat{x}_\text{cav} \rangle$ and $ \langle \hat{x}_\text{mat} \rangle$:
\begin{subequations}
\begin{equation}
\langle \ddot{\hat{x}}_\text{cav} \rangle + (\omega_\text{cav}^2 + 4D\omega_\text{cav}) \langle \hat{x}_\text{cav} \rangle + 2g_{\scriptscriptstyle{\text{QED}}} \sqrt{\omega_\text{cav}\omega_\text{mat}}  \langle \hat{x}_\text{mat} \rangle=0, 
\end{equation}
\begin{equation}
\langle \ddot{\hat{x}}_\text{mat} \rangle + \omega_\text{mat}^2 \langle \hat{x}_\text{mat} \rangle + 2g_{\scriptscriptstyle{\text{QED}}} \sqrt{\omega_\text{cav}\omega_\text{mat}}  \langle \hat{x}_\text{cav} \rangle = 0.
\end{equation}
\label{equations_motion_quantum_position}
\end{subequations}

These are not the only classical equations that could describe the spectra of the coupled system.  Any Hamiltonian $\hat{H}_2$  related to the Hopfield Hamiltonian $\hat{H}_1$ by a unitary transformation will have the same eigenfrequencies but will lead to different Heisenberg equations of motion. We perform a unitary transformation to $\hat{H}_1$ with the operator $\hat{U} = e^{-i\frac{\pi}{2}\hat{b}^\dagger \hat{b}}$. In the new reference frame, $\hat{H}_2 = \hat{U}\hat{H}_1\hat{U}^\dagger + i\hbar \frac{\partial \hat{U}}{\partial t}\hat{U}^\dagger$ is expressed as:  
\begin{equation}
\hat{H}_2 = \hbar \omega_\text{cav} \left( \hat{a}^{\dagger} \hat{a} +\frac{1}{2} \right) + \hbar \omega_\text{mat} \left( \hat{b}^{\dagger \prime} \hat{b}' + \frac{1}{2} \right) + i \hbar g_{\scriptscriptstyle{\text{QED}}} (\hat{a} + \hat{a}^{\dagger})(\hat{b}' - \hat{b}^{\dagger \prime} ) + \hbar D (\hat{a} + \hat{a}^{\dagger})^2,
\label{quantum_hamiltonian_2}
\end{equation}
where the prime $'$ denotes that the matter operators are transformed ($\hat{b} \rightarrow i\hat{b}'$ and $\hat{b}^\dagger \rightarrow -i\hat{b}^{\dagger \prime} $). In the representation of position and momentum operators,  this transformation can be understood as a rotation in phase space so that the canonical variables transform as
\begin{subequations}
\begin{align}
 \hat{x}_\text{mat} & \rightarrow -\frac{\hat{p}_\text{mat}'}{\omega_\text{mat}}, \\
 \hat{p}_\text{mat} & \rightarrow  \omega_\text{mat} \hat{x}_\text{mat}' .
\end{align}
\end{subequations}
In this new reference frame, we can calculate the equations of motion for the expectation values $\langle \hat{x}_\text{cav} \rangle$ and $\langle \hat{x}'_\text{mat} \rangle$:
\begin{subequations}
\begin{equation}
\langle \ddot{\hat{x}}_\text{cav} \rangle + \left(\omega_\text{cav}^2 + 4D\omega_\text{cav} - 4 g_{\scriptscriptstyle{\text{QED}}}^2 \frac{\omega_\text{cav}}{\omega_\text{mat}}  \right) \langle \hat{x}_\text{cav} \rangle  - 2g_{\scriptscriptstyle{\text{QED}}}  \sqrt{\frac{\omega_\text{cav}}{\omega_\text{mat}}} \langle \dot{\hat{x}}'_\text{mat} \rangle=0, 
\end{equation}
\begin{equation}
\langle \ddot{\hat{x}}_\text{mat}' \rangle + \omega_\text{mat}^2 \langle \hat{x}_\text{mat}' \rangle  + 2g_{\scriptscriptstyle{\text{QED}}}\sqrt{\frac{\omega_\text{cav}}{\omega_\text{mat}}}   \langle \dot{\hat{x}}_\text{cav} \rangle  = 0.
\end{equation}
\label{equations_motion_quantum_velocities}
\end{subequations}
We find that, in contrast to Eq. \eqref{equations_motion_quantum_position},
the coupling term is now proportional to the time derivative of the oscillation amplitudes.

To obtain the classical harmonic oscillator models, it is just necessary to associate the expectation values of the quantum operators to classical oscillation amplitudes, e.g., $\langle {\hat{x}}_\text{cav} \rangle\rightarrow x_\text{cav}$, so that Eq. \eqref{equations_motion_quantum_position} becomes

\begin{subequations}
\begin{equation}
\ddot{x}_\text{cav}  + (\omega_\text{cav}^2 + 4D\omega_\text{cav}) x_\text{cav} + 2g_{\scriptscriptstyle{\text{QED}}} \sqrt{\omega_\text{cav}\omega_\text{mat}}  x_\text{mat} =0, \label{equations_motion_classical_position1}
\end{equation}
\begin{equation}
\ddot{x}_\text{mat} + \omega_\text{mat}^2 x_\text{mat}  + 2g_{\scriptscriptstyle{\text{QED}}} \sqrt{\omega_\text{cav}\omega_\text{mat}}  x_\text{cav}  = 0, \label{equations_motion_classical_position2}
\end{equation}
\label{equations_motion_classical_position}
\end{subequations}
and Eq. \eqref{equations_motion_quantum_velocities} becomes
\begin{subequations}
\begin{equation}
\ddot{x}_\text{cav}  + \left(\omega_\text{cav}^2 + 4D\omega_\text{cav} - 4 g_{\scriptscriptstyle{\text{QED}}}^2 \frac{\omega_\text{cav}}{\omega_\text{mat}} \right) x_\text{cav}   - 2g_{\scriptscriptstyle{\text{QED}}}\sqrt{\frac{\omega_\text{cav}}{\omega_\text{mat}}}   \dot{x}_\text{mat}=0, \label{equations_motion_classical_velocities1}
\end{equation}
\begin{equation}
\ddot{x}_\text{mat}  + \omega_\text{mat}^2 x_\text{mat}   + 2g_{\scriptscriptstyle{\text{QED}}}\sqrt{\frac{\omega_\text{cav}}{\omega_\text{mat}}}  \dot{x}_\text{cav}   = 0, \label{equations_motion_classical_velocities2}
\end{equation}
\label{equations_motion_classical_velocities}
\end{subequations}
where we do not make an explicit distinction between ${x}_\text{mat} \equiv \langle \hat{x}_\text{mat}' \rangle$ used in Eqs. \eqref{equations_motion_quantum_velocities}, \eqref{equations_motion_classical_velocities}  and ${x}_\text{mat}\equiv \langle \hat{x}_\text{mat} \rangle$ in Eqs. \eqref{equations_motion_quantum_position}, \eqref{equations_motion_classical_position}. However, the physical interpretation of the expectation values  $\langle \hat{x}_\text{mat} \rangle$ and  $\langle \hat{x}_\text{mat}' \rangle $ (or the oscillation amplitudes $x_\text{mat}$ in each set of equations) is different, as discussed in more detail in Sec. \ref{sec_examples} when applying each equation to specific coupled systems.   Loss mechanisms are not included in these equations (friction terms proportional to the time derivatives $\dot{x}_\text{cav}$ and $\dot{x}_\text{mat}$), because they were derived from Hermitian cavity-QED Hamiltonians. Neglecting losses is usually an excellent approximation for calculating the eigenfrequencies and eigenvectors of the system in the ultrastrong coupling regime, where the coupling strength can be much larger than the system losses (the inclusion of dissipation in cavity-QED descriptions is discussed in Refs. \cite{hughes24, zueco19}).

\begin{figure}
\centering
\includegraphics[scale=0.92]{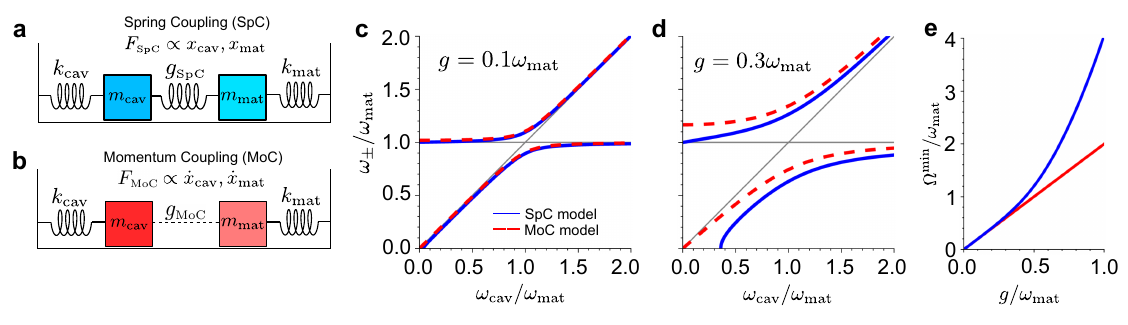}
\caption{Comparison of the Spring Coupling (SpC) and Momentum Coupling (\MC{}) models. a) Schematics of the SpC model in analogy to an oscillator model in classical mechanics. The coupling mechanism of strength $g_{\scriptscriptstyle{\text{SpC}}}$ is analogous to a  force $F_{\scriptscriptstyle{\text{SpC}}}$ exerted by a spring and proportional to the oscillator displacements $x_\text{cav}$ and $x_\text{mat}$. b) Schematics of the \MC{} model. The coupling mechanism of strength $g_{\scriptscriptstyle{\text{\MC{}}}}$ is analogous to a force $F_{\scriptscriptstyle{\text{\MC{}}}}$ proportional to the time derivatives of the oscillator displacements $\dot{x}_\text{cav}$ and $\dot{x}_\text{mat}$. We represent the coupling with dashed lines to highlight the different coupling mechanism compared with the SpC model, but we are not aware of any system described by the \MC{} model in classical mechanics. c) Eigenfrequencies $\omega_\pm$ of the hybrid states calculated from the bare values $\omega_\text{cav}$ and  $\omega_\text{mat}$, with $\omega_\text{mat}$ fixed and $\omega_\text{cav}/\omega_\text{mat}$ changing. $\omega_\pm$ are obtained from the SpC model (blue solid line, corresponding to Eq. \eqref{eigenfrequencies_SpC}) and the \MC{} model (red dashed line, Eq. \eqref{eigenfrequencies_MC}) for coupling strength $g=g_{\scriptscriptstyle{\text{SpC}}}=g_{\scriptscriptstyle{\text{\MC{}}}}= 0.1 \, \omega_\text{mat}$. The thin gray lines correspond to the bare cavity frequency $\omega_\text{cav}$ and the bare frequency of the matter excitation, $\omega_\text{mat}$.    d) Same as panel (c), for coupling strength $g=g_{\scriptscriptstyle{\text{SpC}}}=g_{\scriptscriptstyle{\text{\MC{}}}}=0.3 \, \omega_\text{mat}$. 
e) Minimum splitting between the hybrid modes $\Omega^\text{min} = \omega_+ - \omega_-$, as a function of the coupling strength $g$ for the SpC model (blue solid line) and the \MC{} model (red solid line). All frequencies in panels (c-e) are normalized with respect to the fixed frequency of the matter excitation $\omega_\text{mat}$, so that the results do not depend on the particular value of $\omega_\text{mat}$, only on the $\omega_\text{cav}/\omega_\text{mat}$ and $g/\omega_\text{mat}$ ratios. } \label{figure_models}
\end{figure}

Importantly, once the different interpretation of $\langle \hat{x}_\text{mat} \rangle$ and  $\langle \hat{x}_\text{mat}' \rangle $ is accounted for,  the two sets of coupled harmonic oscillator equations can be used to obtain the same result for any physical magnitude of a given system, as they correspond to Hamiltonians related by a unitary transformation.
In particular, the eigenfrequencies of Eq. \eqref{equations_motion_classical_position} and Eq. \eqref{equations_motion_classical_velocities} are identical. 

Thus, it is always possible to obtain the optical response of the coupled system by considering the coupling to be proportional to either the oscillation amplitudes or their time derivatives. 
An important point to notice is that, in both Eq. \eqref{equations_motion_classical_position2} and Eq. \eqref{equations_motion_classical_velocities2}, the ``matter resonant frequency'' (square root of the term proportional to $x_\text{mat}$) is the bare frequency $\omega_\text{mat}$.  However,  the ``cavity resonant frequency'' (square root of the term proportional to $x_\text{cav}$) is different in the different oscillator models.   In the model characterized  by Eq. \eqref{equations_motion_classical_position1} the shifted cavity frequency is $\sqrt{\omega_\text{cav}^2 + 4D\omega_\text{cav}}$, while in Eq. \eqref{equations_motion_classical_velocities1}  the shifted cavity frequency is $\sqrt{\omega_\text{cav}^2 + 4D\omega_\text{cav} - 4 g_{\scriptscriptstyle{\text{QED}}}^2\frac{\omega_\text{cav}}{\omega_\text{mat}}}$ (shifts of the matter excitation are discussed in Sec. \ref{Appendix_alternative_models} of the Supplementary Material). In the following, we use the term dressed cavity/excitation (or dressed/renormalized frequency) when the shift is not zero and thus the value of the shifted frequency does not coincide with the original value $\omega_\text{cav}$ before coupling (notice that $\sqrt{\omega_\text{cav}^2 + 4D\omega_\text{cav}}= \omega_\text{cav}$ when using Eq. \eqref{equations_motion_classical_position1} with $D= 0$ and $\sqrt{\omega_\text{cav}^2 + 4D\omega_\text{cav} - 4 g_{\scriptscriptstyle{\text{QED}}}^2\frac{\omega_\text{cav}}{\omega_\text{mat}}}= \omega_\text{cav}$ when using Eq. \eqref{equations_motion_classical_velocities1} with $D= g_{\scriptscriptstyle{\text{QED}}}^2/\omega_\text{mat} $, so that in these two cases we will refer to bare cavity frequencies).

 In nanophotonics, coupled harmonic oscillator equations have been widely used to fit data without considering frequency renormalization so we adhere to this procedure, i.e. we consider harmonic oscillator models where the frequency of the cavity and matter excitations are the bare ones.
 This approach gives preference to the model with coupling constant proportional to the oscillation amplitude (Eq. \eqref{equations_motion_classical_position}) or to its derivative (Eq. \eqref{equations_motion_classical_velocities}), depending on the value of $D$, as discussed next. Thus, throughout the remaining of this paper (including the Supplementary Material unless otherwise stated), we analyze these two preferred models using bare frequencies. We denote these models the Spring Coupling (SpC) model and the Momentum Coupling (\MC{}) model, respectively. Other models are discussed in Sec. \ref{Appendix_alternative_models} and summarized in {Sec. \ref{Appendix_Comparison} of the Supplementary Material. Additionally, Sec. \ref{Appendix_equationsofmotion} of the Supplementary Material details how to obtain the classical coupled harmonic oscillator equations directly from the classical electromagnetic Lagrangian.

\subsection{Spring Coupling (SpC) model} \label{subsec_spring_coupling}

We consider first a system without diamagnetic term, $D = 0$. This choice is appropriate, for example, when the interaction between the emitter and cavity excitations is mediated by Coulomb coupling, as discussed in more detail in Sec. \ref{subsec_molecule_nanoparticle}.
Eq. \eqref{equations_motion_classical_position} then becomes

\begin{subequations}
\begin{equation}
\ddot{x}_\text{cav}  + \omega_\text{cav}^2  x_\text{cav}  + 2g_{\scriptscriptstyle{\text{SpC}}} \sqrt{\omega_\text{cav}\omega_\text{mat}}  x_\text{mat} =0, 
\end{equation}
\begin{equation}
 \ddot{x}_\text{mat}  + \omega_\text{mat}^2 x_\text{mat}  + 2g_{\scriptscriptstyle{\text{SpC}}} \sqrt{\omega_\text{cav}\omega_\text{mat}} x_\text{cav}  = 0,
\end{equation}
\label{equations_motion_classical_position_D0}
\end{subequations}
where the coupling is proportional to the classical oscillation amplitudes $x_\text{cav}$ and $x_\text{mat}$ and we have changed the notation $g_{\scriptscriptstyle{\text{SpC}}}=g_{\scriptscriptstyle{\text{QED}}}$ (using a different symbol for the coupling strength in the classical and quantum descriptions becomes useful in Sec. \ref{eq:MCmodel}).  The $\sqrt{\omega_\text{cav}\omega_\text{mat}}$ prefactor appears directly from the Hamiltonian, and ensures that $g$ have units of frequency. Other choices of prefactor have been used (such as using the arithmetic mean of the bare frequencies instead of the geometric mean \cite{autore18}) which are equivalent in the strong coupling regime but not in the ultrastrong one. In the frequency ($\omega$) domain these equations are transformed to
\begin{subequations}
\begin{equation}
-\omega^2 {x}_\text{cav}  + \omega_\text{cav}^2  x_\text{cav}  + 2g_{\scriptscriptstyle{\text{SpC}}} \sqrt{\omega_\text{cav}\omega_\text{mat}}  x_\text{mat} =0, 
\end{equation}
\begin{equation}
-\omega^2 x_\text{mat} + \omega_\text{mat}^2 x_\text{mat}  + 2g_{\scriptscriptstyle{\text{SpC}}} \sqrt{\omega_\text{cav}\omega_\text{mat}} x_\text{cav}  = 0.
\end{equation}
\label{equations_motion_classical_position_D0_w}
\end{subequations}
We refer to Eqs. \eqref{equations_motion_classical_position_D0} and \eqref{equations_motion_classical_position_D0_w} as the Spring Coupling (SpC) model because they are analogous to the equations describing the movement of two coupled springs (sketch in Fig. \ref{figure_models}a) (we emphasize that we could also describe the same physics of ultrastrongly coupled systems by setting $D=0$ in  Eq. \eqref{equations_motion_classical_velocities}, but, in this case, the dressed frequency $\sqrt{\omega_\text{cav}^2  - 4 g_{\scriptscriptstyle{\text{SpC}}}^2\frac{\omega_\text{cav}}{\omega_\text{mat}}}$ would appear in the equations instead of the bare one, contrary to our previous choice). The eigenfrequencies $\omega_{\pm,\scriptscriptstyle{\text{SpC}}}$ of the SpC model are obtained by diagonalizing the matrix associated with Eq. \eqref{equations_motion_classical_position_D0_w}, which leads to 

\begin{equation}
\omega_{\pm,\scriptscriptstyle{\text{SpC}}} = \frac{1}{\sqrt{2}}  \sqrt{\omega_\text{cav}^2 + \omega_\text{mat}^2 \pm \sqrt{(\omega_\text{cav}^2-\omega_\text{mat}^2)^2+16 g_{\scriptscriptstyle{\text{SpC}}}^2 \omega_\text{cav}\omega_\text{mat}}}. \label{eigenfrequencies_SpC}
\end{equation}

We note that the frequencies given by Eq. \eqref{eigenfrequencies_SpC} correspond to the energy difference between the first excited and ground state, and not to the absolute values of the eigenfrequencies themselves. This distinction is not necessary in classical descriptions that set the energy of the ground state to zero (or a fixed value). However, the cavity-QED model indicates a $g_{\scriptscriptstyle{\text{QED}}}$-dependent shift of the ground-state energy from zero, which is a fully quantum phenomenon. The information of this shift is lost when we take the expectation value of the operators $ \langle \hat{x}_\text{cav} \rangle$ and $ \langle \hat{x}_\text{mat} \rangle$ in Eq. \eqref{equations_motion_quantum_position}. The $g_{\scriptscriptstyle{\text{QED}}}$ dependence of this shift can be found, for instance, in Fig. 2f of Ref. \cite{kockum19}.

\subsection{Momentum Coupling (\MC{}) model} \label{subsec_MoC_coupling}
\label{eq:MCmodel}
For a diamagnetic term with $D = \frac{g_{\scriptscriptstyle{\text{QED}}}^2}{\omega_\text{mat}}$ (this value normally appears in atomic physics and in cavity-QED models\cite{ciuti05} in the Coulomb Gauge and is discussed in Ref. \cite{kockum19} and Sec. \ref{subsec_molecule_dielectric}),  Eq. \eqref{equations_motion_classical_velocities} takes the form

\begin{subequations}
\begin{equation}
\ddot{x}_\text{cav}  + \omega_\text{cav}^2  x_\text{cav}  - 2g_{\scriptscriptstyle{\text{\MC{}}}}    \dot{x}_\text{mat} =0, 
\label{equations_motion_classical_velocities_Da}
\end{equation}
\begin{equation}
\ddot{x}_\text{mat}  + \omega_\text{mat}^2 x_\text{mat} + 2g_{\scriptscriptstyle{\text{\MC{}}}}   \dot{x}_\text{cav}   = 0,
\label{equations_motion_classical_velocities_Db}
\end{equation}
\label{equations_motion_classical_velocities_D}
\end{subequations}
 with the coupling term proportional to the time derivative of the oscillation amplitudes (the 'velocities') so that we call this model the Momentum coupling (\MC{}) model (sketch in Fig. \ref{figure_models}b).  The coupling strength $g_{\scriptscriptstyle{\text{\MC{}}}}$ in these equations is related to the constant $g_{\scriptscriptstyle{\text{QED}}}$ in the cavity-QED Hamiltonian as  $g_{\scriptscriptstyle{\text{\MC{}}}}=\sqrt{\frac{\omega_\text{cav}}{\omega_\text{mat}}}g_{\scriptscriptstyle{\text{QED}}}$ (and thus $D = \frac{g_{\scriptscriptstyle{\text{QED}}}^2}{\omega_\text{mat}} = \frac{g_{\scriptscriptstyle{\text{\MC{}}}}^2}{\omega_\text{cav}}$). We introduce this new coupling strength because, in this way, i) Eqs. \eqref{equations_motion_classical_velocities_Da} and \eqref{equations_motion_classical_velocities_Db} take the same form as in previous work \cite{wu10,liu09,barraburillo21} and ii) $g_{\scriptscriptstyle{\text{\MC{}}}}$ becomes independent of the resonant frequencies $\omega_\text{mat}$ and $\omega_\text{cav}$ for the systems studied in Sec. \ref{sec_examples}. However, it is also possible to write  Eqs. \eqref{equations_motion_classical_velocities_Da} and \eqref{equations_motion_classical_velocities_Db} in terms of $g_{\scriptscriptstyle{\text{QED}}}$ as long as one is consistent in all the derivation. We further emphasize that $g_{\scriptscriptstyle{\text{\MC{}}}}=g_{\scriptscriptstyle{\text{QED}}}$ in resonance ($\omega_\text{cav}=\omega_\text{mat}$), and these two parameters only take significantly different values for strong detuning. In the frequency domain Eq. \eqref{equations_motion_classical_velocities_D} becomes
 \begin{subequations}
\begin{equation}
-\omega^2 {x}_\text{cav}  + \omega_\text{cav}^2  x_\text{cav}  + 2i \omega g_{\scriptscriptstyle{\text{\MC{}}}}    {x}_\text{mat} =0, 
\label{equations_motion_classical_velocities_D_wa}
\end{equation}
\begin{equation}
-\omega^2 {x}_\text{mat}  + \omega_\text{mat}^2 x_\text{mat} - 2i\omega g_{\scriptscriptstyle{\text{\MC{}}}}   {x}_\text{cav}   = 0,
\label{equations_motion_classical_velocities_D_wb}
\end{equation}
\label{equations_motion_classical_velocities_D_w}
\end{subequations}
 and the corresponding eigenfrequencies are
\begin{equation}
\omega_{\pm,\scriptscriptstyle{\text{\MC{}}}} = \frac{1}{\sqrt{2}}\sqrt{\omega_\text{cav}^2 + \omega_\text{mat}^2 + 4g_{\scriptscriptstyle{\text{\MC{}}}}^2 \pm \sqrt{(\omega_\text{cav}^2 + \omega_\text{mat}^2 + 4g_{\scriptscriptstyle{\text{\MC{}}}}^2)^2 - 4\omega_\text{cav}^2\omega_\text{mat}^2}}. \label{eigenfrequencies_MC}
\end{equation}

Although the \MC{} is used to describe the coupling between matter excitations and cavity modes (Fig. \ref{figure_models}b), we are not aware of any equivalent mechanical system in classical mechanics that follows the equations of motion in Eqs. \eqref{equations_motion_classical_velocities_Da} and \eqref{equations_motion_classical_velocities_Db} (with coupling terms proportional to the time derivatives of the oscillation amplitude, similarly to friction terms but describing the interaction between two different oscillators). This is in contrast to the SpC model where the equivalent system, composed of masses and springs, is shown in Fig. \ref{figure_models}a.

\subsection{Comparison of the \MC{} and SpC models}
\label{sec:comparison}

 As mentioned above, the \MC{} and SpC models are appropriate when $D = g_{\scriptscriptstyle{\text{QED}}}^2/\omega_\text{mat}$ and $D =0$, respectively (we emphasize again that the resonant frequencies $\omega_\text{cav}$, $\omega_\text{mat}$ in these models are the bare resonant frequencies). Further, regardless of whether the diamagnetic term should be included in the description or not, these models have been used in the past as phenomenological tools for extracting coupling parameters by fitting the spectra of the coupled system obtained from experimental data or simulations \cite{vasa10,symonds08,chikkaraddy16,wang16,zheng17,stuhrenberg18,autore18, wu10,liu09,rodriguez14,barraburillo21}. In this section, we compare the results provided by both models as a function of the coupling strength.
 
The \MC{} and SpC models are known to give very different results for $g \gg 0.1 \omega_\text{mat}$, as we briefly illustrate in this section (we use $g$ in this subsection  to refer to $g_{\scriptscriptstyle{\text{SpC}}}$ or $g_{\scriptscriptstyle{\text{\MC{}}}}$ in discussions that are valid for both models). 
Figure \ref{figure_models}c compares the eigenfrequencies of the SpC (blue solid line) and \MC{} (red dashed line) models for $g=0.1\omega_\text{mat}$, as given by Eqs. \eqref{eigenfrequencies_SpC} and \eqref{eigenfrequencies_MC}, respectively. For simplicity, we consider that the coupling strength is the same for all values of $\omega_\text{cav}$ (a different parameter choice is discussed in Sec. \ref{Appendix_eigenvalues-nonconstantg} of the Supplementary Material). The eigenfrequencies of the hybrid modes $\omega_\pm$ are calculated as a function of the bare cavity frequency  $\omega_\text{cav}$, with the bare $\omega_\text{mat}$ frequency fixed (all frequencies are normalized by  $\omega_\text{mat}$, so that the figures are independent of the value of this parameter).   The eigenfrequencies obtained within the \MC{} and SpC models follow a nearly identical dependence on $\omega_\text{cav}$, and the agreement is even better for $g < 0.1 \, \omega_\text{mat}$. Thus, when analyzing systems not in the ultrastrong coupling regime, the two models can generally be used interchangeably with minimal impact on the results, although exceptions can exist \cite{nodar23}.

In contrast, the choice of the model is crucial for even larger coupling strengths, such that the system is well into the ultrastrong coupling regime.  The differences between the two models are illustrated in Fig. \ref{figure_models}d for coupling strength $g=0.3 \, \omega_\text{mat}$. In this case, the two models predict significantly different eigenfrequencies of the coupled system. The difference is smaller for larger cavity frequencies, $\omega_\text{cav} \gg \omega_\text{mat}$, because the oscillators become uncoupled and the eigenfrequencies approach the bare frequencies $\omega_\text{cav}$ and $\omega_\text{mat}$ in the two models. However, even for a relatively large $\frac{\omega_\text{cav}}{\omega_\text{mat}}  = 1.5$, the difference between the values of $\omega_\pm$ according to the two models is around 10$\%$. 

We compare next the splitting $\Omega = \omega_+ - \omega_-$  between the two eigenmodes at zero detuning, $\omega_\text{cav} = \omega_\text{mat}$. In the \MC{} model, the splitting equals twice the coupling strength, i.e., $\Omega = 2g$, which is the minimum splitting in this model \cite{khitrova06,bellessa04}. On the other hand, in the SpC model, the relation between $\Omega$ and the coupling strength for zero detuning is
\begin{equation}
\Omega_{\scriptscriptstyle{\text{SpC}}} = \omega_{+,\scriptscriptstyle{\text{SpC}}} - \omega_{-,\scriptscriptstyle{\text{SpC}}} = \omega_\text{mat} \left( \sqrt{ 1 + \frac{2g_{\scriptscriptstyle{\text{SpC}}}}{\omega_\text{mat}}} - \sqrt{ 1 - \frac{2g_{\scriptscriptstyle{\text{SpC}}}}{\omega_\text{mat}}}\right).
\end{equation}
We find $\Omega_{\scriptscriptstyle{\text{SpC}}} = 2.11g_{\scriptscriptstyle{\text{SpC}}} $ for the values used in Fig. \ref{figure_models}d. Further,  according to the SpC model, the minimum splitting between the branches does not happen at zero detuning but at cavity frequencies larger than the matter excitation frequencies. To further emphasize the difference between the models, Fig. \ref{figure_models}e shows the minimum splitting as a function of coupling strength, with a linear dependence for the \MC{}  (red solid line) model, $\Omega^\text{min} = 2g$, in contrast with the strong deviation from nonlinearity of the SpC model results (blue line) for  $g \gg 0.1 \omega_\text{mat}$. As a consequence, close to the so-called deep strong coupling regime $\left( \frac{g}{\omega_\text{mat} } \approx 1 \right)$, $\Omega^\text{min}_{\scriptscriptstyle{\text{SpC}}}$ is approximately twice that of the \MC{} model.

Last, Fig. \ref{figure_models}d shows important differences at small cavity frequencies, $\omega_\text{cav} \ll \omega_\text{mat}$. The dispersion of the \MC{} model shows two hybrid modes for all values of the detuning,  with the lower mode frequency $\omega_{-,\scriptscriptstyle{\text{\MC{}}}}$ tending towards  $\omega_\text{cav}$ for decreasing value of $\omega_\text{cav}$. In contrast, for the SpC model the lower mode ceases to exist ($\omega_{-,\scriptscriptstyle{\text{SpC}}}$  becomes imaginary) under the condition $\frac{\omega_\text{cav}}{\omega_\text{mat}}< \left(\frac{2g_{\scriptscriptstyle{\text{SpC}}}}{\omega_\text{mat}}\right)^2$ (for fixed $g_{\scriptscriptstyle{\text{SpC}}}=0.3\omega_\text{mat}$; see Sec. \ref{Appendix_eigenvalues-nonconstantg} of the Supplementary Material where a different choice is discussed). Further, in the SpC description, the upper branch approaches the bare matter frequency at $\omega_\text{cav}\rightarrow 0$, but this is not the case in the \MC{} model, where the corresponding asymptotic limit is
$\omega_{+,\scriptscriptstyle{\text{\MC{}}}} = \sqrt{\omega_\text{mat}^2+4g_{\scriptscriptstyle{\text{\MC{}}}}^2}$. Thus, in the \MC{} model, the coupling affects the upper hybrid mode even in this highly detuned situation.

The two models' different asymptotic limits of the upper branch determine the predicted range of energies where hybrid modes can exist. The \MC{} results show a frequency  band between $\omega_\text{mat}$ and $\sqrt{\omega_\text{mat}^2+4g_{\scriptscriptstyle{\text{\MC{}}}}^2}$ with no modes available. This forbidden band is not present in the SpC dispersion. In Sec. \ref{sec_examples} and Sec. \ref{Appendix_reststrahlen} of the Supplementary Material,  we connect this result with the Reststrahlen band of polar materials and  show that we can reproduce the experimental dispersion of these materials by using the \MC{} \cite{yoo21} and alternative models but not the SpC model.

\begin{figure}[H]
\centering
\includegraphics[scale=0.26]{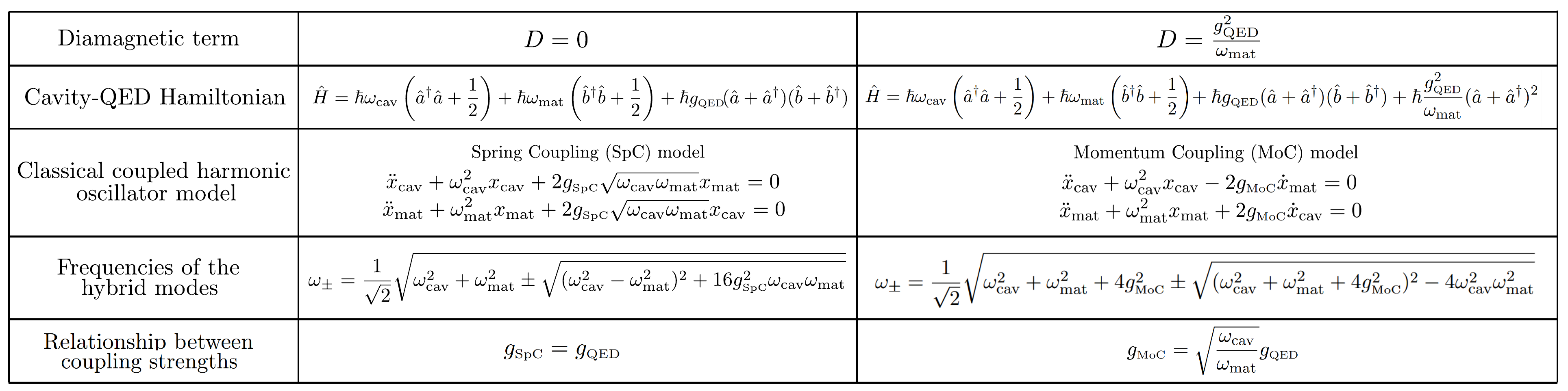}
\caption*{Table 1: Summary of the correspondences of the classical  SpC and \MC{} models with the cavity-QED description without diamagnetic term $D=0$ (second column) and with diamagnetic term and $D=\frac{g_{\scriptscriptstyle{\text{\MC{}}}}^2}{\omega_\text{cav}}$ (third column). The second row shows the two considered cavity-QED Hamiltonians. The third row indicates the equations of motion of the oscillation amplitudes $x_\text{cav}$ and $x_\text{mat}$ obtained with the classical SpC (second column) and \MC{} (third column)  harmonic oscillator models. The fourth row provides the frequencies of the two resulting hybrid modes, which are the same for the cavity-QED and classical models for the value of $D$ and choice of coupled harmonic oscillator model indicated in each column. The last row indicates the relationship between the coupling constant $g_{\scriptscriptstyle{\text{QED}}}$ in the cavity-QED Hamiltonian and those in the classical coupled harmonic oscillator models ($g_{\scriptscriptstyle{\text{\MC{}}}}$ and $g_{\scriptscriptstyle{\text{SpC}}}$). For the system in Sec. \ref{subsec_molecule_dielectric}, $g_{\scriptscriptstyle{\text{\MC{}}}}$ is constant and thus $g_{\scriptscriptstyle{\text{QED}}}\propto\sqrt{\omega_\text{mat}/\omega_\text{cav}}$. } 
\end{figure}

The connection between classical and quantum models is summarized in Table 1. The classical SpC and \MC{} models result in the same eigenfrequencies as cavity-QED  Hamiltonians without the diamagnetic term ($D=0$) and with  $D = \frac{g_{\scriptscriptstyle{\text{QED}}}^2}{\omega_\text{mat}}=\frac{g_{\scriptscriptstyle{\text{\MC{}}}}^2}{\omega_\text{cav}}$, respectively. Other classical coupled harmonic oscillator models where dressed frequencies are used instead of the bare ones (with an associated change of the coupling term) are discussed in Sec. \ref{Appendix_alternative_models} of the SI. For completeness, we also discuss in Sec. \ref{Appendix_linearized} of the Supplementary Material an often-used linearized model that is a good approximation to the \MC{} and SpC models for low coupling strengths (especially for the anticrossing region of the spectrum corresponding to small detunings). However, this linearized model is not appropriate in the ultrastrong coupling regime.

At this point, we have discussed the connections between a general quantum description and classical equations of motion. However, we still need to determine how to choose between the \MC{} and SpC models for a given system (or equivalently, whether the Hamiltonian has $D\neq0$ or $D=0$).
In the next section, we consider three representative systems to explore this question and highlight the key role played by the nature of the matter-cavity interaction (Coulomb coupling or coupling with transverse electromagnetic modes in dielectric cavities).

Additionally, we have focused thus far on the eigenfrequencies, which can be extracted directly from the equations of coupled harmonic oscillators without needing an exact understanding of what the oscillation amplitudes $x_\text{cav}$ and $x_\text{mat}$  represent. However, a clear physical interpretation of these parameters is necessary to evaluate magnitudes of interest, such as the electric field at a given location inside or outside the optical cavity. In Sec. \ref{sec_examples},  we also address how $x_\text{cav}$ and $x_\text{mat}$  relate to relevant physical quantities in the representative systems of choice. 

\section{Physical observables from classical models in configurations of interest} \label{sec_examples}

 We analyze in this section the three canonical nanophotonics systems introduced in Fig. \ref{figure_intro}, for which different cavity-QED Hamiltonians (with and without the diamagnetic term) are appropriate. In Sec. \ref{subsec_molecule_dielectric}, we focus on the textbook case of a single molecular emitter (or another quantum emitter) interacting with transverse electromagnetic modes of the dielectric Fabry-Pérot cavity in Fig. \ref{figure_intro}a (in transverse modes, the fields are perpendicular to the wavevector in all Fourier components). As a second example, we analyze in Sec. \ref{subsec_molecule_nanoparticle} a molecular emitter close to a small metallic nanoparticle (Fig. \ref{figure_intro}b), where the coupling is governed by Coulomb interactions (the fields mediating this interaction are longitudinal, i.e. parallel to the wavevector in all Fourier components). The last example (Sec. \ref{subsec_bulk_dielectric}) consists of an ensemble of molecular emitters (representing a bulk material)  inside a Fabry-Pérot cavity (Fig. \ref{figure_intro}c), where the molecules couple with a transverse electromagnetic mode of the cavity, and also interact with each other through Coulomb coupling.

\subsection{A quantum emitter interacting with a transverse mode of a dielectric cavity} \label{subsec_molecule_dielectric}

We consider first a dipole interacting with a single transverse mode of a resonant dielectric cavity (Figs. \ref{figure_intro}a and \ref{figure_dipole_cavity}a). The dipole is associated with matter excitations, and it can represent an excitonic transition of a molecule or quantum dot or a transition between vibrational states, for example. For concreteness, we consider the coupling with a molecular emitter in the following. Cavity-QED models of this system have successfully described phenomena such as the modification of the spontaneous emission rate of the emitter \cite{purcell46,pelton15}, of the photon statistics of the emitted light \cite{mckeever03,saezblazquez18,nodar23} or of the coherence time of the quantum states \cite{thorwart00}.

The whole derivation of the equations of motion of the classical variables within the Coulomb gauge is discussed in the Supplementary Material (Sec. \ref{Appendix_equationsofmotion}), but we summarize it in the following. We represent the molecular emitter as two point charges with relative position $\mathbf{l}$ (forming a dipole), which couple through Coulomb interactions determined by the potential $V_\text{Cou}(l)$ approximated as a harmonic one,  $V_\text{Cou}(l) = \frac{1}{2} m_\text{red} \omega_\text{mat}^2 l^2 $, with $l = |\mathbf{l}|$ the distance and $m_\text{red}$ the reduced mass of this two-body system. The dipole moment induced in the molecular emitter is $\mathbf{d} = q \mathbf{l}$, where $q$ is the absolute value of the charge of the particles in the dipole. On the other hand, the cavity mode is characterized by the vector potential $\mathbf{A}$, which is the canonical position variable of the transverse electromagnetic fields \cite{cohentannoudji97}. This description does not include non-linear effects,  being thus valid for harmonic molecular vibrations, and also for anharmonic vibrations or excitonic transitions under weak illumination.

In Cavity-QED models, the standard approach to describe light-matter interactions in this system is to use the minimal-coupling classical Hamiltonian in the Coulomb gauge of the form $H_\text{min-c} = \frac{q^2(\mathbf{p}-\mathbf{A})^2}{2m_\text{red}}$, where $\mathbf{p}$ is the classical canonical momentum of the dipolar matter excitation. To obtain the quantum Hamiltonian, we use the following quantization relations \cite{hopfield58,cohentannoudji97}:
\begin{equation}
\hat{A}(\mathbf{r}) = \sqrt{\frac{\hbar}{2\omega_\text{cav} \varepsilon_0 V_\text{eff}}} \Xi(\mathbf{r})(\hat{a} + \hat{a}^\dagger), \label{quantization_A}
\end{equation}
\begin{equation}
\hat{\Pi}(\mathbf{r}) = -i\sqrt{\frac{\hbar \omega_\text{cav} \varepsilon_0 V_\text{eff}}{2}} \Xi(\mathbf{r})(\hat{a} - \hat{a}^\dagger), \label{quantization_Pi}
\end{equation}
\begin{equation}
\hat{d} = \sqrt{\frac{\hbar f_\text{mat}}{2\omega_\text{mat}}}(\hat{b}+\hat{b}^\dagger), \label{quantization_d}
\end{equation} 
\begin{equation}
\hat{p} = -i\sqrt{\frac{\hbar  \omega_\text{mat} }{2 f_\text{mat}}}(\hat{b}-\hat{b}^\dagger), \label{quantization_pmat}
\end{equation}
where $\hat{\Pi}(\mathbf{r})  $ is the canonical momentum associated to the vector potential $\hat{A}(\mathbf{r})$ (see Secs. \ref{Appendix_equationsofmotion}, \ref{Appendix_alternative_models} of the Supplementary Material). The function $\Xi(\mathbf{r})$ accounts for the spatial distribution of the vector potential of the cavity mode and is normalized so that its maximum value is 1. Further, we have introduced the effective mode volume of the cavity field\cite{esteban14}, $V_\text{eff}$, and the oscillator strength of the dipolar excitation $f_\text{mat} = \frac{q^2}{m_\text{red}}$. From the minimal-coupling Hamiltonian $H_\text{min-c}$, the light-matter interaction term is $H_\text{int} = \frac{-q^2 \mathbf{p}\cdot \mathbf{A}}{m_\text{red}}$. Considering that the induced dipole moment and the vector potential form an angle $\theta$, and using Eqs. \eqref{quantization_A} and \eqref{quantization_pmat}, the interaction term of the quantized Hamiltonian becomes
\begin{equation}
\hat{H}_\text{int} = i \hbar \frac{1}{2}  \sqrt{\frac{f_\text{mat}}{\varepsilon_0 V_\text{eff}}}\sqrt{\frac{\omega_\text{mat}}{\omega_\text{cav}}} \Xi(\mathbf{r})  \cos \theta (\hat{a} + \hat{a}^\dagger)(\hat{b}-\hat{b}^\dagger).
\end{equation}
Comparing this expression with the third term of the Hopfield Hamiltonian (Eq. \eqref{quantum_hamiltonian_2}), we directly obtain that the coupling strength in the cavity-QED formalism is $g_{\scriptscriptstyle{\text{QED}}} = \frac{1}{2}\sqrt{\frac{f_\text{mat}}{\varepsilon_0 V_\text{eff}}}\sqrt{\frac{\omega_\text{mat}}{\omega_\text{cav}}}  \Xi(\mathbf{r})  \cos \theta$. We consider from now on the maximum coupling strength  $g_{\scriptscriptstyle{\text{QED}}} = \frac{1}{2}\sqrt{\frac{f_\text{mat}}{\varepsilon_0 V_\text{eff}}}\sqrt{\frac{\omega_\text{mat}}{\omega_\text{cav}}}$, which is achieved in the position of the maximum field $(\Xi(\mathbf{r}) = 1)$ for optimal orientation ($\theta = 0$). Further, the $\mathbf{A}^2$ term in the minimal-coupling Hamiltonian leads to a diamagnetic term (fourth term on the right-hand side of Eq. \eqref{quantum_hamiltonian_2}) with $D = \frac{g_{\scriptscriptstyle{\text{QED}}}^2}{\omega_\text{mat}}$. Following the discussion of Sec. \ref{sec_quantum},  the presence of the diamagnetic term in the cavity-QED Hamiltonian with this exact value of $D$ indicates that this system can be described by the classical \MC{} model in Eq. \eqref{equations_motion_classical_velocities_D}.

Next, we use the connection between the classical and cavity-QED approaches to illustrate the procedure to obtain the value of physical observables from the classical oscillation amplitudes of the cavity $x_\text{cav}$ and of the molecular excitation $x_\text{mat}$. The classical coupling strength $g_{\scriptscriptstyle{\text{\MC{}}}}$ is directly obtained from the quantum value as
$g_{\scriptscriptstyle{\text{\MC{}}}} = \sqrt{\frac{\omega_\text{cav}}{\omega_\text{mat} }} g_{\scriptscriptstyle{\text{QED}}} =  \frac{1}{2}\sqrt{\frac{f_\text{mat}}{\varepsilon_0 V_\text{eff}}} $. Further, the quantum position operators ($\propto \hat{a} + \hat{a}^\dagger$ and $\propto \hat{b} + \hat{b}^\dagger$)  and the classical oscillation amplitudes  ($x_\text{cav}$ and  $x_\text{mat}$) are related by the standard quantization relationship
\begin{subequations}
\begin{equation}
\text{Re}(x_\text{cav}) = \langle \hat{x}_\text{cav} \rangle = \sqrt{\frac{\hbar}{2\omega_\text{cav}}}\langle \hat{a}+\hat{a}^\dagger\rangle, \label{quantization_xcav}
\end{equation}
\begin{equation}
\text{Re}(x_\text{mat}) = \langle \hat{x}_\text{mat} \rangle = \sqrt{\frac{\hbar}{2\omega_\text{mat}}}\langle \hat{b}+\hat{b}^\dagger\rangle, \label{quantization_xmat}
\end{equation}
\label{quantization_xcav_xmat}
\end{subequations}
where, for an appropriate comparison between classical amplitudes and quantum operators, the real part of the oscillator amplitudes must be taken: $\Re(x_\text{cav})=\Re(|x_\text{cav}| e^{-i\omega t+\phi}) \propto |x_\text{cav}| \cos(\omega t+\phi)$, with $\phi$ a phase. 
Equation \eqref{quantization_xcav} and Eq. \eqref{quantization_A} indicate that the oscillation amplitude $x_\text{cav}$ in the \MC{} model (Eq. \eqref{equations_motion_classical_velocities_D}) is given by $x_\text{cav} = \mathcal{A} \sqrt{\varepsilon_0 V_\text{eff}}$, where $\mathcal{A}$ is the maximum amplitude of the classical vector potential (i.e. in the position where $\Xi(\mathbf{r}) = 1$). Therefore, the oscillation amplitude $x_\text{cav}$ can be used to calculate the spatial distribution of this potential as $A(\mathbf{r}) = \mathcal{A} \Xi(\mathbf{r}) = \frac{x_\text{cav}}{\sqrt{\varepsilon_0 V_\text{eff}}} \Xi(\mathbf{r})$ ($A(\mathbf{r}) = \langle \hat{A}(\mathbf{r}) \rangle$ is the classical counterpart of the quantum operator of the vector potential).  Equivalently, from Eqs. \eqref{quantization_xmat} and \eqref{quantization_d}, the amplitude of the oscillator corresponding to the matter excitation is directly connected with the induced classical dipole moment ($d = |\mathbf{d}|) $ as $x_\text{mat} = \frac{d}{\sqrt{f_\text{mat}}}$. These relations are schematically shown in Fig. \ref{figure_dipole_cavity}a.

\begin{figure}[t!]
\centering
\includegraphics[scale=0.66]{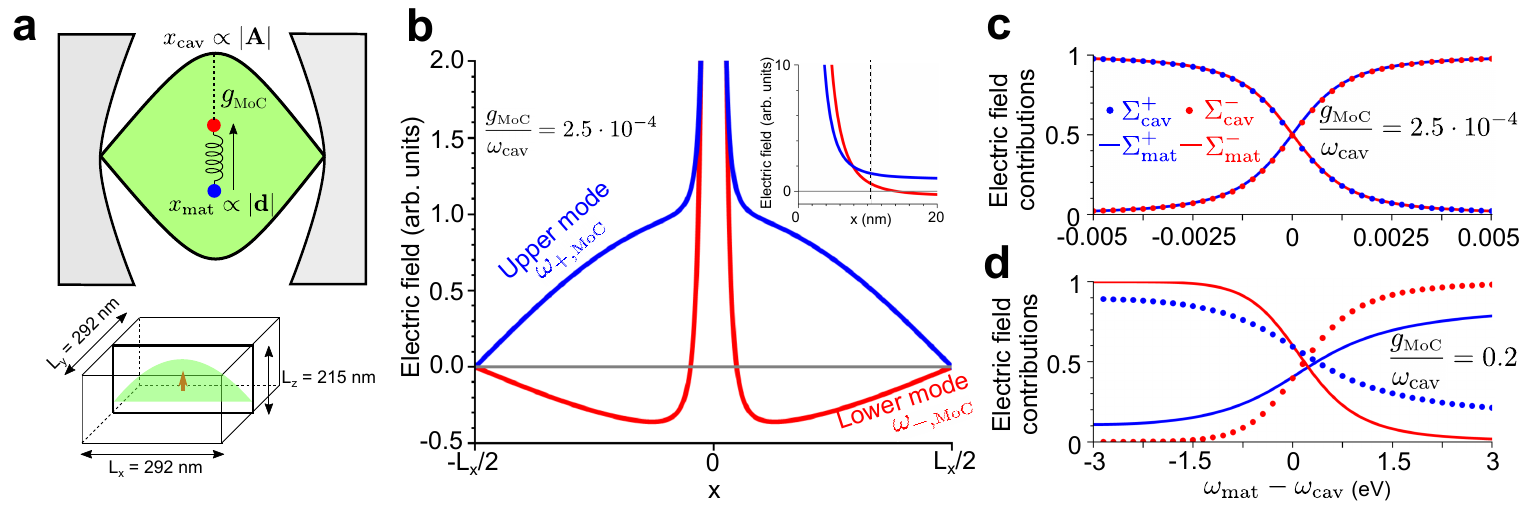}
\caption{Interaction of a quantum emitter with a transverse cavity mode within the classical \MC{} model. a) Schematics of the system. The two oscillators are associated with the vector potential
$\mathbf{A}$ of the cavity mode and the induced dipole moment $\mathbf{d}$ of the excitation in the quantum emitter, which we consider to be a molecule. 
The oscillators are coupled with each other with strength
$g_{\scriptscriptstyle{\text{\MC{}}}}$. The bottom sketch indicates the cavity dimensions that we analyze in the rest of the panels. The emitter is placed at the center of the cavity. The green shaded areas in the sketches represent the field distribution of the cavity mode. b) Spatial distribution of the electric field for the upper (blue) and the lower  (red) hybrid modes at frequencies $\omega_{+,\scriptscriptstyle{\text{\MC{}}}}$ and $\omega_{-,\scriptscriptstyle{\text{\MC{}}}}$, respectively, for coupling strength $g_{\scriptscriptstyle{\text{\MC{}}}} = 2.5 \cdot 10^{-4} \omega_\text{cav}$. The electric field is calculated along the cavity axis (along the $x$ direction in panel (a), with $x = y = z = 0$ corresponding to the cavity center). The inset is a zoom of the region near the emitter. c) Contribution to the electric field from the cavity $\Sigma_\text{cav}^\pm$ (dots) and from the emitter $\Sigma_\text{mat}^\pm$ (solid lines), for the hybrid mode at frequency $\omega_{+,\scriptscriptstyle{\text{\MC{}}}}$ (blue) and the hybrid mode at frequency $\omega_{-,\scriptscriptstyle{\text{\MC{}}}}$ (red), as a function of the detuning $\omega_\text{mat} - \omega_\text{cav}$. The fields are real and are evaluated at the position $(x,y,z)$ = (10.5 nm, 0, 0), i.e., at 10.5 nm distance from the center of the cavity where the molecular emitter is located (see sketch in (a) for directions), which corresponds to the position indicated by the dashed line in the inset of panel (b). The coupling strength is $g_{\scriptscriptstyle{\text{\MC{}}}} = 2.5 \cdot 10^{-4} \omega_\text{cav}$. d) Same as in (c), for $g_{\scriptscriptstyle{\text{\MC{}}}} = 0.2 \, \omega_\text{cav}$.
\label{figure_dipole_cavity} }
\end{figure} 

We are finally in conditions to obtain the value of physical observables such as the electric field from the classical harmonic \MC{} model. We first consider the spatial distribution of the electric fields of each hybrid mode.
The transverse cavity mode field (given by $\mathbf{A}(\mathbf{r},t)$) must be added to the longitudinal near field induced by the induced dipole\footnote{To satisfy the boundary conditions in a closed cavity, additional terms due to image dipoles should be included. However,  we neglect these terms for simplicity since their contribution is typically small compared to the near field of the dipole $\propto \frac{1}{r^3}$ and of the field of the cavity mode. }, which is obtained from the scalar Coulomb potential
\begin{equation}
V_\text{Cou}(\mathbf{r},t) = \frac{1}{4\pi\varepsilon_0} \frac{d(t) \mathbf{n_d}\cdot \mathbf{n_r}}{|\mathbf{r}|^2},
\end{equation}
with unit vectors $\mathbf{n_d} = \frac{\mathbf{d}}{|\mathbf{d}|}$ and $\mathbf{n_r} = \frac{\mathbf{r}}{|\mathbf{r}|}$. The total electric field is therefore given as 
\begin{equation}
\mathbf{E}(\mathbf{r},t) = -\nabla V_\text{Cou}(\mathbf{r},t) - \frac{\partial \mathbf{A}(\mathbf{r},t)}{\partial t},
\label{eq_relation_electric_field_potentials}
\end{equation}
and the electric field at frequencies $\omega_{\pm,\scriptscriptstyle{\text{\MC{}}}}$ of each hybrid mode (given by Eq. \eqref{eigenfrequencies_MC}) corresponds to
\begin{align}
\mathbf{E}(\mathbf{r},\omega_{\pm,\scriptscriptstyle{\text{\MC{}}}}) =& \frac{3(\mathbf{n_d}\cdot \mathbf{n_r}) \mathbf{n_r} - \mathbf{n_d}}{4\pi \varepsilon_0 r^3} d(\omega_{\pm,\scriptscriptstyle{\text{\MC{}}}}) + i\omega_{\pm,\scriptscriptstyle{\text{\MC{}}}} A(\mathbf{r},\omega_{\pm,\scriptscriptstyle{\text{\MC{}}}}) \mathbf{n_A} \nonumber  \\
=& \underbrace{\frac{3(\mathbf{n_d}\cdot \mathbf{n_r}) \mathbf{n_r} - \mathbf{n_d}}{4\pi \varepsilon_0 r^3} \sqrt{f_\text{mat}} x_\text{mat}(\omega_{\pm,\scriptscriptstyle{\text{\MC{}}}})}_{\mathbf{E}_\text{mat}(\mathbf{r},\omega_{\pm,\scriptscriptstyle{\text{\MC{}}}})} + \underbrace{\frac{i\Xi(\mathbf{r})}{\sqrt{\varepsilon_0 \mathcal{V}_\text{eff}}}\omega_{\pm,\scriptscriptstyle{\text{\MC{}}}} x_\text{cav}(\omega_{\pm,\scriptscriptstyle{\text{\MC{}}}}) \mathbf{n_A}}_{\mathbf{E}_\text{cav}(\mathbf{r},\omega_{\pm,\scriptscriptstyle{\text{\MC{}}}}) },
\label{electric_field_dipole_cavity}
\end{align}
with $\mathbf{n_A} = \frac{\mathbf{A}(\mathbf{r})}{|\mathbf{A}(\mathbf{r})|}$. This equation indicates the contribution of the cavity  $\mathbf{E}_\text{cav}(\mathbf{r},\omega_{\pm,\scriptscriptstyle{\text{\MC{}}}})\propto x_\text{cav}$ and of the matter excitation $\mathbf{E}_\text{mat }(\mathbf{r},\omega_{\pm,\scriptscriptstyle{\text{\MC{}}}})\propto x_\text{mat}$ to the electric field.
Further, we use the eigenvectors (Eq. \eqref{equations_motion_classical_velocities_D_wa}) to obtain the ratio between the amplitudes $x_\text{cav}$ and $x_\text{mat}$ of the classical harmonic oscillators:
\begin{equation}
\frac{x_\text{cav}(\omega_{\pm,\scriptscriptstyle{\text{\MC{}}}})}{x_\text{mat}(\omega_{\pm,\scriptscriptstyle{\text{\MC{}}}})} = \frac{-2i\omega_{\pm,\scriptscriptstyle{\text{\MC{}}}} g_{\scriptscriptstyle{\text{\MC{}}}}}{\omega_\text{cav}^2-\omega_{\pm,\scriptscriptstyle{\text{\MC{}}}}^2}. \label{ratio_xcav_xmat}
\end{equation}
Inserting Eq. \eqref{ratio_xcav_xmat} into Eq. \eqref{electric_field_dipole_cavity}, we obtain the ratio between the contributions of the cavity electric field and the matter excitation.

Equations  \eqref{electric_field_dipole_cavity} and \eqref{ratio_xcav_xmat} are a main result of this subsection and can be used to obtain the electric field at any position and for an arbitrary transverse mode with field distribution given by $\Xi(\mathbf{r})$. We consider for illustration the particular case of a molecule (as an example of quantum emitter) introduced in the center of a dielectric cavity consisting in a rectangular vacuum box enclosed in the three dimensions by perfect mirrors, as sketched in Fig. \ref{figure_dipole_cavity}a. The cross-section of the box is square, with size $L_x = L_y =$ 292 nm and its height is  $L_z =$ 215 nm, which results in a fundamental lowest-order cavity mode at frequency $\omega_\text{cav}$ = 3 eV and an effective volume $V_\text{eff} = 4.483 \cdot 10^6 \text{ nm}^3$ (for an easier comparison between classical frequencies $\omega$ and quantum energies $\hbar\omega$, in this paper we use eV as a unit for both of them). This value of $V_\text{eff}$ is calculated from the general expression of dielectric structures \cite{foresi97}
\begin{equation}
V_\text{eff} = \frac{\int \varepsilon(\mathbf{r}) |\Xi(\mathbf{r})|^2 d\mathbf{r}}{\max[\varepsilon(\mathbf{r}) |\Xi(\mathbf{r})|^2]},
\end{equation}
where  $\varepsilon(\mathbf{r})$ refers to the permittivity of the system at position $\mathbf{r}$, and in this particular case we consider $\varepsilon(\mathbf{r})=1$ inside the cavity.
The molecular excitation is nearly resonant with the cavity, $\omega_\text{mat}  \approx \omega_\text{cav} =$ 3 eV, but its exact frequency is changed to study the effects of detuning. The transition dipole moment $\mu_\text{mat} = \sqrt{\frac{\hbar f_\text{mat}}{2\omega_\text{mat}}}$ (associated with the transition from the ground state to the first excited state)  is parallel to the $z$ axis and is relatively strong,  $\mu_\text{mat}$ = 15 Debye, achievable with organic molecules such as nonacene, for example, \cite{kuisma22}. This value of the transition dipole moment implies that this molecular emitter has an oscillator strength of $f_\text{mat}  = \frac{(118.74e)^2}{m_\text{p}}$, where $e$ is the electron charge and $m_\text{p}$ the mass of the proton. By placing the molecular emitter in the center of the cavity where the electric field of the mode is maximum, this choice of parameters leads to a coupling strength $g_{\scriptscriptstyle{\text{\MC{}}}} \approx 2.5 \cdot 10^{-4} \omega_\text{cav}$, far from the ultrastrong coupling regime (a larger value of $g_{\scriptscriptstyle{\text{\MC{}}}}$ is considered at the end of this subsection).

We show in Fig. \ref{figure_dipole_cavity}b the distribution of the $z$ component of the electric field inside this cavity for the upper hybrid mode $E_z(x,\omega_{+,\scriptscriptstyle{\text{\MC{}}}})$ and for the lower hybrid mode $E_z(x,\omega_{-,\scriptscriptstyle{\text{\MC{}}}})$, as obtained from Eq. \eqref{electric_field_dipole_cavity}. We plot the fields as a function of the position in the $x$ direction with respect to the location of the molecular emitter at the center of the cavity. To highlight the differences between the contributions of the cavity and the induced dipole in the two modes, we choose a slight detuning of $\omega_\text{cav} - \omega_\text{mat} = 1.5$ meV. Since the classical \MC{} model does not give the absolute value of the eigenmode fields, we choose arbitrary units so that the contribution of the cavity mode to the electric field of the upper hybrid mode ($\mathbf{E}_\text{cav}(\mathbf{r},\omega_{+,\scriptscriptstyle{\text{\MC{}}}})$ in Eq. \eqref{electric_field_dipole_cavity}) has a maximum absolute value of 1. This choice fixes all the other values according to  Eq. \eqref{ratio_xcav_xmat}\footnote{The eigenstates of the  Hopfield Hamiltonian from Eq. \eqref{quantum_hamiltonian_2} have a symmetry where the cavity contribution of one hybrid mode is the same as the matter contribution of the other mode and vice versa, satisfying the equality $\langle \hat{a} + \hat{a}^\dagger \rangle (\omega_{\pm,\scriptscriptstyle{\text{\MC{}}}}) = \langle \hat{b} + \hat{b}^\dagger \rangle (\omega_{\mp,\scriptscriptstyle{\text{\MC{}}}})$. This property allows us to connect the amplitudes of the classical oscillators of the two hybrid eigenmodes as $\sqrt{\omega_\text{cav}} x_\text{cav} (\omega_{\pm,\scriptscriptstyle{\text{\MC{}}}}) = \sqrt{\omega_\text{mat}} x_\text{mat} (\omega_{\mp,\scriptscriptstyle{\text{\MC{}}}})$ (from Eq. \eqref{quantization_xcav_xmat}).  }. The fields are dominated by the cavity mode far from the cavity center and by the contribution from the molecular dipole close to $x=0$. The field distribution shows a clear difference in the behavior of the two hybrid modes. For the upper mode the induced dipole points in the same direction as the cavity field ($\frac{x_\text{cav}(\omega_{+,\scriptscriptstyle{\text{\MC{}}}})}{x_\text{mat}(\omega_{+,\scriptscriptstyle{\text{\MC{}}}})} > 0$), but in the inverse direction for the lower mode ($\frac{x_\text{cav}(\omega_{-,\scriptscriptstyle{\text{\MC{}}}})}{x_\text{mat}(\omega_{-,\scriptscriptstyle{\text{\MC{}}}})} < 0$). Further, at the detuning considered, the relative contribution of the cavity to the fields is larger for the upper than for the lower mode, as indicated by the values of the electric field far from the molecular emitter at $\omega_{+,\scriptscriptstyle{\text{\MC{}}}}$ and $\omega_{-,\scriptscriptstyle{\text{\MC{}}}}$. In contrast, as shown in the inset, the relative contribution from the molecular dipole to the field close to the molecule ($x=0$) is stronger for the lower mode. Figure \ref{figure_dipole_cavity}b thus confirms that the classical harmonic oscillator model allows for calculating the relative contribution of cavity and matter for each mode, as desired.

Further, Eqs.  \eqref{electric_field_dipole_cavity} and \eqref{ratio_xcav_xmat} also enable to examine the dependence of the field $\mathbf{E}(\mathbf{r},\omega_{\pm,\scriptscriptstyle{\text{\MC{}}}})$ inside the cavity with detuning $\omega_\text{mat} - \omega_\text{cav}$.  Figure \ref{figure_dipole_cavity}c shows the contributions to this electric field of the cavity and the molecular emitter for each hybrid mode, normalized with respect to the sum of both contributions, according to $\Sigma_\text{cav}^\pm = \frac{|\mathbf{E}_{\text{cav}}(\omega_{\pm,\scriptscriptstyle{\text{\MC{}}}})|^2}{|\mathbf{E}_{\text{cav}}(\omega_{\pm,\scriptscriptstyle{\text{\MC{}}}})|^2+|\mathbf{E}_{\text{mat}}(\omega_{\pm,\scriptscriptstyle{\text{\MC{}}}})|^2}$ (dots) and $\Sigma_\text{mat}^\pm = \frac{|\mathbf{E}_{\text{mat}}(\omega_{\pm,\scriptscriptstyle{\text{\MC{}}}})|^2}{|\mathbf{E}_{\text{cav}}(\omega_{\pm,\scriptscriptstyle{\text{\MC{}}}})|^2+|\mathbf{E}_{\text{mat}}(\omega_{\pm,\scriptscriptstyle{\text{\MC{}}}})|^2}$ (solid lines). These ratios play a similar role as the Hopfield coefficients from cavity-QED descriptions. The blue (red) dots and solid lines correspond to the upper (lower) hybrid mode. We obtain $\mathbf{E}_{\text{cav}}(\omega_{\pm,\scriptscriptstyle{\text{\MC{}}}})$ and $\mathbf{E}_{\text{mat}}(\omega_{\pm,\scriptscriptstyle{\text{\MC{}}}})$ by replacing Eq. \eqref{ratio_xcav_xmat} into  Eq. \eqref{electric_field_dipole_cavity}, for a fixed coupling strength $g_{\scriptscriptstyle{\text{\MC{}}}} = 2.5 \cdot 10^{-4} \omega_\text{cav}$ and for a distance of 10.5 nm from the molecular emitter in the $x$ direction. This position (indicated by the dashed line in the inset of Fig. \ref{figure_dipole_cavity}b) is chosen because it is where the matter and cavity contributions have the same weight for the two hybrid modes at zero detuning and very small coupling strengths ($\Sigma_\text{cav}^\pm = \Sigma_\text{mat}^\pm = 0.5$. For $\omega_\text{cav} > \omega_\text{mat}$ the field of the lower mode is predominantly given by the matter excitation  ($\Sigma_\text{mat}^- > \Sigma_\text{cav}^-$ as indicated by the red dots and the red solid line). In contrast, for the upper mode, the cavity contribution dominates ($\Sigma_\text{cav}^+ > \Sigma_\text{mat}^+$, blue). Further, already at detunings as small as $\omega_\text{cav}-\omega_\text{mat} \gtrsim 15 \text{ meV} = 5 \cdot 10^{-3} \omega_\text{cav}$,  the modes are essentially uncoupled for this small coupling strength ($\Sigma_\text{mat}^+ \ll \Sigma_\text{cav}^+$ and $\Sigma_\text{mat}^- \gg \Sigma_\text{cav}^-$).

The coupling strength we have considered in this subsection corresponds to the strong coupling regime (we have neglected losses) but is far from the ultrastrong coupling regime so that the phenomena studied can also be explained with the classical linearized model (Sec. \ref{Appendix_linearized} of the Supplementary Material). On the other hand, we consider again in Fig. \ref{figure_dipole_cavity}d the contributions to the electric field $\Sigma_\text{cav}^\pm$ and $\Sigma_\text{mat}^\pm$ as a function of the detuning, but in this case for a considerably larger coupling strength $g_{\scriptscriptstyle{\text{\MC{}}}} = 0.2 \omega_\text{cav}$.  This value of $g_{\scriptscriptstyle{\text{\MC{}}}}$ is not currently achievable with dielectric cavities at the single molecule or single emitter level  it would correspond to a transition dipole moment $\mu_\text{mat}$ = $1.2 \cdot 10^4$ Debye), but we choose it to illustrate the analysis of ultrastrongly-coupled systems within the classical \MC{} model. Further, such large $g_{\scriptscriptstyle{\text{\MC{}}}}$ can be achieved in dielectric cavities fully filled by a material or many molecular emitters, as discussed in Sec. \ref{subsec_bulk_dielectric}. For zero detuning $\omega_\text{cav} = \omega_\text{mat}$, the contributions of the induced dipole and the cavity are no longer identical in the ultrastrong coupling regime, with $\Sigma_\text{cav}^+ \approx 0.6$ and $\Sigma_\text{mat}^+ \approx 0.4$ for the upper hybrid mode at frequency $\omega_{+,\scriptscriptstyle{\text{\MC{}}}}$ (and the opposite for the lower hybrid mode).  More strikingly, the results in Fig. \ref{figure_dipole_cavity}d  indicate a very different tendency of the modes at large detunings as compared to strong coupling, especially in the case of the upper hybrid mode at frequency $\omega_{+,\scriptscriptstyle{\text{\MC{}}}}$. In the ultrastrong coupling regime, in the $\omega_\text{mat} \rightarrow 0$ limit ($\omega_\text{mat}-\omega_\text{cav} \rightarrow -3$  eV), this mode  (blue solid line and dots) has significant contributions from both the cavity and the matter ($\Sigma_\text{cav}^+ \approx 0.9$ and $\Sigma_\text{mat}^+ \approx 0.1$). Thus these two excitations do not decouple in this limit. This behavior is consistent with the discussion of the dispersion in Fig. \ref{figure_models}d, where at large detunings, the upper mode frequency does not reach the bare frequency $\omega_\text{mat}$. The SpC model (not shown) does not reproduce this behavior because the modes become uncoupled ($\Sigma_\text{cav}^+ \approx 1$ and $\Sigma_\text{mat}^+ \approx 0$).

The described methodology thus enables obtaining results equivalent to those of the cavity-QED description (Hopfield Hamiltonian with the diamagnetic term) by using an intuitive classical model of coupled harmonic oscillators. In summary, we have shown in this section how to use the classical \MC{} model to characterize the fields in a hybrid system composed of a molecular emitter coupled to a transverse mode of a cavity.

\subsection{A quantum emitter interacting with the longitudinal field of a metallic nanoparticle through Coulomb coupling} \label{subsec_molecule_nanoparticle}

Next, we consider a quantum emitter placed close to a metallic nanoparticle to analyze  how to model  an alternative system and obtain physical observables in the strong and ultrastrong coupling regimes. These nanoparticles are attractive in nanophotonics because they support localized surface plasmon modes characterized by very low effective volumes \cite{pelton19,waks10,trugler08,kuisma22,chikkaraddy16}. Since the coupling strength is inversely proportional to the square root of the effective mode volume, very large coupling strengths can be obtained even when the nanoparticle interacts with a single molecule or quantum dot. We consider again a molecule as a representative quantum emitter.

In order to analyze the interaction of the dipolar plasmonic mode of the nanoparticle with a molecular (harmonic) excitation of dipole moment $d_\text{mat}$, we consider that the size of the nanoparticle and the molecule-nanoparticle distance are much smaller than the light wavelength and treat the system within the quasistatic approximation. Under this approximation, the temporal variation of the vector potential $\mathbf{A}$ in Eq. \eqref{eq_relation_electric_field_potentials} is negligible. Therefore the coupling between the nanoparticle and the molecular emitter is governed by Coulomb interactions expressed by a scalar potential $V_\text{Cou}$. The coupling is then mediated by longitudinal fields, in contrast to the coupling with transverse fields in Sec. \ref{subsec_molecule_dielectric}.

  In this context, the  emitter-nanoparticle coupling cannot be modeled with the minimal coupling Hamiltonian as in Sec. \ref{subsec_molecule_dielectric}, and it is described instead through the interaction Hamiltonian \cite{gerryknight}
\begin{equation}
\hat{H}_\text{int2} = -\hat{\mathbf{d}}_\text{mat} \cdot \hat{\mathbf{E}}^\parallel_\text{cav}(\mathbf{r}_\text{mat}).
\label{eq_hamiltonian_dipole_dipole}
\end{equation} 
$\hat{\mathbf{E}}^\parallel_\text{cav}$ is the electric field associated with the dipolar mode of the nanocavity, which in the quasistatic approximation is completely longitudinal (we indicate this explicitly with the symbol $\parallel$) and $d_\text{cav}$ is the induced plasmonic dipole moment (operator $\hat{\mathbf{d}}_\text{cav}$). For simplicity, we consider small spherical particles of radius $R_\text{cav}$ composed by a Drude metal with plasma frequency $\omega_\text{p}$, but this approach could be generalized to other systems. The spherical particles present a dipolar plasmonic resonance of Lorentzian lineshape at frequency $\omega_\text{cav} = \frac{\omega_\text{p}}{\sqrt{3}}$, and oscillator strength $f_\text{cav} = 4\pi\varepsilon_0 R_\text{cav}^3 \omega_\text{cav}^2$ \cite{barnes16}. The quasistatic field outside them is directly determined by $d_\text{cav}$ according to  $\hat{\mathbf{E}}^\parallel_\text{cav}(\mathbf{r}) = \frac{3 (\hat{\mathbf{d}}_\text{cav} \cdot \mathbf{n}_{\mathbf{r}\text{cav}})\mathbf{n}_{\mathbf{r}\text{cav}}- \hat{\mathbf{d}}_\text{cav}}{4\pi\varepsilon_0|\mathbf{r}_\text{cav}-\mathbf{r}|^3} $, where $\mathbf{r}_\text{cav}$ is the center of the nanoparticle, $|\mathbf{r}-\mathbf{r}_\text{cav}|>R_\text{cav}$, and we define the unit vector $\mathbf{n}_{\mathbf{r}\text{cav}} = \frac{\mathbf{r}-\mathbf{r}_\text{cav}}{|\mathbf{r}-\mathbf{r}_\text{cav}|}$.

 We insert the quantized expressions of the induced dipole moments $\hat{d}_\text{cav}$ and  $\hat{d}_\text{mat}$ of Eq. \eqref{quantization_d} into the Hamiltonian in Eq. \eqref{eq_hamiltonian_dipole_dipole} and the expression of $\hat{\mathbf{E}}^\parallel_\text{cav}(\mathbf{r})$ and
obtain
\begin{equation}
\hat{H}_\text{int2} = \hbar g_{\scriptscriptstyle{\text{SpC}}} (\hat{a}+\hat{a}^\dagger)(\hat{b}+\hat{b}^\dagger),
\end{equation}
with coupling strength
\begin{equation}
g_{\scriptscriptstyle{\text{SpC}}} = \frac{1}{2} \frac{\sqrt{f_\text{cav}}\sqrt{f_\text{mat}}}{4\pi\varepsilon_0 |\mathbf{r}_\text{cav}-\mathbf{r}_\text{mat}|^3 \sqrt{\omega_\text{cav}\omega_\text{mat}}}[\mathbf{n}_{\mathbf{d}\text{cav}} \cdot \mathbf{n}_{\mathbf{d}\text{mat}} - 3(\mathbf{n}_{\mathbf{d}\text{cav}} \cdot \mathbf{n}_{\mathbf{r}\text{rel}})(\mathbf{n}_{\mathbf{d}\text{mat}} \cdot \mathbf{n}_{\mathbf{r}\text{rel}})],
\label{g_dipole_dipole}
\end{equation}
where we have defined the unit vectors as $\mathbf{n}_{\mathbf{d}\text{cav}} = \frac{\mathbf{d}_\text{cav}}{|\mathbf{d}_\text{cav}|}$, $\mathbf{n}_{\mathbf{d}\text{mat}} = \frac{\mathbf{d}_\text{mat}}{|\mathbf{d}_\text{mat}|}$ and $\mathbf{n}_{\mathbf{r}\text{rel}}= \frac{\mathbf{r}_\text{cav}-\mathbf{r}_\text{mat}}{|\mathbf{r}_\text{cav}-\mathbf{r}_\text{mat}|}$. The total Hamiltonian is thus the sum of $\hat{H}_\text{int2}$ and the terms related to the energy of the uncoupled plasmon and molecular excitation, corresponding to the  Hopfield Hamiltonian of Eq. \eqref{hopfield_hamiltonian_1}  without the diamagnetic term $(D=0)$. Thus,  the corresponding classical model to be adopted is the SpC model in Sec. \ref{subsec_spring_coupling}, with the equations of motion in Eq. \eqref{equations_motion_classical_position_D0}. Additional details can be found in Section \ref{Appendix_equationsofmotion} of the Supplementary Material.

The representation of the plasmon-molecule system with the SpC model is schematically shown in Fig. \ref{figure_dipole_dipole}a. To obtain the observables in this system, we use the equivalence of the oscillation amplitudes $x_\text{cav}$ and $x_\text{mat}$ with the induced dipole moments of the cavity and the molecular (or matter) excitation. This equivalence can be obtained from Eqs.  \eqref{quantization_d} and  \eqref{quantization_xcav_xmat} and it follows  $x_\text{cav} = \frac{d_\text{cav}}{\sqrt{f_\text{cav}}}$ and $x_\text{mat} = \frac{d_\text{mat}}{\sqrt{f_\text{mat}}}$.
 Further, this treatment can be extended to other dipole-dipole interactions beyond the coupling of a molecular emitter with a plasmon (direct dipole-dipole interactions between molecules are considered in Sec. \ref{subsec_bulk_dielectric}).

 \begin{figure}[t!]
\centering
\includegraphics[scale=0.68]{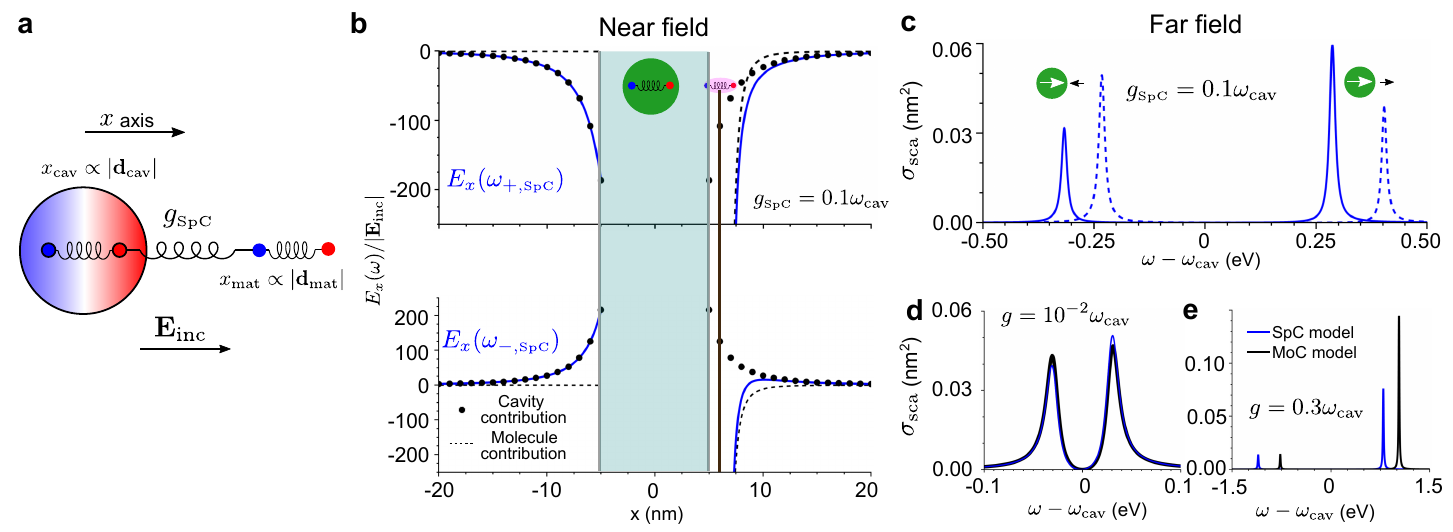}
\caption{Modelling of the coupling between a quantum emitter and a metallic spherical nanoparticle (a plasmonic nanocavity)  within the classical SpC model. a) Schematics of the system. The quantum emitter is considered to be a molecule. The molecular excitation (of induced dipole moment $\mathbf{d}_\text{mat}$) and the dipolar mode of the plasmonic nanocavity (of induced dipole moment $\mathbf{d}_\text{cav}$) are described as two harmonic oscillators (of oscillation amplitudes $x_\text{mat}$ and $x_\text{cav}$) that are coupled with strength $g_{\scriptscriptstyle{\text{SpC}}}$. The system is excited by a laser of electric field amplitude $\mathbf{E}_\text{inc}$. The radius of the spherical nanoparticle is 5 nm, and the molecular emitter is placed at a 1 nm distance from the nanoparticle surface along the $x$ axis (the center of the nanoparticle corresponds to $x = y = z = 0$). $\mathbf{d}_\text{cav}$, $\mathbf{d}_\text{mat}$ and $\mathbf{E}_\text{inc}$ are polarized along $x$.  b) Electric field distribution along the $x$ axis ($y = z = 0$) when the system is excited at the frequency of the upper hybrid mode $\omega_{+,\scriptscriptstyle{\text{SpC}}}$ (top panel) and of the lower hybrid mode $\omega_{-,\scriptscriptstyle{\text{SpC}}}$ (bottom panel). The fields are real and are evaluated only outside the nanocavity, with the positions inside highlighted by the green-shaded area. The position of the molecular emitter is indicated by the vertical brown line. We evaluate the fields for coupling strength $g_{\scriptscriptstyle{\text{SpC}}} = 0.1 \, \omega_\text{cav}$, and $\omega_\text{cav} = \omega_\text{mat} =$ 3 eV. For each hybrid mode, the cavity contribution to the field is indicated by dots, the contribution from the emitter by dashed lines and the total field by blue solid lines. c) Scattering cross-section of the same system, with $g_{\scriptscriptstyle{\text{SpC}}}= 0.1 \, \omega_\text{cav}$, as a function of the detuning of the laser $\omega - \omega_\text{cav}$.  Solid lines:  tuned system with frequencies $\omega_\text{cav} = \omega_\text{mat} =$ 3 eV. Dashed lines:  detuned system with frequencies $\omega_\text{cav} =$ 3 eV and $\omega_\text{mat} =$ 3.2 eV. d) Scattering cross-section of the tuned system ($\omega_\text{cav} = \omega_\text{mat} =$ 3 eV), comparing the result of the SpC model (blue line) to the results of the \MC{} model (black line), in the strong coupling regime,  $g = 10^{-2} \omega_\text{cav}$. e) Same as in (d)  for the ultrastrong coupling regime,  $g = 0.3 \, \omega_\text{cav}$. In all results $f_\text{cav} = (4345e)^2/m_\text{p}$ (where $m_\text{p}$ is the mass of the proton), $F_\text{cav}=\sqrt{f_\text{cav}}|\mathbf{E}_\text{inc}|$, $f_\text{mat}=(118.74e)^2/m_\text{p}$, $F_\text{mat}=\sqrt{f_\text{mat}}|\mathbf{E}_\text{inc}|$, $\kappa=$ 20 meV and $\gamma=$ 10 meV  (except that we modify $f_\text{cav}$ in panels (d) and (e) to achieve the desired values of $g_{\scriptscriptstyle{\text{SpC}}}$). }
\label{figure_dipole_dipole}
\end{figure} 

We consider next that the dipolar mode of the metallic nanoparticle is illuminated by an external field of amplitude $\mathbf{E}_\text{inc}$ and frequency $\omega$. We introduce this field in the SpC model as a forcing term that acts both onto the nanoparticle and onto the molecular emitter. Specifically, this is done by adding terms $F_\alpha e^{-i\omega t} =  \sqrt{f_\alpha}|\mathbf{E}_\text{inc}| e^{-i\omega t}$ ($\alpha$ = 'cav' or $\alpha$ = 'mat') on the right-hand side of Eq. \eqref{equations_motion_classical_position_D0}, i.e., the amplitude $F_\alpha$ of the time-dependent force is proportional to the induced dipole moments $d_\alpha$ and the electric field of the illumination (see Sec. \ref{Appendix_equationsofmotion} in the Supplementary Material for further details). By solving the equations of motion of the SpC model (Eq. \eqref{equations_motion_classical_position_D0}) in the frequency domain with this external force included, we can calculate the induced dipole moments of the cavity plasmon and matter excitation:
\begin{subequations}
\begin{equation}
d_\text{cav}(\omega) = \sqrt{f_\text{cav}} x_\text{cav}(\omega) =  \sqrt{f_\text{cav}} \frac{F_\text{cav}(\omega_\text{mat}^2-\omega^2)-F_\text{mat}2g_{\scriptscriptstyle{\text{SpC}}} \sqrt{\omega_\text{cav} \omega_\text{mat}} }{(\omega_\text{cav}^2-\omega^2)(\omega_\text{mat}^2-\omega^2)-4g_{\scriptscriptstyle{\text{SpC}}}^2 \omega_\text{cav} \omega_\text{mat}}, \label{amplitude_dcav}
\end{equation}
\begin{equation}
d_\text{mat}(\omega) = \sqrt{f_\text{mat}} x_\text{mat}(\omega) = \sqrt{f_\text{mat}} \frac{F_\text{mat}(\omega_\text{cav}^2-\omega^2)-F_\text{cav}2g_{\scriptscriptstyle{\text{SpC}}} \sqrt{\omega_\text{cav} \omega_\text{mat}}}{(\omega_\text{cav}^2-\omega^2)(\omega_\text{mat}^2-\omega^2)-4g_{\scriptscriptstyle{\text{SpC}}}^2 \omega_\text{cav} \omega_\text{mat}}.
\end{equation}
\label{amplitudes_dipole_dipole}
\end{subequations}
These expressions are consistent with an alternative classical model that describes the nanocavity and the molecular emitter as dipolar polarizable objects (Sec. \ref{Appendix_equationsofmotion} of the Supplementary Material), supporting the validity of the general approach presented here. In the absence of losses\cite{hughes24} the induced dipole moments $d_\text{cav}$ and $d_\text{mat}$ diverge at the eigenfrequencies $\omega_{\pm,\scriptscriptstyle{\text{SpC}}}$ of the SpC model (Eq. \eqref{eigenfrequencies_SpC}). To avoid these divergences, we add an imaginary part to the bare cavity and matter frequencies in this section. These imaginary parts are related to the decay rates  of the cavity, $\kappa$, and of the matter excitation, $\gamma$, as $\Im(\omega_\text{cav}) = -\frac{\kappa}{2}$  and $\Im(\omega_\text{mat}) = -\frac{\gamma}{2}$,  respectively.

As an example, we consider a metallic spherical nanoparticle of radius $R_\text{cav}$ = 5 nm and with a cavity mode of frequency $\omega_\text{cav}$ = 3 eV. We consider the same molecular emitter of Sec. \ref{subsec_molecule_dielectric}, with a strong transition dipole moment of magnitude $\mu_\text{mat} = 15$ Debye.  As indicated by Eq. \eqref{g_dipole_dipole}, the coupling strength of the system can be adjusted based on the position and orientation of the molecular emitter. We choose that the dipolar molecular transition is polarized perpendicularly to the surface of the nanoparticle and parallel to the amplitude of the incident field $\mathbf{E}_\text{inc}$, to maximize the coupling strength (as a consequence $\mathbf{d}_\text{cav}$, $\mathbf{d}_\text{mat}$, $\hat{\mathbf{E}}^\parallel_\text{cav}(\mathbf{r}_\text{mat})$ and $\mathbf{n}_{\mathbf{r}\text{rel}}$ are all oriented in the same direction in e.g. Eqs. \eqref{eq_hamiltonian_dipole_dipole}, \eqref{g_dipole_dipole}).  With this choice, and placing the molecular emitter at 1 nm from the surface of the nanoparticle, we obtain a coupling strength $g_{\scriptscriptstyle{\text{SpC}}} \approx 300 \text{ meV} = 0.1 \, \omega_\text{cav}$ and thus reach the limit of the ultrastrong coupling regime. This large value of $g_{\scriptscriptstyle{\text{SpC}}}$ is possible due to the small size of the nanoparticle (large field confinement) and to the strong transition dipole moment considered for the molecular emitter, which lies slightly beyond the values of $\mu_\text{mat} = 3 - 5$ Debyes corresponding to typical molecules used in combination with plasmonic systems. Even larger field confinement may be possible in non-spherical experimental nanostructure configurations that exploit very narrow gaps \cite{chikkaraddy16, wu21}.To ensure that the system is also in the strong coupling regime when considering lower values of $g_{\scriptscriptstyle{\text{SpC}}}$ below, we choose $\gamma = 10$ meV and a damping rate of the plasmonic cavity $\kappa = 20$ meV that is small compared to those of usual plasmonic metals.

The induced dipole moments obtained from Eq. \eqref{amplitudes_dipole_dipole} can be used, for example, to calculate the near-field distribution under excitation at frequency $\omega$. The total electric field is the sum of the cavity $\mathbf{E}^\parallel_\text{cav}$ and molecular or matter contribution $\mathbf{E}^\parallel_\text{mat}$. Under the quasistatic approximation, with $d_\text{cav}(\omega) = \sqrt{f_\text{cav}} x_\text{cav}(\omega)$ and $d_\text{mat}(\omega) = \sqrt{f_\text{mat}} x_\text{mat}(\omega)$ we obtain that
the fields at position $\mathbf{r}$ outside the metallic nanoparticle, $|\mathbf{r}-\mathbf{r}_\text{cav}| > R_\text{cav}$, depend on the amplitude of the harmonic oscillators as 
\begin{align}
&\mathbf{E}^\parallel(\mathbf{r},\omega) = \underbrace{\frac{3( \mathbf{n}_{\mathbf{d}\text{cav}} \cdot \mathbf{n}_{\mathbf{r}\text{cav}}) \mathbf{n}_{\mathbf{r}\text{cav}} - \mathbf{n}_{\mathbf{d}\text{cav}}}{4\pi \varepsilon_0 |\mathbf{r}-\mathbf{r}_\text{cav}|^3} \sqrt{f_\text{cav}} x_\text{cav}(\omega)}_{\mathbf{E}^\parallel_\text{cav}(\mathbf{r},\omega)} + \underbrace{\frac{3(\mathbf{n}_{\mathbf{d}\text{mat}}\cdot \mathbf{n}_{\mathbf{r}\text{mat}}) \mathbf{n}_{\mathbf{r}\text{mat}} - \mathbf{n}_{\mathbf{d}\text{mat}}}{4\pi \varepsilon_0 |\mathbf{r}-\mathbf{r}_\text{mat}|^3} \sqrt{f_\text{mat}} x_\text{mat}(\omega)}_{\mathbf{E}^\parallel_\text{mat}(\mathbf{r},\omega)}.
\label{electric_field_dipole_dipole}
\end{align}
From this expression, the fields at the frequency of each hybrid mode are calculated by replacing into Eq. \eqref{electric_field_dipole_dipole} the oscillation amplitudes in Eq. \eqref{amplitudes_dipole_dipole} induced at the mode frequencies $\omega_{\pm,\scriptscriptstyle{\text{SpC}}}$.

The electric fields along the $x$-axis associated with the upper and lower mode frequencies are plotted in the top and bottom panels of Fig. \ref{figure_dipole_dipole}b (blue lines), respectively. These fields are real and polarized along the $x$ direction. We further show the decomposition of the fields into the contribution of the cavity (black dots) and the molecular emitter (black dashed line) as given by the first and second terms on the right-hand side of Eq. \eqref{electric_field_dipole_dipole}, respectively. It can be appreciated from Fig. \ref{figure_dipole_dipole}b that, for example, when the upper hybrid mode is excited, the dipoles associated with the cavity and the molecular emitter are oriented in the same direction (same sign). In contrast, for the lower mode, the dipoles point towards the opposite direction.

The near field plotted in Fig. \ref{figure_dipole_dipole}b is useful for analyzing the behavior of the hybrid modes. Still, it is difficult to measure, and most experiments focus on far-field spectral information, such as in the scattering cross-section spectral $\sigma_\text{sca}$. Due to the small emitter-nanocavity distance, we neglect retardation effects so that $\sigma_\text{sca}$ is related  to the total induced dipole moment of the system  as \cite{novotnyhecht} 
\begin{align}
\sigma_\text{sca}(\omega) &= \frac{\omega^4}{6\pi \varepsilon_0^2 c^4}  \left|\frac{\mathbf{d}_\text{cav}(\omega)}{|\mathbf{E}_\text{inc}|} + \frac{\mathbf{d}_\text{mat}(\omega)}{|\mathbf{E}_\text{inc}|} \right|^2 \nonumber \\ &= \frac{\omega^4}{6\pi \varepsilon_0^2 c^4}  \left| \frac{\sqrt{f_\text{cav}}x_\text{cav}(\omega)}{|\mathbf{E}_\text{inc}|}    \mathbf{n}_{\mathbf{d} \text{cav}} + \frac{\sqrt{f_\text{mat}}x_\text{mat}(\omega)}{|\mathbf{E}_\text{inc}|}  \mathbf{n}_{\mathbf{d} \text{mat}} \right|^2. \label{scattering_cross_section}
\end{align}
We show in Fig. \ref{figure_dipole_dipole}c the scattering cross section for the same nanoparticle-molecular emitter system in the outset of the ultrastrong coupling regime ($g_{\scriptscriptstyle{\text{SpC}}} = 0.1 \, \omega_\text{cav}$). Since the oscillator strength of the cavity is much larger than that of the  emitter ($f_\text{cav} \gg f_\text{mat}$), the spectrum is entirely dominated by the cavity contribution, obtained from Eq. \eqref{amplitude_dcav} (however, in other systems, where both oscillator strengths are similar, $f_\text{cav} \approx f_\text{mat}$, it is crucial to consider both contributions in  Eq. \eqref{scattering_cross_section}).  The scattering cross-section spectra are shown for two different detunings between the nanocavity and the molecular emitter. At zero detuning ($\omega_\text{cav} = \omega_\text{mat}$ = 3 eV, solid lines in Fig. \ref{figure_dipole_dipole}c) the upper hybrid mode has a (moderately) larger cross section than the lower hybrid mode, mostly due to the $\omega^4$ factor in Eq. \eqref{scattering_cross_section}. However, when the molecular excitation is blue detuned with respect to the cavity ($\omega_\text{cav}$ = 3 eV and  $\omega_\text{mat}$ = 3.2 eV, dashed line), the strength of the peak in the cross-section spectra associated with the lower hybrid mode increases and the upper hybrid mode becomes weaker. This behavior occurs because, for this detuning, the lower hybrid mode acquires a larger contribution of the cavity resonance that dominates the scattering spectra, while the predominantly emitter-like behavior of the upper mode results in a smaller cross section due to $f_\text{mat} \ll f_\text{cav}$.

To assess the importance of using the classical SpC model to describe this system, we compare the results of the scattering cross-section spectra calculated with this model with those obtained using the \MC{} model. For this purpose, it is necessary to obtain the expressions of the scattering cross section for the latter model under external illumination. By introducing forcing terms in the equations of motion of the \MC{} model (Eq. \eqref{equations_motion_classical_velocities_D}) to account for the external field, we obtain the corresponding oscillation amplitudes
\begin{subequations}
\begin{equation}
x_\text{cav,\MC{}}(\omega) = \frac{F_\text{cav}(\omega_\text{mat}^2-\omega^2) - F_\text{mat} 2i g_{\scriptscriptstyle{\text{\MC{}}}} \omega}{(\omega_\text{cav}^2-\omega^2)(\omega_\text{mat}^2-\omega^2)-4g_{\scriptscriptstyle{\text{\MC{}}}}^2 \omega^2},
\end{equation}
\begin{equation}
x_\text{mat,\MC{}}(\omega) = \frac{F_\text{cav}2i g_{\scriptscriptstyle{\text{\MC{}}}} \omega + F_\text{mat}(\omega_\text{mat}^2-\omega^2)}{(\omega_\text{cav}^2-\omega^2)(\omega_\text{mat}^2-\omega^2)-4g_{\scriptscriptstyle{\text{\MC{}}}}^2 \omega^2}.
\end{equation}
\end{subequations}

We calculate the scattering cross section according to each classical model by introducing these oscillations amplitudes in Eq. \eqref{scattering_cross_section}. Figure \ref{figure_dipole_dipole}d shows the spectra for the system at zero detuning ($\omega_\text{cav} = \omega_\text{mat}$ = 3 eV) in the strong coupling regime but far from the ultrastrong coupling regime, with $g = 10^{-2} \omega_\text{cav}$. As expected, the spectra calculated from the two models overlap almost perfectly (black line: \MC{} model; blue line: SpC model). The difference between the two calculations is less than $10\%$ at the hybrid mode frequencies $\omega_{\pm}$. This small error is consistent with the good agreement of the eigenfrequencies in Sec. \ref{sec_quantum} for this relatively low value of $g$.  

In contrast, if the system is well into the ultrastrong coupling regime, with coupling strength $g = 0.3 \, \omega_\text{cav}$, the spectra obtained with the two models are very different (Fig. \ref{figure_dipole_dipole}e). There is a clear disagreement in the peak positions due to the difference in the eigenfrequencies of the two models (see Fig. \ref{figure_models}d). Further, the \MC{} model predicts that the strength of the peak corresponding to the excitation of the upper hybrid mode is twice larger than the equivalent value from the SpC model. These significant differences emphasize the importance of the choice of the model in this regime. However, we note that for such large coupling, higher-order modes of the nanocavity are likely to play an important role  in the coupling, which would need to be considered in realistic systems \cite{delga14}. Further, examining how this analysis is modified when going beyond the quasistatic description would be of interest.

\subsection{An ensemble of interacting molecules in a Fabry-Pérot cavity} \label{subsec_bulk_dielectric}

The previous two examples illustrate the procedure for connecting the variables in the SpC and \MC{} models to physical observables. In both cases, the optical cavity was coupled to a single quantum emitter, a very challenging situation for experimentally reaching the ultrastrong coupling regime.  An alternative approach to access the necessary coupling strengths consists in filling a cavity with many molecules or with a material supporting a well-defined excitation (such as a phononic resonance) \cite{kenacohen13,brodbeck17,barraburillo21}. We consider in this section a homogeneous ensemble of molecular emitters as the material that interacts with resonant transverse electromagnetic modes of a Fabry-Pérot cavity (left sketch in Fig. \ref{figure_bulk_cavity}a), a system of significant relevance in experiments \cite{thomas19,george16,coles14,ribeiro18}. Each molecule presents a vibrational excitation that is modeled as a dipole of induced dipole moment  $\mathbf{d}_i$ (we focus here on the case of molecular emitters for specificity, but the same derivation can also be applied to phononic or similar materials by focusing on the induced dipole moment associated to each unit cell).  We consider that all molecular emitters are identical and thus have the same oscillator strength $f_\text{dip}$ and resonant frequency $\omega_\text{dip}$. We use the subindex $dip$ to emphasize that, at this stage, we are considering the individual molecular dipoles and not the whole  material (the full ensemble of molecular emitters) involved in the coupling. For simplicity, we assume that there are $N_\text{dip}$ molecular emitters distributed homogeneously.  The electromagnetic modes of the Fabry-Pérot cavity are standing waves with vector potential $\mathbf{A}_\alpha$ and frequency $\omega_{\text{cav},\alpha}$, where all $\alpha$ modes are orthogonal.

Following the relations between the observables and oscillators given in Sec. \ref{subsec_molecule_dielectric}, we represent each vibrational dipole as a harmonic oscillator with oscillation amplitude $x_{\text{dip},i} = \frac{|\mathbf{d}_i|}{\sqrt{f_\text{dip}}}$ and each cavity mode with the variable $x_{\text{cav},\alpha} = \sqrt{\varepsilon_0 V_\text{eff}} \mathcal{A}_\alpha$, where $ \mathcal{A}_\alpha$ is the maximum amplitude of the vector potential of the $\alpha$ mode. Notably, this system encompasses the two types of interaction discussed in the previous subsections: (i) each induced dipole $i$ is coupled to all other dipoles $j$ (through the direct Coulomb molecule-molecule interaction) following the SpC model, where the coupling strength $g_{\scriptscriptstyle{\text{SpC}}}^{(i,j)} $ is given by Eq. \eqref{g_dipole_dipole}; (ii) each induced dipole $i$ is coupled to all transverse cavity modes $\alpha$ according to the \MC{} model with coupling strength $g^{(\alpha,i)}_{\scriptscriptstyle{\text{\MC{}}}} = \frac{1}{2}\sqrt{\frac{f_\text{dip}}{\varepsilon_0 V_\text{eff}}} \Xi_\alpha(\mathbf{r}_i) \cos \theta_{\alpha,i} $ (see Sec. \ref{subsec_molecule_dielectric}), where $\Xi_\alpha(\mathbf{r}_i)$ is the normalized amplitude value of the cavity field at the position of molecular emitter $i$ and $\theta_{(\alpha,i)}$ is the angle between the orientation of the induced dipole moment of the $i^\text{th}$ molecular emitter and the polarization of each cavity mode. We assume that all molecular emitters are oriented in the same direction as the cavity field,  and thus $\cos \theta_{\alpha,i} = 1$ for all $\alpha$ and $i$.  All the interactions present in this system are shown schematically in the left panel of Fig. \ref{figure_bulk_cavity}a. To combine all couplings in a single model, we just include in the harmonic oscillator equations the coupling terms associated with the longitudinal dipole-dipole interactions (SpC model, Eq. \eqref{equations_motion_classical_position_D0}) and with the interaction of the molecular emitters with the transverse cavity modes (\MC{} model, Eq. \eqref{equations_motion_classical_velocities_D}). The resulting equations are
\begin{subequations}
\begin{equation}
\ddot{x}_{\text{dip},i} + \omega^2_\text{dip} x_{\text{dip},i} + \sum_\alpha 2g_{\scriptscriptstyle{\text{\MC{}}}}^{(\alpha,i)}  \dot{x}_{\text{cav},\alpha}  + \sum_{j \neq i} 2 \omega_\text{dip} g_{\scriptscriptstyle{\text{SpC}}}^{(i,j)} x_{\text{dip},j} = 0,
\end{equation}
\begin{equation}
\ddot{x}_{\text{cav},\alpha}  + \omega^2_{\text{cav},\alpha} x_{\text{cav},\alpha}  - \sum_i 2g_{\scriptscriptstyle{\text{\MC{}}}}^{(\alpha,i)*} \dot{x}_{\text{dip},i} = 0,
\end{equation}
\label{equations_motion_bulk}
\end{subequations}
where the sum extends over all cavity modes ($\sum_\alpha$) and molecular emitters ($\sum_i$ and $\sum_j$).

The direct calculation of the dynamics of the entire system requires solving  $N_\text{dip} \cross N_\text{cav}$ equations, where $N_\text{cav}$ is the number of cavity modes. However, due to the homogeneity of the material and the orthogonality of the cavity modes, each cavity mode $\alpha$ only couples with a collective matter excitation, which is represented by an oscillator of oscillation amplitude $x_{\text{mat},\alpha} \propto \sum_i \Xi_\alpha (\mathbf{r}_i) x_{\text{dip},i}$, i.e. the amplitude of the individual molecular oscillators in the collective mode $\alpha$ is weighted by the cavity mode field at the same position. $x_{\text{mat},\alpha}$ thus captures the response of the whole material formed by the ensemble of molecules, as highlighted by the use of the $mat$ subindex. The motion of each cavity mode $\alpha$ and the associated collective mode can then be obtained by solving the following two coupled equations (see Sec. \ref{Appendix_collectivetransformation} in Supplementary Material for the full derivation and the value of the different parameters)
\begin{subequations}
\begin{equation}
\ddot{x}_{\text{mat},\alpha } + (\omega_\text{dip}^2 + 2\omega_\text{dip} g_{\text{shift}})x_{\text{mat},\alpha } + 2 g_{\scriptscriptstyle{\text{\MC{}}}}^\text{max}\sqrt{N_\text{eff}} \dot{x}_{\text{cav},\alpha}  = 0, \label{equations_motion_bulk_transformed_1}
\end{equation}
\begin{equation}
\ddot{x}_{\text{cav},\alpha}  + \omega_{\text{cav},\alpha}^2 x_{\text{cav},\alpha} - 2g_{\scriptscriptstyle{\text{\MC{}}}}^\text{max} \sqrt{N_\text{eff}}          \dot{x}_{\text{mat},\alpha } = 0.
\label{equations_motion_bulk_transformed_2}
\end{equation}
\label{equations_motion_bulk_transformed}
\end{subequations}
In these equations, $g_{\text{shift}}$ is a parameter that depends on the values $g_{\scriptscriptstyle{\text{SpC}}}^{(i,j)}$ and that effectively describes the effect of the molecule-molecule dipolar interactions on the frequency of the $\alpha$ collective matter excitation, and $g_{\scriptscriptstyle{\text{\MC{}}}}^\text{max}$ is the maximum coupling strength between a single molecular emitter and the transverse cavity mode, obtained for a molecular emitter placed at the antinodes of the mode. $N_\text{eff}$ is the effective number of molecular emitters that are coupled to the mode ($N_\text{eff} = N_\text{dip} /2$ for a Fabry-Pérot mode). Equation \eqref{equations_motion_bulk_transformed} indicates that it is possible to describe the coupling between a cavity mode and a collective molecular excitation by considering only two harmonic oscillators, which are independent of the other cavity and collective molecular modes. The coupling strength between each collective matter excitation and the corresponding cavity mode increases with  $N_\text{eff}$  as $  G_{\scriptscriptstyle{\text{\MC{}}}}=g_{\scriptscriptstyle{\text{\MC{}}}}^\text{max} \sqrt{N_\text{eff}}$. This scaling with $\sqrt{N_\text{eff}}$ is consistent with the quantum Dicke model \cite{dicke54}, and explains the large coupling strengths that have been demonstrated in these systems \cite{barraburillo21,schwartz11,barachati18}. Further, the dipole-dipole interaction between the molecular emitters shifts the frequency of the collective excitation from $\omega_\text{dip}$ to $\Omega_\text{mat} = \sqrt{\omega_\text{dip}^2 + 2\omega_\text{dip}g_{\text{shift}}}$ (except when the cavity mode presents extremely fast spatial variations, where more complex effects can occur\cite{ribeiro23}). This shift corresponds to that described by the Clausius-Mossotti model of the permittivity of a material, where the resonances in the permittivity do not occur at the same frequency as that of the individual microscopic polarizable units. $\Omega_\text{mat}$ can  be considered as either the result of dressing the excitation of the individual molecular emitters, or as the bare resonance of the  whole material formed by the ensemble of molecular emitters. In this paper, we adopt the latter convention, as we are interested in the coupling of cavity photons with the material itself, and not with the individual constituent molecules. Thus, $\Omega_\text{mat}$ is considered as a bare frequency. After the change of variables, we obtain
\begin{subequations}
\begin{equation}
\ddot{x}_{\text{mat},\alpha } + \Omega_\text{mat}^2 x_{\text{mat},\alpha } + 2 G_{\scriptscriptstyle{\text{\MC{}}}} \dot{x}_{\text{cav},\alpha}  = 0,
\end{equation}
\begin{equation}
\ddot{x}_{\text{cav},\alpha}  + \omega_{\text{cav},\alpha}^2 x_{\text{cav},\alpha} - 2G_{\scriptscriptstyle{\text{\MC{}}}}   \dot{x}_{\text{mat},\alpha } = 0.
\end{equation}
\label{equations_motion_bulk_transformed_changedterminology}
\end{subequations}

In this description, each cavity mode $\alpha$ only couples to the collective molecular mode where the induced dipoles are polarized following the orientation and spatial distribution $\Xi_\alpha(\mathbf{r})$ of the cavity mode field. This collective molecular mode thus has a total induced dipole moment $\mathbf{d}_\alpha = \frac{1}{\sqrt{N_\text{eff}}} \sum_i \Xi_\alpha(\mathbf{r}_i) \mathbf{d}_i$, where $\mathbf{d}_i$ are the single-molecule induced dipole moments (see Sec. \ref{Appendix_collectivetransformation} in Supplementary Material). Importantly,  Eq. \eqref{equations_motion_bulk_transformed_changedterminology} indicates that the interaction between each cavity mode with the corresponding collective matter mode is described classically within the \MC{} model. As a consequence, the description of this coupling is fully equivalent to the analysis of the coupling between the same cavity mode and an individual dipole of frequency $\Omega_\text{mat}$ and increased coupling strength $G_{\scriptscriptstyle{\text{\MC{}}}}$, as indicated schematically in Fig. \ref{figure_bulk_cavity}a, so that the response of the cavity filled by a large number of molecular emitters can be described by adapting the analysis and conclusions in Sec. \ref{subsec_molecule_dielectric}. For example, the expression of the eigenvectors as a function of the contributions from the cavity and collective molecular modes can be obtained in principle using Eq. \eqref{ratio_xcav_xmat}. The electric field inside the cavity corresponding to each hybrid mode could be obtained by noticing that i) $x_{\text{cav},\alpha}$ gives the amplitude of the vector potential $\mathcal{A}_\alpha$, ii) the oscillator $x_{\text{mat},\alpha}$ is proportional to the induced dipole moment $\mathbf{d}_\alpha$, which enables to calculate the individual induced dipole moments $\mathbf{d}_i$ by inverting the relation $\mathbf{d}_\alpha = \frac{1}{\sqrt{N_\text{eff}}} \sum_i \Xi_\alpha(\mathbf{r}_i) \mathbf{d}_i$ for each $\alpha$ and iii) these single-molecule quantities lead to the polarization density $\mathbf{P}(\mathbf{r}) = \frac{\mathbf{d}_i(\mathbf{r})}{\Delta V}$, where $\Delta V$ is the volume that each individual dipole occupies ($\Delta V$ is the same for all dipoles).

\begin{figure}
\centering
\includegraphics[scale=0.8]{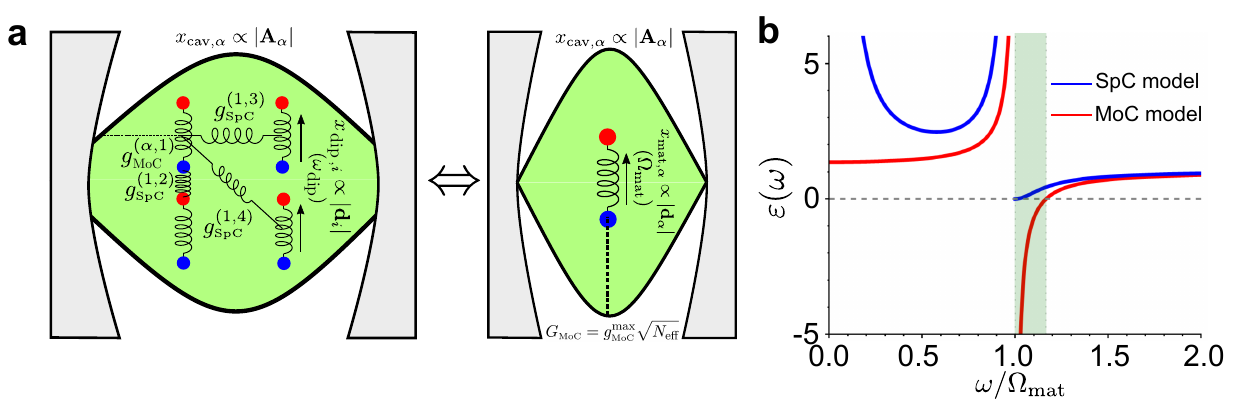}
\caption{ Interaction between matter excitations within a homogeneous material  and the transverse modes of a dielectric cavity. a) (Left) Schematic of the system. The homogeneous material is modelled as an ensemble of dipolar molecular emitters with a vibration at frequency $\omega_{\text{dip},i}$ . 
The oscillators $x_{\text{cav},\alpha}$ represent the vector potential $\mathbf{A}_{\alpha}$ associated with all  modes $\alpha$ in the cavity, and the individual matter oscillators $x_{\text{dip},i}$ represent the induced dipole moments $\mathbf{d}_i$ of each molecular emitter. The cavity-molecular emitter interactions are modeled with the \MC{} model and coupling strength $g_{\scriptscriptstyle{\text{\MC{}}}}^{(\alpha,i)}$, and the molecule-molecule dipolar interactions with the SpC model and coupling strength $g_{\scriptscriptstyle{\text{SpC}}}^{(i,j)}$. We indicate all the interactions of the molecular emitter with index $i=1$. (Right) Schematic indicating that the description of the full system is equivalent to the coupling, within the \MC{} model, of the cavity mode $\alpha$ with a single molecular excitation of induced dipole moment $\mathbf{d}_\alpha$, modified frequency $\Omega_\text{mat}$ and modified coupling strength $G_{\scriptscriptstyle{\text{\MC{}}}}$. b) Permittivity of the material inside the cavity, obtained from the classical SpC model (blue solid line, Eq. \eqref{eq_permittivity_SpC}) and the \MC{} model (red solid line, Eq. \eqref{eq_permittivity_MC}), for the collective coupling strength $G = 0.3 \, \omega_\text{cav}$. } \label{figure_bulk_cavity}
\end{figure}

We have thus shown that the \MC{} model constitutes the proper description of the coupling between transverse cavity modes and collective matter excitations in homogeneous materials. We further confirm the validity of this model to describe the system by demonstrating that it allows for recovering the typical bulk permittivity of phononic materials or ensembles of molecules and that this cannot be captured by the SpC model. We first note that, according to recent work \cite{barraburillo21, canales21,abujetas19}, the dispersion of the cavity-matter system is exactly the same as the bulk dispersion of the material. This enables to relate the spectrum of the \MC{} model with the bulk permittivity  $\varepsilon(\omega)$ of the material in the following manner:  the cavity modes of the bare cavity (without molecular emitters) follow the dispersion of free photons as $\omega_{\text{cav},\alpha} = ck_\alpha$, with  $c$ the light speed in vacuum and  $k_\alpha$ the wavevector that is determined by the length $L_\text{cav}$ of the Fabry-Pérot cavity (for perfect mirrors) as $k_\alpha = n_\alpha \pi  / L_\text{cav}$, for an integer $n_\alpha$ and normal incidence. For the cavity filled with molecular emitters,  the frequency of each cavity mode of wavevector $k_\alpha$ is modified from $\omega_{\text{cav},\alpha}$  to $\omega = \frac{ck_\alpha}{\sqrt{\varepsilon(\omega)}} = \frac{\omega_{\text{cav},\alpha}}{\sqrt{\varepsilon(\omega)}} $ due to the permittivity of the material. According to the discussion above, these $\omega$ values must be equal to the eigenfrequencies $\omega_{\pm,\scriptscriptstyle{\text{\MC{}}}}$ of the \MC{} model. From Eq. \eqref{equations_motion_classical_velocities_D_w}, we know that the \MC{} eigenfrequencies and the bare cavity frequencies are related as $(\omega_{\text{cav},\alpha}^2 - \omega_{\pm,\scriptscriptstyle{\text{\MC{}}}}^2)(\Omega_\text{mat}^2-\omega^2) - 4G_{\scriptscriptstyle{\text{\MC{}}}}^2\omega_{\pm,\scriptscriptstyle{\text{\MC{}}}}^2=0$. We can rewrite this relation as 
\begin{equation}
\omega_{\pm,\scriptscriptstyle{\text{\MC{}}}}^2 = \frac{\omega_{\text{cav},\alpha}^2}{1 + \frac{4G_{\scriptscriptstyle{\text{\MC{}}}}^2}{\Omega_\text{mat}^2 - \omega^2}} .
\label{eq_permittivity_MC_2}
\end{equation}
By comparing Eq. \eqref{eq_permittivity_MC_2} with the previous relation  $\omega = \frac{ck_\alpha}{\sqrt{\varepsilon(\omega)}} = \frac{\omega_{\text{cav},\alpha}}{\sqrt{\varepsilon(\omega)}}$ it is possible to identify the permittivity of the material in the cavity as
\begin{equation}
\varepsilon_{\scriptscriptstyle{\text{\MC{}}}}(\omega) =  1 + \frac{4G_{\scriptscriptstyle{\text{\MC{}}}}^2}{\Omega_\text{mat}^2 - \omega^2}.
\label{eq_permittivity_MC}
\end{equation}

Eq. \eqref{eq_permittivity_MC} is the same that was discussed by Hopfield \cite{hopfield58} and can be compared with the permittivity of polar materials. The latter can often be described in a range of infrared frequencies as  
\begin{equation}
\varepsilon(\omega) = \varepsilon_\infty \left( 1 + \frac{\omega_\text{LO}^2 - \omega_\text{TO}^2}{\omega_\text{TO}^2-\omega^2} \right),
\label{eq_permittivity_polar_materials}
\end{equation}
where $\omega_\text{TO}$ and $\omega_\text{LO}$ are the frequencies of the transverse optical and longitudinal optical phonons, respectively \cite{ashcroftmermin}.
Thus, the \MC{} model recovers the permittivity of a polar material or an ensemble of molecules, with the correspondences $\Omega_\text{mat} = \omega_\text{TO}$ and  $G_{\scriptscriptstyle{\text{\MC{}}}} = \sqrt{\frac{\omega_\text{LO}^2-\omega_\text{TO}^2}{4}}$. The only difference is that Eq. \eqref{eq_permittivity_MC} does not include the high-frequency permittivity $\varepsilon_\infty$ because this parameter originates from additional molecular excitations that are not considered in our model. In order to show that  the  \MC{} model is the only model with bare frequencies that correctly describes the permittivity of these materials, we derive the permittivity $\varepsilon_{\scriptscriptstyle{\text{SpC}}}(\omega)$ obtained within the SpC model by repeating the procedure with Eq. \eqref{equations_motion_classical_position_D0_w}. We obtain:
\begin{equation}
\varepsilon_{\scriptscriptstyle{\text{SpC}}}(\omega) =   \left( \frac{}{} \frac{2G_{\scriptscriptstyle{\text{SpC}}}^2 \Omega_\text{mat}}{  \omega(\Omega_\text{mat}^2-\omega^2)} + \sqrt{1 + \left(\frac{2G_{\scriptscriptstyle{\text{SpC}}}^2 \Omega_\text{mat}}{  \omega(\Omega_\text{mat}^2-\omega^2)} \right)^2} \right)^2,
\label{eq_permittivity_SpC}
\end{equation}
\noindent which does not follow the standard form of the permittivity (Eq. \eqref{eq_permittivity_polar_materials}). 

For comparison,  we plot in Fig. \ref{figure_bulk_cavity}b the permittivities obtained with the \MC{} model (red solid line, Eq. \eqref{eq_permittivity_MC}) and the SpC model (blue solid line, Eq. \eqref{eq_permittivity_SpC}), as a function of the normalized frequency $\omega/\Omega_\text{mat}$, with $G = 0.3 \, \Omega_\text{mat}$. The distinct behavior of permittivity predicted by the two models becomes evident when comparing their Reststrahlen bands. The Reststrahlen band represent the frequency range where electromagnetic waves cannot propagate in the bulk material (and also correspond to the maximum polaritonic gap achievable through the coupling of matter excitations with optical modes in dielectric resonators \cite{barraburillo21,chen15}).
The Reststrahlen band is delimited in polar materials by the phonon frequencies $\omega_\text{TO}$ and $\omega_\text{LO}$. The \MC{} model describes the Reststrahlen band appropriately, because the permittivity is negative in the range  $\omega \in \left(\Omega_\text{mat}, \sqrt{\Omega_\text{mat}^2 + 4G_{\scriptscriptstyle{\text{\MC{}}}}^2} \right) = (\omega_\text{TO},\omega_\text{LO})$ (highlighted by the green area in Fig. \ref{figure_bulk_cavity}b). In contrast, the permittivity $\varepsilon_{\scriptscriptstyle{\text{SpC}}}$ associated with the SpC model is non-negative for all frequencies and thus is unable to describe the presence of a Reststrahlen band.

As an additional difference between both models, only the \MC{} model results in a permittivity that does not diverge in the $\omega \rightarrow 0$ limit, in agreement with the expected behavior (Eq. \eqref{eq_permittivity_polar_materials}). We further  discuss the classical modeling of the Reststrahlen band in Sec. \ref{Appendix_reststrahlen} of the Supplementary Material, where we do not require the use in the coupled harmonic oscillator equations of the resonant frequency  of the $\it bare$ excitation of the material ($\Omega_\text{mat}=\omega_\text{TO}$) and cavity (which is the choice that defines the \MC{}, see discussion at the end of Sec. \ref{sec_derivation_classical}  and before Eq. \eqref{equations_motion_bulk_transformed_changedterminology}). We show that, without this constraint, i.e., by using a $\it dressed$  excitation of the material or the cavity, the Reststrahlen band can also be accurately described by alternative models where the coupling term is proportional to the oscillation amplitude.

In this subsection, we have focused on the coupling with (harmonic) vibrations and phonons. Still, the discussion holds validity for other dipolar matter excitations, independent of their physical origin, such as molecular excitons. The main difference between excitons and vibrations is that the former are two-level systems (fermionic transitions), which, when the number of coupled molecules is low enough, introduces many non-linear effects not included in classical harmonic oscillator models.  However, when many molecules are present,  the collective excitation is bosonic according to the Holstein-Primakoff transformation \cite{holstein40}.  Therefore, while the discussions in Secs. \ref{subsec_molecule_dielectric} and \ref{subsec_molecule_nanoparticle} are valid for harmonic excitations or for obtaining properties such as eigenvalues and electric field distribution under weak illumination, the discussion in this subsection is applicable more broadly.

\section{Conclusions} \label{sec_conclusions}

We have analyzed the application of classical coupled harmonic oscillator models to describe nanophotonic systems under ultrastrong coupling and the connection of these models with quantum descriptions. This study focuses on the two classical models typically used in this context, here referred to as the Spring Coupling (SpC) and Momentum Coupling (\MC{}) models, where the difference relies on whether the coupling term is proportional to the oscillation amplitudes (SpC model) or to their time derivatives (\MC{} model). The choice between these models typically does not have significant consequences in the weak and strong coupling regimes, where both can be approximated to the same linearized model (this approximation is discussed in the Supplementary Material and is equivalent to the rotating-wave approximation in quantum models). However, the SpC and \MC{} models result in very different eigenvalues in the ultrastrong coupling regime. We show that the SpC model describes light-matter coupling induced by Coulomb interactions, such as those governing the interaction between different quantum emitters and between quantum emitters and small plasmonic nanoparticles, and that this model results in the same eigenvalue spectra as the quantum Hopfield Hamiltonian without diamagnetic term. On the other hand, the \MC{} model reproduces the spectra of systems for which the diamagnetic term should be present in the Hamiltonian, corresponding to systems where matter excitations interact with transverse electromagnetic fields (for example, in conventional dielectric cavities). The SpC and \MC{} models thus result in the same spectra of ultrastrongly-coupled nanophotonic systems as a cavity-QED description without and with diamagnetic term, respectively, but using a simpler framework. These two classical models consider the bare cavity and matter frequencies, but we generalize the discussion in the Supplementary Material (Sec. \ref{Appendix_alternative_models})  to alternative models of classical oscillators. This generalized analysis indicates that dressing the frequencies allows us to transform coupled harmonic oscillator models where the coupling is proportional to the oscillation amplitudes to equivalent equations with coupling proportional to their time derivatives and vice versa.

Additionally, classical oscillator models are typically used to calculate the eigenvalues of the system, but we discuss how they also provide other experimentally measurable magnitudes in three canonical systems of nanophotonics. We first show that the \MC{} model can be applied to calculate the electric field distribution of the two hybrid modes of a dielectric cavity filled by a single quantum emitter. Next,  we use the SpC model to calculate the near-field distribution and the far-field scattering spectra of a quantum emitter located near a metallic nanoparticle. Last, the two models are combined to consider an ensemble of molecules inside a dielectric cavity. The molecules that interact with each other through Coulomb interactions (SpC model) and also with the transverse electromagnetic modes of the dielectric cavity (\MC{} model). In this case, we show that the system response can be obtained by considering that each transverse cavity mode interacts with a collective molecular excitation. The only effect of the molecule-molecule dipolar interactions is to modify the effective frequency of these collective excitations, and the \MC{} model describes the ultrastrong coupling between these collective excitations and the cavity modes. Interestingly, the \MC{} model enables to recover correctly the permittivity and bulk dispersion of the material filling the cavity, and thus also the Reststrahlen band observed in polar materials, which is not the case for the SpC model. Alternative coupled harmonic oscillator models of the bulk dispersion are discussed in Sec. \ref{Appendix_reststrahlen} of the Supplementary Material. Our work hence advances the exploration of classical descriptions of the ultrastrong coupling regime. It opens the possibility of simplifying the analysis of a wide variety of complex systems often described with quantum models.

\section*{Funding}
U. M., J. A., and R. E. acknowledge grant PID2022-139579NB-I00 funded by MICIU/AEI/10.13039/ 501100011033 and by ERDF, EU, as well as grant no. IT 1526-22 from the Basque Government for consolidated groups of the Basque University and project 4usmart Elkartek funded by the Department of Economy of the Basque Government. R.H acknowledges Grant CEX2020-001038-M funded by MICIU/AEI/10.13039/501100011033 and Grant PID2021-123949OB-I00 funded by MICIU/AEI/10.13039/501100011033 and by ERDF/EU. L. M. M. acknowledges projects PID2020-115221GB-C41 and CEX2023-001286-S (financed by MICIU/AEI/10.13039/501100011033) and the Government of Aragon through Project Q-MAD.

\vspace{2 cm}

\noindent The datasets used to generate the figures in this paper are available in https://digital.csic.es/handle/ 10261/380579/.

\newpage
\clearpage
\renewcommand{\theequation}{S\arabic{equation}}
\renewcommand{\thesection}{S\arabic{section}}
\renewcommand{\thefigure}{S\arabic{figure}}

\renewcommand{\thepage}{S\arabic{page}}  
\setcounter{page}{1}
\setcounter{equation}{0}
\setcounter{figure}{0}
\setcounter{section}{0}

\begin{center}
\begin{LARGE}

\textbf{Supplementary Material for:}
\medskip

\textbf{Description of ultrastrong light-matter interaction through coupled harmonic oscillator models and their connection with cavity-QED Hamiltonians}
\end{LARGE}

\bigskip

\begin{large}
Unai Muniain$^{*1}$, Javier Aizpurua$^{1,2,3}$, Rainer Hillenbrand$^{2,3, 4}$, Luis Martín-Moreno$^{5, 6}$ and Ruben Esteban$^{*1, 7}$
\end{large}
\begin{small}

$^1$\textit{Donostia International Physics Center, Paseo Manuel de Lardizabal 4, 20018 Donostia-San Sebastián, Spain}

$^2$\textit{IKERBASQUE, Basque Foundation for Science, María Díaz de Haro 3, 48013 Bilbao, Spain}

$^3$\textit{Department of Electricity and Electronics, University of the Basque Country (UPV/EHU), 48940 Leioa, Spain }

$^4$\textit{CIC nanoGUNE BRTA, Tolosa Hiribidea 76, 20018 Donostia-San Sebastián, Spain }

$^5$\textit{Instituto de Nanociencia y Materiales de Aragón (INMA), CSIC-Universidad de Zaragoza, 50009 Zaragoza, Spain }

$^6$\textit{Departamento de Física de la Materia Condensada, Universidad de Zaragoza, 50009 Zaragoza, Spain }

$^7$\textit{Centro de Física de Materiales (CFM-MPC), CSIC-UPV/EHU, Paseo Manuel de Lardizabal 5, 20018 Donostia-San Sebastián, Spain}

Corresponding authors: Unai Munian (unaimuni@gmail.com); Ruben Esteban (ruben.esteban@ehu.eus)
\end{small}
\medskip
\end{center}
\addtocontents{toc}{\setcounter{tocdepth}{2}}
\tableofcontents

\section{Derivation of the equations of motion in the classical coupled harmonic oscillator models}
\label{Appendix_equationsofmotion}

In the main article, we derive the classical models of coupled harmonic oscillators from the cavity Quantum Electrodynamics (QED) Hamiltonians. In this supplementary section, we derive in detail the equations of motion of the classical harmonic oscillators within a classical electromagnetic description that departs from the classical Lagrangian (Sec. \ref{Appendix_quantumHamiltonians} shows how to use this approach to obtain also the cavity-QED Hamiltonians). 

We start this derivation from the general classical Lagrangian representing charges and electromagnetic fields, which we then particularize for the specific systems we analyze in the main article.  Afterward, we show that the Spring Coupling (SpC) and the Momentum Coupling (\MC{}) models defined in the main article are obtained from the Euler-Lagrange equations of motion of these Lagrangians. Thus, a fully classical description is enough to model ultrastrong coupling in different nanophotonic systems without the need to use any quantum model. Last, we discuss how to introduce laser illumination into the SpC Model (necessary for Sec.  \ref{subsec_molecule_nanoparticle} of the main text) and confirm  the validity of the SpC model by comparing it with a an alternative description based on classical polarizabilities. 

The form of the electromagnetic Lagrangian depends on the gauge. We choose the Coulomb gauge, which leads to the following expression \cite{cohentannoudji}:
\begin{equation}
L_\text{Cou} = \sum_j \frac{1}{2}m_j \dot{\mathbf{r}}_j^2 - \sum_{i,j>i} \frac{q_i q_j}{4\pi\varepsilon_0 |\mathbf{r}_i-\mathbf{r}_j|} + \int \left[ \frac{\varepsilon_0}{2}(|\dot{\mathbf{A}}(\mathbf{r})|^2 - c^2 |\nabla \cross \mathbf{A}(\mathbf{r})|^2) + \mathbf{j}(\mathbf{r}) \cdot \mathbf{A} (\mathbf{r})  \right] d\mathbf{r}. \label{general_EM_lagrangian}
\end{equation}
In this Lagrangian, the electromagnetic degrees of freedom are encapsulated in the dynamical field variable $\mathbf{A}(\mathbf{r})$, which represents the vector potential of the fields, with the condition $\nabla \cdot \mathbf{A} = 0$ due to the choice of gauge. The energy of these fields is scaled by the vacuum permittivity $\varepsilon_0$ and the light speed in vacuum $c$ (for simplicity, we assume in this section that the material filling the cavity is vacuum). On the other hand, all the dynamics related to the matter structure are expressed by the spatial positions $\mathbf{r}_i$, mass $m_i$, and charge $q_i$ of each point-like charge indexed by $i$. Each point charge interacts with all the others according to the Coulomb potential energy (second term on the right-hand side) and with the transverse electromagnetic fields (according to the $\int \mathbf{j}(\mathbf{r})\cdot \mathbf{A}(\mathbf{r})d\mathbf{r}$ term, where $\mathbf{j}(\mathbf{r}) = \sum_i q_i \dot{\mathbf{r}}_i \delta(\mathbf{r}-\mathbf{r}_i)$ is the current density at any position $\mathbf{r}$). 

The equations of motion obtained from the Lagrangian in Eq. \eqref{general_EM_lagrangian} for the variables $\mathbf{A}(\mathbf{r})$ and $\mathbf{r}_i$ are equivalent to Maxwell's equation for a general system. We are interested in obtaining the equations of motion that describe the dynamics of systems formed by molecules or similar quantum emitters interacting with cavity modes in the strong and the ultrastrong coupling regimes. First, we focus on the terms of the Lagrangian related to the electromagnetic field (which in the Coulomb gauge is entirely described with the vector potential $\mathbf{A}$). The vector potential is separated into the components $\mathbf{A}_\alpha(\mathbf{r})$ of all transverse modes $\alpha$ of the cavity as $\mathbf{A}(\mathbf{r}) = \sum_\alpha \mathbf{A}_\alpha (\mathbf{r}) = \sum_\alpha \mathcal{A}_\alpha \Xi_\alpha(\mathbf{r}) \mathbf{n}_\alpha(\mathbf{r})$. For each $\alpha$ index, the field is polarized at any position in the direction determined by the unit vector $\mathbf{n}_\alpha(\mathbf{r})$, the maximum scalar amplitude is given by $\mathcal{A}_\alpha$ and the fields have spatial distribution $\Xi_\alpha(\mathbf{r})$, normalized so that  $\Xi_\alpha(\mathbf{r}) = 1$ in the position where the field is maximum. Further, we consider that the $\alpha$ modes form an orthogonal basis, and the integral of the field distribution over space gives the effective volume of the mode, i.e.
\begin{equation}
\label{eq:volume}
\int \Xi_\alpha (\mathbf{r}) \Xi^*_{\alpha'} (\mathbf{r}) \mathbf{n}_\alpha(\mathbf{r}) \cdot \mathbf{n}_{\alpha'}(\mathbf{r}) \; d\mathbf{r} = V_{\text{eff},\alpha} \delta_{\alpha,\alpha'}.
\end{equation}
By taking into account the decomposition of the modes and their orthogonality, the terms of the Lagrangian of Eq. \eqref{general_EM_lagrangian} only related to the electromagnetic fields are written as
\begin{align}
\int \frac{\varepsilon_0}{2} \left( \left| \sum_\alpha \dot{\mathcal{A}}_\alpha \Xi_\alpha(\mathbf{r}) \mathbf{n}_\alpha   \right|^2  - c^2 \left| \nabla \cross \sum_\alpha \mathcal{A}_\alpha \Xi_\alpha(\mathbf{r}) \mathbf{n}_\alpha   \right|^2     \right) d\mathbf{r} = \sum_\alpha \frac{\varepsilon_0 V_{\text{eff},\alpha}}{2} \left( \dot{\mathcal{A}}_\alpha \dot{\mathcal{A}}^*_\alpha -   \omega_{\text{cav},\alpha}^2 \mathcal{A}_\alpha \mathcal{A}^*_\alpha  \right). \label{lagrangian_cavity}
\end{align} 

We now focus on the terms of the Lagrangian associated with the matter degrees of freedom to describe the matter excitations. We model the material as an ensemble of dipoles indexed by $j$, each formed by two point charges that have the same mass $m_j$ and opposite charges and are placed in positions $\mathbf{r}_{j+}$ and $\mathbf{r}_{j-}$ (representing e.g. the simplest description of a quantum emitter).  At equilibrium, $\mathbf{r}_{j+}-\mathbf{r}_{j-} = \mathbf{r}_{j}^\text{eq}$, where $\mathbf{r}_{j}^\text{eq}$ can take into account the coupling with other dipoles. For example, when modeling a complex molecule $\mathbf{r}_{j}^\text{eq}$ would be obtained  including the interaction between all charges forming the molecule.  We make the harmonic approximation to the Coulomb potential experienced by each dipole with respect to the equilibrium position: $\approx \frac{1}{2}m_\text{red} \omega_\text{mat}^2 |\mathbf{r}_{j+} - \mathbf{r}_{j-} - \mathbf{r}_{j}^\text{eq}|^2$, where $m_{\text{red}}$ is the reduced mass of the dipole. We also assume that the mass center of the dipole is static at position  $\mathbf{r}_j = \frac{\mathbf{r}_{j+}+\mathbf{r}_{j-}}{2}$, while the distance between point charges from the equilibrium position, i.e., $\mathbf{l}_j = \mathbf{r}_{j+} - \mathbf{r}_{j-} - \mathbf{r}_\text{eq}$ and, equivalently, the induced dipole moment $\mathbf{d}_j = q_j \mathbf{l}_j$, evolve in time. From these assumptions, the Coulomb potential energy in the second term in Eq. \eqref{general_EM_lagrangian} includes the harmonic potential energy corresponding to the charges in each dipole and the potential energy due to the interaction between different dipoles.  Accordingly, the terms related to the matter degrees of freedom in the Lagrangian transform as
\begin{align}
\sum_{j} \frac{1}{2}m_j \dot{\mathbf{r}}_j^2 - \sum_{i,j>i} \frac{q_i q_j}{4\pi\varepsilon_0 |\mathbf{r}_i-\mathbf{r}_j|} =& \sum_j \left( \frac{1}{2} \frac{m_{\text{red},j}}{q_j^2} \dot{d}_j^2 - \frac{1}{2}\frac{m_{\text{red},j}}{q_j^2} \omega_{\text{mat},j}^2 d_j^2 \right) \nonumber \\
&- \sum_{i,j>i} \frac{1}{4\pi \varepsilon_0 |\mathbf{r}_i - \mathbf{r}_j|^3} \left[ \mathbf{d}_i \cdot \mathbf{d}_j - 3 (\mathbf{d}_i \cdot \mathbf{n}_{\mathbf{r}ij})(\mathbf{d}_j \cdot \mathbf{n}_{\mathbf{r}ij}) \right], \label{lagrangian_matter}
\end{align}
with $d_j = |\mathbf{d}_j|$ and the unit vector $\mathbf{n}_{\mathbf{r}ij}  = \frac{\mathbf{r}_j - \mathbf{r}_i}{|\mathbf{r}_j - \mathbf{r}_i|}$. Equation \eqref{lagrangian_matter} has been derived using the harmonic approximation of the dipolar potential and, as a consequence, all terms of the Lagrangian that do not account for light-matter interaction are quadratic with respect to the amplitudes of the vector potential and their time derivatives (Eq. \eqref{lagrangian_cavity}), or with respect to the induced dipole moments and their time derivatives (Eq. \eqref{lagrangian_matter}). Therefore, if light and matter were uncoupled, the dynamical evolution of these variables would be the same as that of free harmonic oscillators. We now discuss how the interaction between the cavity modes and the dipoles affects the equations of motion.  The coupling of each dipole with the transverse fields of the cavity appears in the Lagrangian as
\begin{align}
\int \mathbf{j} \cdot \mathbf{A} \; d\mathbf{r} &= \int \left(\sum_j q_j \dot{\mathbf{r}}_{j+} \delta(\mathbf{r}-\mathbf{r}_{j+}) - q_j \dot{\mathbf{r}}_{j-} \delta(\mathbf{r}-\mathbf{r}_{j-})\right)\left( \sum_\alpha \mathcal{A}_\alpha \Xi_\alpha(\mathbf{r}) \mathbf{n}_\alpha \right) d\mathbf{r} \nonumber \\
&= \sum_{j,\alpha} q_j [\mathbf{r}_{j+}\Xi_\alpha(\dot{\mathbf{r}}_{j+}) - \dot{\mathbf{r}}_{j-}\Xi_\alpha(\mathbf{r}_{j-})] \mathcal{A}_\alpha \mathbf{n}_\alpha \approx \sum_{j,\alpha}  \mathcal{A}_\alpha \Xi_\alpha(\mathbf{r}_{j}) \dot{\mathbf{d}}_j \cdot \mathbf{n}_\alpha
\end{align}
In the last step, we have performed the long-wavelength approximation, so that the fields do not vary in the length scale of each dipole, i.e., $\Xi(\mathbf{r}_{j+}) \approx \Xi(\mathbf{r}_{j-})$ for any $j$. The total Lagrangian of the system in the Coulomb gauge reads 
\begin{align}
L_\text{Cou}(d_j,\dot{d}_j, \mathcal{A}_\alpha, \dot{\mathcal{A}}_\alpha, \mathcal{A}^*_\alpha, \dot{\mathcal{A}}^*_\alpha ) =&   \sum_\alpha \frac{\varepsilon_0 V_{\text{eff},\alpha}}{2} \left( \dot{\mathcal{A}}_\alpha \dot{\mathcal{A}}^*_\alpha -   \omega_{\text{cav},\alpha}^2 \mathcal{A}_\alpha \mathcal{A}^*_\alpha  \right) +  \sum_j \frac{1}{2} \frac{1}{f_{\text{mat},j}} \left(\dot{d}_j^2 - \omega_{\text{mat},j}^2 d_j^2 \right)\nonumber \\
+& \sum_{j,\alpha}  \mathcal{A}_\alpha \dot{d}_j \Xi_\alpha(\mathbf{r}_{j}) \cos \theta_{\alpha,j} - \sum_{i,j} d_i d_j \frac{\mathbf{n}_{\mathbf{d}i} \cdot \mathbf{n}_{\mathbf{d}j} - 3(\mathbf{n}_{\mathbf{d}i} \cdot \mathbf{n}_{\mathbf{r}ij})(\mathbf{n}_{\mathbf{d}j} \cdot \mathbf{n}_{\mathbf{r}ij})}{4\pi \varepsilon_0 |\mathbf{r}_i - \mathbf{r}_j|^3}, \label{lagrangian_final}
\end{align}
where  $\mathbf{n}_{\mathbf{d}j} = \frac{\mathbf{d}_j}{|\mathbf{d}_j|}$,  $\theta_{\alpha,j}$ is the angle between the induced  dipole moment $\mathbf{d}_j$ and the direction $\mathbf{n}_\alpha$ of the electric field in the mode $\alpha$, and $f_\text{mat} = \frac{q_j^2}{m_\text{red}}$ is the oscillator strength of the $j^\text{th}$ dipole.

From the Lagrangian $L_\text{Cou}$ of Eq. \eqref{lagrangian_final}, we can derive the equations of motion of the classical coupled harmonic oscillators by calculating the Euler-Lagrange equations, $\frac{d}{dt}\frac{\partial L_\text{Cou}}{\partial \dot{x}} - \frac{\partial L_\text{Cou}}{\partial x} = 0$, for $x \in\{d_j, \mathcal{A}^*_\alpha \}$.  The resulting equations of motion are
\begin{subequations}
\begin{align}
&\ddot{\mathcal{A}}_\alpha + \omega_{\text{cav},\alpha}^2 \mathcal{A}_\alpha - \sum_j \dot{d}_j \frac{ \Xi_\alpha(\mathbf{r}_{j}) \cos \theta_{\alpha,j} }{\varepsilon_0 V_{\text{eff},\alpha}} = 0, \\
&\ddot{d}_j + \omega_{\text{mat},j}^2 d_j + f_{\text{mat},j}\sum_{i \neq j} \frac{\mathbf{n}_{\mathbf{d}i} \cdot \mathbf{n}_{\mathbf{d}j} - 3(\mathbf{n}_{\mathbf{d}i} \cdot \mathbf{n}_{\mathbf{r}ij})(\mathbf{n}_{\mathbf{d}j} \cdot \mathbf{n}_{\mathbf{r}ij})}{4\pi \varepsilon_0 |\mathbf{r}_i - \mathbf{r}_j|^3} d_i + \sum_\alpha \dot{\mathcal{A}}_\alpha f_{\text{mat},j} \Xi^*_\alpha(\mathbf{r}_{j}) \cos \theta_{\alpha,j}   = 0.
\end{align}
\label{equations_motion_eulerlagrange}
\end{subequations}
These equations account for all dipole-cavity and dipole-dipole interactions, as analyzed in Sec. \ref{subsec_bulk_dielectric} of the main article. To show how to obtain the \MC{} and SpC models, we focus on the two canonical examples analyzed in Secs. \ref{subsec_molecule_dielectric} and \ref{subsec_molecule_nanoparticle} of the main article:

\begin{itemize}
    \item \underline{Coupling between a quantum emitter and a transverse mode of a dielectric cavity} (Sec. \ref{subsec_molecule_dielectric}): By considering a single transverse mode $\alpha$ of the cavity interacting with one molecular emitter with induced dipole moment $d$, all Coulomb interactions in Eq.  \eqref{equations_motion_eulerlagrange} are eliminated. The equations of motion become
    \begin{subequations}
    \begin{align}
    &\ddot{\mathcal{A}} + \omega_{\text{cav}}^2 \mathcal{A} -  \dot{d} \frac{ \Xi(\mathbf{r}_\text{mat}) \cos \theta }{\varepsilon_0 V_{\text{eff}}} = 0, \\
    &\ddot{d} + \omega_{\text{mat}}^2 d +  \dot{\mathcal{A}} f_{\text{mat}} \Xi^*(\mathbf{r}_\text{mat}) \cos \theta   = 0.
    \end{align}
    \label{equations_motion_eulerlagrange_MC}
    \end{subequations}
    By replacing here the oscillation amplitudes $x_\text{cav} = \mathcal{A} \sqrt{\varepsilon_0 V_\text{eff}}$ and $x_\text{mat} = \frac{d}{\sqrt{f_\text{mat}}}$ and introducing the coupling strength
    \begin{equation}
        g_{\scriptscriptstyle{\text{\MC{}}}} = \frac{1}{2}\sqrt{\frac{f_\text{mat}}{\varepsilon_0 V_\text{eff}}} \Xi(\mathbf{r}_\text{mat}) \cos \theta,
        \label{eq_coupling_strength_MC}
    \end{equation}we recover the equations of motion of the \MC{} model (Eq. \eqref{equations_motion_classical_velocities_D} in the main article).
    
    \item \underline{Coupling between a quantum emitter and a plasmonic nanoparticle via Coulomb interactions}  (Sec. \ref{subsec_molecule_nanoparticle}): We consider that the emitter (a molecule) and the nanoparticle have induced dipole moments $d_\text{mat}$ and $d_\text{cav}$, respectively. Under the quasistatic approximation of the plasmonic response, the vector potential components of all transverse modes are neglected. 
    With this approximation and for only two dipoles, Eq. \eqref{equations_motion_eulerlagrange} is written as
    \begin{subequations}
    \begin{align}
    &\ddot{d}_\text{cav} + \omega_\text{cav}^2 d_\text{cav} + f_{\text{cav}} \frac{\mathbf{n}_{\mathbf{d}\text{cav}} \cdot \mathbf{n}_{\mathbf{d}\text{mat}} - 3(\mathbf{n}_{\mathbf{d}\text{cav}} \cdot \mathbf{n}_{\mathbf{r}\text{rel}})(\mathbf{n}_{\mathbf{d}\text{mat}} \cdot \mathbf{n}_{\mathbf{r}\text{rel}})}{4\pi \varepsilon_0 |\mathbf{r}_\text{cav} - \mathbf{r}_\text{mat}|^3} d_\text{mat}   = 0, \\
    &\ddot{d}_\text{mat} + \omega_{\text{mat}}^2 d_\text{mat} + f_{\text{mat}}  \frac{\mathbf{n}_{\mathbf{d}\text{cav}} \cdot \mathbf{n}_{\mathbf{d}\text{mat}} - 3(\mathbf{n}_{\mathbf{d}\text{cav}} \cdot \mathbf{n}_{\mathbf{r}\text{rel}})(\mathbf{n}_{\mathbf{d}\text{mat}} \cdot \mathbf{n}_{\mathbf{r}\text{rel}})}{4\pi \varepsilon_0 |\mathbf{r}_\text{cav} - \mathbf{r}_\text{mat}|^3} d_\text{cav}   = 0,
    \end{align}
    \label{equations_motion_eulerlagrange_SpC}
    \end{subequations}
    where $\mathbf{n}_{\mathbf{r}\text{rel}} = \frac{\mathbf{r}_\text{cav}-\mathbf{r}_\text{mat}}{|\mathbf{r}_\text{cav}-\mathbf{r}_\text{mat}|}$ is the unitary vector of the relative direction between the nanocavity and the molecular emitter.
    By replacing $x_\text{cav} = \frac{d_\text{cav}}{\sqrt{f_\text{cav}}}$ and $x_\text{mat} = \frac{d_\text{mat}}{\sqrt{f_\text{mat}}}$, and defining the coupling strength $g_{\scriptscriptstyle{\text{SpC}}}$  as
    \begin{equation}
    g_{\scriptscriptstyle{\text{SpC}}} = \frac{1}{2} \frac{\sqrt{f_\text{cav}}\sqrt{f_\text{mat}}}{4\pi\varepsilon_0 |\mathbf{r}_\text{cav}-\mathbf{r}_\text{mat}|^3 \sqrt{\omega_\text{cav}\omega_\text{mat}}}[\mathbf{n}_{\mathbf{d}\text{cav}} \cdot \mathbf{n}_{\mathbf{d}\text{mat}} - 3(\mathbf{n}_{\mathbf{d}\text{cav}} \cdot \mathbf{n}_{\mathbf{r}\text{rel}})(\mathbf{n}_{\mathbf{d}\text{mat}} \cdot \mathbf{n}_{\mathbf{r}\text{rel}})],
    \label{g_dipole_dipole_SI}
    \end{equation}
    we recover the equations of the SpC model (Eq. \eqref{equations_motion_classical_position_D0} in the main article).
\end{itemize}

\subsubsection*{Spring coupling model with external laser illumination}

In Sec. \ref{subsec_molecule_nanoparticle} of the main text, the dipolar mode of a metallic nanoparticle is excited by an external laser. We now discuss briefly how to introduce the incident laser field in the model of the interaction of this metallic nanocavity with a quantum emitter, e.g. a molecule. The incident field is treated as a planewave of wavevector $\mathbf{k}_\text{inc}$, amplitude $\mathcal{A}_\text{inc}$ and frequency $\omega$, with an associated vector potential of the form  $\mathbf{A}_\text{inc}(\mathbf{r},t) =  \mathcal{A}_\text{inc} e^{i\mathbf{k}_\text{inc}\cdot \mathbf{r}}e^{-i\omega t}$. Under the quasistatic approximation, all transverse modes $\alpha$ of the system are neglected, and thus the only component of the vector potential considered in the Lagrangian of Eq. \eqref{lagrangian_final} corresponds to the external laser $\mathbf{A}_\text{inc}(\mathbf{r},t)$. With these considerations, the Lagrangian of Eq.  \eqref{lagrangian_final} becomes
\begin{align}
&L_\text{Cou}^\text{dip-dip} (d_\text{cav},\dot{d}_\text{cav}, d_\text{mat},\dot{d}_\text{mat} ) =  \frac{1}{2} \frac{1}{f_{\text{cav}}} \left(\dot{d}_{\text{cav}}^2 - \omega_{\text{cav}}^2 d_\text{cav}^2 \right) + \frac{1}{2} \frac{1}{f_{\text{mat}}} \left(\dot{d}_\text{mat}^2 - \omega_{\text{mat}}^2 d_\text{mat}^2 \right)  \nonumber \\
& - d_\text{cav} d_\text{mat} \frac{\mathbf{n}_{\mathbf{d}\text{cav}} \cdot \mathbf{n}_{\mathbf{d}\text{mat}} - 3(\mathbf{n}_{\mathbf{d}\text{cav}} \cdot \mathbf{n}_{\mathbf{r}\text{rel}})(\mathbf{n}_{\mathbf{d}\text{mat}} \cdot \mathbf{n}_{\mathbf{r}\text{rel}})}{4\pi \varepsilon_0 |\mathbf{r}_\text{cav} - \mathbf{r}_\text{mat}|^3} +  \mathcal{A}_\text{inc} e^{-i\omega t}(\dot{d}_\text{cav}  \cos \theta_{\text{inc},\text{cav}} + \dot{d}_\text{mat}  \cos \theta_{\text{inc},\text{mat}} ), 
\label{lagrangian_dipole_dipole}
\end{align}
where $\theta_{\text{inc},\text{cav}}$ and $\theta_{\text{inc},\text{mat}}$ are the angles between the incident field and the induced dipole moments of the cavity and molecular emitter, respectively. The superscript "dip-dip" emphasizes that we only consider dipole-dipole interactions for this system (under the quasistatic approximation).
The dynamics of the variables $d_\text{cav}$ and $d_\text{mat}$ are obtained within the Euler-Lagrange equations of Eq. \eqref{lagrangian_dipole_dipole}. By calculating these equations of motion and transforming the variables into the oscillation amplitudes  $x_\text{cav} = \frac{d_\text{cav}}{\sqrt{f_\text{cav}}}$ and $x_\text{mat} = \frac{d_\text{mat}}{\sqrt{f_\text{mat}}}$, the resulting equations are
\begin{subequations}
\begin{equation}
\ddot{x}_\text{cav} + \omega_\text{cav}^2 x_\text{cav} + \frac{\mathbf{n}_{\mathbf{d}\text{cav}} \cdot \mathbf{n}_{\mathbf{d}\text{mat}} - 3(\mathbf{n}_{\mathbf{d}\text{cav}} \cdot \mathbf{n}_{\mathbf{r}\text{rel}})(\mathbf{n}_{\mathbf{d}\text{mat}} \cdot \mathbf{n}_{\mathbf{r}\text{rel}})}{4\pi \varepsilon_0 |\mathbf{r}_\text{cav} - \mathbf{r}_\text{mat}|^3}  x_\text{mat} = - \sqrt{f_\text{cav}} \cos\theta_{\text{inc},\text{cav}} \frac{d}{dt}\left(\mathcal{A}_\text{inc} e^{-i\omega t} \right), 
\end{equation}
\begin{equation}
\ddot{x}_\text{mat} + \omega_\text{mat}^2 x_\text{mat} + \frac{\mathbf{n}_{\mathbf{d}\text{cav}} \cdot \mathbf{n}_{\mathbf{d}\text{mat}} - 3(\mathbf{n}_{\mathbf{d}\text{cav}} \cdot \mathbf{n}_{\mathbf{r}\text{rel}})(\mathbf{n}_{\mathbf{d}\text{mat}} \cdot \mathbf{n}_{\mathbf{r}\text{rel}})}{4\pi \varepsilon_0 |\mathbf{r}_\text{cav} - \mathbf{r}_\text{mat}|^3}  x_\text{cav}  =  -\sqrt{f_\text{mat}} \cos\theta_{\text{inc},\text{mat}} \frac{d}{dt}\left(\mathcal{A}_\text{inc} e^{-i\omega t} \right).
\end{equation}
\label{eq_supplementary_SpC_with_laser}
\end{subequations} 
Therefore, the incident field is incorporated into the SpC equations of motion (Eq. \eqref{equations_motion_classical_position_D0} in the main article) by adding time-dependent force-like terms of amplitude $F_\text{cav} = i\omega \mathcal{A}_\text{inc} \sqrt{f_\text{cav}} \cos\theta_{\text{inc},\text{cav}} $ and $F_\text{mat} = i\omega \mathcal{A}_\text{inc} \sqrt{f_\text{mat}} \cos\theta_{\text{inc},\text{mat}} $ to the nanocavity and the molecular emitter, respectively.

\subsubsection*{Classical description of the coupling between a molecular emitter and a plasmonic nanocavity based on their polarizability}

The interaction of a small metallic nanoparticle with a molecular emitter (or another quantum emitter) can also be described classically by using polarizabilities $\alpha_\text{cav}$ and $\alpha_\text{mat}$ for both particles so that the dipole moment induced by the electric field at each position $\mathbf{r}_\text{cav}$ and $\mathbf{r}_\text{mat}$ is given by $\mathbf{d}_\text{cav} = \alpha_\text{cav} \mathbf{E}(\mathbf{r}_\text{cav})$ and $\mathbf{d}_\text{mat} = \alpha_\text{mat} \mathbf{E}(\mathbf{r}_\text{mat})$, respectively. We briefly show here that this approach leads to the same equations as the SpC model obtained from the electromagnetic Lagrangian, which supports the validity of the general approach used in the main text. For the cavity mode (plasmon in metallic nanoparticle) and the molecular excitation (or any matter excitation in general), we consider the polarizability given by the Lorentz oscillator model. In the case of the molecular emitter, we assume a single molecular excitation with Lorentzian polarizability centered at resonant frequency $\omega_\text{mat}$,  linewidth determined by the damping frequency $\gamma$, and oscillator strength $f_\text{mat}$. Similarly, we also model the nanocavity response as given by a single plasmonic resonance that follows a Lorentzian-like lineshape (for a Drude permittivity), which is the typical lineshape in the quasistatic regime. This resonance is centered at frequency $\omega_\text{cav}$ and is characterized by losses $\kappa$ and oscillator strength $f_\text{cav}$. The polarizabilities of the plasmonic nanocavity and the molecular emitter are then given by
\begin{subequations}
\begin{equation}
\alpha_\text{cav}(\omega) = \frac{f_\text{cav}}{\omega_\text{cav}^2-\omega^2-i\omega \kappa},
\end{equation}
\begin{equation}
\alpha_\text{mat}(\omega) = \frac{f_\text{mat}}{\omega_\text{mat}^2-\omega^2-i\omega \gamma}.
\end{equation}
\label{eq_polarizabilities}
\end{subequations}
The dipole moment of the molecular emitter and the nanoparticle is induced by the electric field $\mathbf{E}_\text{inc}$ of the external laser and also by the electric field generated by either the plasmonic mode ($\mathbf{E}_\text{cav}$) or the matter excitation in the molecule ($\mathbf{E}_\text{mat}$), respectively. We then have $\mathbf{d}_\text{cav} = \alpha_\text{cav}[\mathbf{E}_\text{mat}(\mathbf{r}_\text{cav}) + \mathbf{E}_\text{inc}]$ and $\mathbf{d}_\text{mat} = \alpha_\text{mat}[\mathbf{E}_\text{cav}(\mathbf{r}_\text{mat}) + \mathbf{E}_\text{inc}]$. By inserting in these expressions the polarizabilities given by Eq. \eqref{eq_polarizabilities} and considering that the quasi-static fields induced by the dipoles excited at the cavity and the molecule follow the dependence,
\begin{subequations}
\begin{align}
   & \mathbf{E}_\text{mat}(\mathbf{r}_\text{cav}) = \frac{\mathbf{n}_{\mathbf{d}\text{mat}} - 3(\mathbf{n}_{\mathbf{d}\text{mat}} \cdot \mathbf{n}_{\mathbf{r}\text{rel}})\mathbf{n}_{\mathbf{r}\text{rel}}}{4\pi \varepsilon_0 |\mathbf{r}_\text{cav} - \mathbf{r}_\text{mat}|^3} d_\text{mat},\\
   & \mathbf{E}_\text{cav}(\mathbf{r}_\text{mat}) = \frac{\mathbf{n}_{\mathbf{d}\text{cav}} - 3(\mathbf{n}_{\mathbf{d}\text{cav}} \cdot \mathbf{n}_{\mathbf{r}\text{rel}})\mathbf{n}_{\mathbf{r}\text{rel}}}{4\pi \varepsilon_0 |\mathbf{r}_\text{cav} - \mathbf{r}_\text{mat}|^3 } d_\text{cav},
\end{align}
\end{subequations}
we obtain the expressions of the induced dipole moments
\begin{subequations}
\begin{equation}
(\omega_\text{cav}^2-\omega^2 - i\omega\kappa)d_\text{cav} = f_\text{cav} \left[ \frac{\mathbf{n}_{\mathbf{d}\text{cav}} \cdot \mathbf{n}_{\mathbf{d}\text{mat}} - 3(\mathbf{n}_{\mathbf{d}\text{cav}} \cdot \mathbf{n}_{\mathbf{r}\text{rel}})(\mathbf{n}_{\mathbf{d}\text{mat}} \cdot \mathbf{n}_{\mathbf{r}\text{rel}})}{4\pi \varepsilon_0 |\mathbf{r}_\text{cav} - \mathbf{r}_\text{mat}|^3} d_\text{mat} + \mathbf{E}_\text{inc} \cdot \mathbf{n}_{\mathbf{d}\text{cav}} \right],
\end{equation}
\begin{equation}
(\omega_\text{mat}^2-\omega^2 - i\omega\gamma)d_\text{mat} = f_\text{mat} \left[ \frac{\mathbf{n}_{\mathbf{d}\text{cav}} \cdot \mathbf{n}_{\mathbf{d}\text{mat}} - 3(\mathbf{n}_{\mathbf{d}\text{cav}} \cdot \mathbf{n}_{\mathbf{r}\text{rel}})(\mathbf{n}_{\mathbf{d}\text{mat}} \cdot \mathbf{n}_{\mathbf{r}\text{rel}})}{4\pi \varepsilon_0 |\mathbf{r}_\text{cav} - \mathbf{r}_\text{mat}|^3} d_\text{cav}  + \mathbf{E}_\text{inc} \cdot \mathbf{n}_{\mathbf{d}\text{mat}} \right].
\end{equation}
\end{subequations}
These equations are equivalent to Eq. \eqref{eq_supplementary_SpC_with_laser} in frequency domain, with $x_\text{cav} = \frac{d_\text{cav}}{\sqrt{f_\text{cav}}}$, $x_\text{mat} = \frac{d_\text{mat}}{\sqrt{f_\text{mat}}}$ and using the relation $|\mathbf{E}_\text{inc}| = |i\omega \mathcal{A}_\text{inc}|$ that follows from the definition of the vector potential.

\section{ Alternative classical models of coupled harmonic oscillators} \label{Appendix_alternative_models}

The discussion of Supplementary Sec. \ref{Appendix_equationsofmotion} concluded that the classical \MC{} model describes the coupling of matter excitations with transverse electromagnetic modes, while the SpC model can express dipole-dipole interactions. Crucially, the bare cavity and matter frequencies appear directly in these models without dressing the energies. 
In this supplementary section, we demonstrate that other classical coupled harmonic oscillator models exist, equivalent to the \MC{} and SpC models, but involving some frequency dressing (this effect is related to the discussion in Ref. \cite{kockum19rep} between the dressing of the frequencies and the presence or absence of diamagnetic term).  The alternative models depend on the gauge chosen for the classical Lagrangian and Hamiltonian descriptions. We discuss oscillator models in two of the most commonly used gauges: the Coulomb and dipole gauges. We also show that the physical interpretation of the oscillation amplitudes depends on the particular coupled harmonic oscillator model that is used.

More specifically, Secs. \ref{Appendix_alternative_models_Coulomb} and \ref{Appendix_alternative_models_dipole} consider the coupling with transverse modes in dielectric cavities. We derive alternative coupled harmonic oscillator equations that use dressed frequencies and coupling terms proportional to the amplitude of the oscillators (in contrast with the equivalent \MC{} model, which uses bare frequencies and coupling terms proportional to the time derivatives of the oscillator amplitudes). 
We first show in Sec. \ref{Appendix_alternative_models_Coulomb} how to derive, within the Coulomb gauge, an alternative coupled harmonic oscillator model in which the cavity mode is dressed.  Then, in Sec. \ref{Appendix_alternative_models_dipole}, the use of the dipole gauge yields a second alternative coupled harmonic oscillator model with dressed matter excitation and coupling terms again proportional to the oscillations amplitudes. 

Afterwards, in Sec S2.3, we consider  Coulomb coupling through longitudinal fields, and obtain coupled harmonic oscillator equations with dressing of the matter excitation and coupling term proportional to the time derivatives of the oscillator amplitudes (for comparison, in the equivalent SpC model, the frequencies are the bare ones and the coupling terms are proportional to the oscillator amplitudes of the oscillation models). This section considers the Coulomb gauge, but the dipole gauge yields identical results.

\subsection{Alternative model of a matter excitation interacting with transverse cavity modes obtained within the Coulomb gauge}
\label{Appendix_alternative_models_Coulomb}

We first describe the coupling between a transverse electromagnetic mode and a dipolar excitation of a molecule (or another quantum emitter), which is the system discussed in Sec. \ref{subsec_molecule_dielectric} of the main article. The aim is to obtain alternative equations of motion of this system. We start with the classical Lagrangian in the Coulomb gauge given by Eq. \eqref{lagrangian_final}, which for the considered system can be expressed as
\begin{equation}
L_\text{Cou}^\text{min-c}(d,\dot{d}, \mathcal{A}, \dot{\mathcal{A}})  = \frac{\varepsilon_0 V_\text{eff}}{2 } (\dot{\mathcal{A}}^2 - \omega_\text{cav}^2 \mathcal{A}^2) + \frac{1}{2f_\text{mat}} (\dot{d}^2 - \omega_\text{mat}^2 d^2) + \mathcal{A} \dot{d}.
\label{lagrangian_coulomb_gauge}
\end{equation}

To simplify the analytical expressions in the following discussion, we consider Eq. \eqref{lagrangian_coulomb_gauge} for a specific case where the molecular emitter is placed in the position of maximum field of the mode and oriented in the same direction as the field polarization so that  $\Xi(\mathbf{r_\text{mat}}) \cos\theta = 1$ (see Sec. \ref{Appendix_equationsofmotion} for the definition of these parameters). However, the discussion of this section remains valid for other values of $\Xi(\mathbf{r_\text{mat}}) \cos\theta.$

It has been shown in Supplementary Sec. \ref{Appendix_equationsofmotion} that the Euler-Lagrange equations derived from Eq. \eqref{lagrangian_coulomb_gauge} lead to the \MC{} model. We use here Hamilton's equations to derive the \MC{} model in an alternative manner and also to obtain another equivalent classical model of harmonic oscillators. To first derive the classical Hamiltonian of the system, we obtain the canonical momenta related to the transverse electromagnetic modes and the induced dipole moment in the Coulomb gauge as
\begin{subequations}
\begin{align}
\Pi_\text{Cou} =& \frac{\partial L_\text{Cou}}{\partial \dot{\mathcal{A}}} = \varepsilon_0 V_\text{eff} \dot{\mathcal{A}}, \\
p_\text{Cou} =& \frac{\partial L_\text{Cou}}{\partial \dot{d}} = \frac{\dot{d}}{f_\text{mat}}+\mathcal{A}.
\end{align}
\label{eq_canonical_momenta_coulomb_gauge}
\end{subequations}
According to these expressions, the dynamical variable $\Pi_\text{Cou}$ expresses the transverse electric field of the cavity modes from the relation $\mathbf{E} = -\frac{\partial \mathbf{A}}{\partial t}$. On the other hand,  the relation between the induced dipole moment $d$ and its canonical momentum $p_\text{Cou}$ is more complicated because $p_\text{Cou}$ depends not only on $d$ but also on the vector potential. Using Eq. \eqref{eq_canonical_momenta_coulomb_gauge}, the calculation of the Hamiltonian $H^\text{min-c}_\text{Cou} =  \dot{\mathcal{A}} \Pi_\text{Cou} +  \dot{d}  p_\text{Cou} - L^\text{min-c}_\text{Cou}$ is straightforward:
\begin{align}
H_\text{Cou}^\text{min-c} =&   \frac{\Pi_\text{Cou}^2}{2\varepsilon_0 V_\text{eff}} + \frac{1}{2}\varepsilon_0 V_\text{eff} \omega_\text{cav}^2 \mathcal{A}^2   + \frac{f_\text{mat}}{2}p_\text{Cou}^2 + \frac{1}{2}\frac{\omega_\text{mat}^2}{f_\text{mat}}d^2      - f_\text{mat} p_\text{Cou} \mathcal{A} + \frac{1}{2} f_\text{mat}\mathcal{A}^2.
\label{hamiltonian_coulomb_gauge}
\end{align}
This expression has the well-known form of the minimal-coupling Hamiltonian. This is the reason why we include the superindex "min-c" in the Lagrangian of Eq. \eqref{lagrangian_coulomb_gauge} and in the Hamiltonian of Eq. \eqref{hamiltonian_coulomb_gauge}. We can directly derive the Hamilton's equations of motion of all canonical variables:
\begin{subequations}
\begin{align}
&\dot{\mathcal{A}} = \frac{\partial H^\text{min-c}_\text{Cou}}{\partial \Pi_\text{Cou}} = \frac{\Pi_\text{Cou}}{\varepsilon_0 V_\text{eff}}, \\
&\dot{\Pi}_\text{Cou}  = -\frac{\partial H^\text{min-c}_\text{Cou}}{\partial \mathcal{A}} = -\varepsilon_0 V_\text{eff} \omega_\text{cav}^2 \mathcal{A} + f_\text{mat} (p_\text{Cou} - \mathcal{A}), \\
&\dot{d} = \frac{\partial H^\text{min-c}_\text{Cou}}{\partial p_\text{Cou}} = f_\text{mat} (p_\text{Cou} - \mathcal{A}), \\
&\dot{p}_\text{Cou} = -\frac{\partial H^\text{min-c}_\text{Cou}}{\partial d} =  -\frac{\omega_{\text{mat}}^2}{f_\text{mat}}   d_j .
\end{align}
\label{equations_motion_hamilton_coulomb}
\end{subequations}

Hamilton's equations can be used to obtain classical harmonic oscillator models by eliminating two variables, leading to two second-order differential equations. By choosing the variables $\mathcal{A}$ and  $d$ to describe the dynamics of the system, we obtain
\begin{subequations}
\begin{align}
&\ddot{\mathcal{A}} + \omega_\text{cav}^2 \mathcal{A} - \frac{\dot{d}}{\varepsilon_0 V_\text{eff}} = 0, \\
&\ddot{d} + \omega_\text{mat}^2 d + f_\text{mat} \dot{\mathcal{A}} = 0. 
\end{align}
\label{equations_motion_A_d}
\end{subequations}
This system of equations can be converted into Eq.  \eqref{equations_motion_classical_velocities_D} in the main text, and thus we recover the \MC{} model. However, there are other possible ways to represent the response of this system with harmonic oscillators. An alternative is to choose the variable $p_\text{Cou}$ for the matter excitation and $\mathcal{A}$ for the cavity mode. By eliminating the rest of the variables in Eq. \eqref{equations_motion_hamilton_coulomb}, the equations of motion for the chosen variables are written as
\begin{subequations}
\begin{align}
& \ddot{\mathcal{A}} + \left( \omega_\text{cav}^2  +  \frac{f_\text{mat}}{\varepsilon_0 V_\text{eff}} \right) \mathcal{A} -  \frac{f_\text{mat}}{\varepsilon_0 V_\text{eff}} p_\text{Cou}  = 0, \\
&\ddot{p}_\text{Cou} + \omega_\text{mat}^2 p_\text{Cou} -  \omega_\text{mat}^2  \mathcal{A} = 0. 
\end{align}
\label{equations_motion_A_p}
\end{subequations}
With the transformation $x_\text{cav} = \sqrt{\varepsilon_0 V_\text{eff}} \mathcal{A}$ used in Sec. \ref{subsec_molecule_dielectric} of the main text, and with the new transformation $x'_\text{mat} = \frac{\sqrt{f_\text{mat}}}{\omega_\text{mat}} p_\text{Cou}$, Eq. \eqref{equations_motion_A_p} becomes
\begin{subequations}
\begin{align}
&\ddot{x}_\text{cav} + (\omega_\text{cav}^2 + 4g_{\scriptscriptstyle{\text{\MC{}}}}^2) x_\text{cav} - 2 g_{\scriptscriptstyle{\text{\MC{}}}} \omega_\text{mat} x'_\text{mat} = 0, \\
&\ddot{x}'_\text{mat} + \omega_\text{mat}^2 x'_\text{mat} - 2 g_{\scriptscriptstyle{\text{\MC{}}}} \omega_\text{mat} x_\text{cav} = 0,
\end{align}
\label{equations_motion_Coulomb_gauge_SpC}
\end{subequations}
with the same coupling strength $g_{\scriptscriptstyle{\text{\MC{}}}} = \frac{1}{2} \sqrt{\frac{f_\text{mat}}{\varepsilon_0 V_\text{eff}}}$ that is used to describe the cavity-dipole coupling within the \MC{} model. 

Equations \eqref{equations_motion_A_d} and \eqref{equations_motion_Coulomb_gauge_SpC} (the former corresponding to the \MC{} model) have been derived for the same system and thus must result in the same response of the system. However, several interesting aspects can be observed. First, in Eq. \eqref{equations_motion_Coulomb_gauge_SpC} $x'_\text{mat}$ is related to $p_\text{Cou}$, while $x_\text{mat}$ is related to $d$ in the \MC{} model. Thus, it is important to consider this difference when calculating physical observables, as in Sec. \ref{subsec_molecule_dielectric} of the main text. Second, Eq. \eqref{equations_motion_Coulomb_gauge_SpC} contains coupling terms proportional to the oscillation amplitudes $x_\text{cav}$ and $x'_\text{mat}$ (as in the SpC model) instead of to their time derivatives $\dot{x}_\text{cav}$ and $\dot{x}_\text{mat}$ (as in the \MC{} model). Last, in Eq. \eqref{equations_motion_Coulomb_gauge_SpC} the frequency of the cavity mode is dressed from $\omega_\text{cav}$ to $\sqrt{\omega_\text{cav}^2 + 4g_{\scriptscriptstyle{\text{\MC{}}}}^2}$. The different coupling terms and the frequency dressing compensate each other, ensuring that Eq. \eqref{equations_motion_Coulomb_gauge_SpC} yields the same result as the \MC{} model. Therefore, the molecule-dielectric cavity system can be equivalently described using coupling terms proportional to the oscillation amplitudes or to their time derivatives, provided that the frequency of the cavity mode and the physical interpretation of the oscillation amplitudes are modified appropriately.

\subsection{Alternative model of a matter excitation interacting with transverse cavity modes obtained within the dipole gauge}
\label{Appendix_alternative_models_dipole}

We have shown that the results of the \MC{} model can be recovered using equations with a different coupling term and a dressed frequency of the cavity mode. Here, we use the dipole gauge to show that we can also obtain equivalent equations by dressing the frequency of the matter excitation. We consider again a single matter excitation and a transverse electromagnetic mode.

The Lagrangian in the Coulomb gauge $L_\text{Cou}$ of Eq. \eqref{lagrangian_coulomb_gauge} can be transformed to any other  Lagrangian $L'$ with the operation $L' = L_\text{Cou} + \frac{d \mathcal{G}(\mathcal{A},d,t)}{dt}$, by using a general function $\mathcal{G}(\mathcal{A},d,t)$. In particular, the transformation to the dipole gauge is done with the  choice $\mathcal{G} = - d \mathcal{A} $. This is equivalent to the Power-Zienau-Woolley transformation \cite{guywoolley20} in cavity-QED descriptions, with the unitary operator
\begin{equation}
\hat{U} = \exp{\frac{i}{\hbar} \int \mathbf{P} \cdot \mathbf{A} \; d\mathbf{r}  },
\end{equation}
where $\mathbf{P}$ is the polarization density. After applying the gauge transformation to Eq. \eqref{lagrangian_coulomb_gauge},  the Lagrangian of the system in the dipole gauge is
\begin{align}
L_\text{Dip}^\text{min-c}(d,\dot{d}, \mathcal{A}, \dot{\mathcal{A}}) =&    \frac{\varepsilon_0 V_\text{eff}}{2 } (\dot{\mathcal{A}}^2 - \omega_\text{cav}^2 \mathcal{A}^2) + \frac{1}{2f_\text{mat}} (\dot{d}^2 - \omega_\text{mat}^2 d^2) - \dot{\mathcal{A}} d. 
\end{align}
\begin{subequations}

We repeat the procedure implemented in the Coulomb gauge in Sec. \ref{Appendix_alternative_models_Coulomb} to obtain the equations of motion of the dynamical variables in the dipole gauge. The canonical momenta are calculated as
\begin{align}
\Pi_\text{Dip} =& \frac{\partial L_\text{Dip}}{\partial \dot{\mathcal{A}}} = \varepsilon_0 V_\text{eff} \dot{\mathcal{A}} - d,  \\
p_\text{Dip} =& \frac{\partial L_\text{Dip}}{\partial \dot{d}} = \frac{\dot{d}}{f_\text{mat}} .
\end{align}
\label{eq_canonical_momenta_dipole_gauge}
\end{subequations}
In the dipole gauge,  $p_\text{Dip}$ is only related to the time derivative of the induced dipole moment. However, the canonical momentum associated with the cavity mode, $\Pi_\text{Dip}$, depends on both $d$ and the vector potential, in contrast to the result of the Coulomb gauge (Eq. \eqref{eq_canonical_momenta_coulomb_gauge}). Thus, in the dipole gauge this variable represents the displacement vector $\Pi_\text{Dip} \propto |\mathbf{D}| = |\varepsilon_0  \mathbf{E} + \mathbf{P}|$ instead of the electric field of the cavity mode as happens in the Coulomb gauge, where $\Pi_\text{Cou} \propto |\mathbf{E}|$. The resulting Hamiltonian in the dipole gauge is 
\begin{align}
H_\text{Dip}^\text{min-c} =&   \frac{\Pi_\text{Dip}^2}{2\varepsilon_0 V_\text{eff}} + \frac{1}{2}\varepsilon_0 V_\text{eff} \omega_\text{cav}^2 \mathcal{A}^2   + \frac{f_\text{mat}}{2}p_\text{Dip}^2 + \frac{1}{2}\frac{\omega_\text{mat}^2}{f_\text{mat}}d^2    + \frac{\Pi_\text{Dip} d}{\varepsilon_0 V_\text{eff}}+ \frac{d^2}{2 \varepsilon_0 V_\text{eff}},
\label{hamiltonian_dipole_gauge}
\end{align}
with corresponding Hamilton's equations of motion:
\begin{subequations}
\begin{align}
&\dot{\mathcal{A}} = \frac{\partial H^\text{min-c}_\text{Dip}}{\partial \Pi_\text{Dip}} = \frac{\Pi_\text{Dip} + d}{\varepsilon_0 V_\text{eff}}, \\
&\dot{\Pi}_\text{Dip}  = -\frac{\partial H^\text{min-c}_\text{Dip}}{\partial \mathcal{A}} = -\varepsilon_0 V_\text{eff} \omega_\text{cav}^2 \mathcal{A},  \\
&\dot{d} = \frac{\partial H^\text{min-c}_\text{Dip}}{\partial p_\text{Dip}} = f_\text{mat} p_\text{Dip}, \\
&\dot{p}_\text{Dip} = -\frac{\partial H^\text{min-c}_\text{Dip}}{\partial d} =  -\frac{\omega_{\text{mat}}^2}{f_\text{mat}}   d - \frac{\Pi_\text{Dip} + d}{\varepsilon_0 V_\text{eff}} .
\end{align}
\label{equations_motion_hamilton_dipole}
\end{subequations}

The choice of variables $\mathcal{A}$ and $d$ to obtain second-order differential equations leads to the transformation from Eq. \eqref{equations_motion_hamilton_dipole} to Eq. \eqref{equations_motion_A_d}. Therefore, the \MC{} model is obtained independently of the considered gauge for these variables. On the other hand, with the choice of the variables $d$ and $\Pi_\text{Dip}$, we obtain
\begin{subequations}
\begin{align}
&\ddot{\Pi}_\text{Dip} + \omega_\text{cav}^2 \Pi_\text{Dip} + \omega_\text{cav}^2  d = 0 \\
&\ddot{d} + \left(\omega_\text{mat}^2 + \frac{f_\text{mat}}{\varepsilon_0 V_\text{eff} }\right)  d +   \frac{f_\text{mat}}{\varepsilon_0 V_\text{eff}}\Pi_\text{Dip} = 0.
\end{align}
\end{subequations}
This equation can be rewritten in terms of oscillation amplitudes. By using the matter oscillator amplitude $x_\text{mat} = \frac{d}{\sqrt{f_\text{mat}}}$ and the new cavity oscillator amplitude $x'_\text{cav} = \frac{\Pi_\text{Dip}}{\sqrt{\varepsilon_0 V_\text{eff}} \omega_\text{cav}}$, the resulting equations are
\begin{subequations}
\begin{align}
&\ddot{x}'_\text{cav} + \omega_\text{cav}^2 x'_\text{cav} + 2 g_{\scriptscriptstyle{\text{\MC{}}}} \omega_\text{cav} x_\text{mat} = 0, \\
&\ddot{x}_\text{mat} + (\omega_\text{mat}^2 + 4 g_{\scriptscriptstyle{\text{\MC{}}}}^2 ) x_\text{mat} + 2 g_{\scriptscriptstyle{\text{\MC{}}}} \omega_\text{cav} x'_\text{cav} = 0,
\end{align}
\end{subequations}
which gives the same results as the \MC{} model, but with the coupling term proportional to the oscillator oscillation amplitudes $x'_\text{cav}$ and $x_\text{mat}$ and with the frequency of the matter excitation dressed, i.e. renormalized, from $\omega_\text{mat}$ to $\sqrt{\omega_\text{mat}^2+4 g_{\scriptscriptstyle{\text{\MC{}}}}^2}$.

\subsection{Alternative model of a molecular emitter interacting with a metallic nanoparticle}
\label{Appendix_alternative_models_metallic}

In Supplementary Secs. \ref{Appendix_alternative_models_Coulomb} and \ref{Appendix_alternative_models_dipole} we have shown that the coupling between a dipolar excitation of a molecular emitter and a transverse cavity mode can be described equivalently with the \MC{} model (coupling terms proportional to the time derivatives $\dot{x}_\text{cav}$ and $\dot{x}_\text{mat}$) or with models where the coupling terms are proportional to the oscillation amplitudes and the frequencies of the oscillators are dressed. Here, we use the Coulomb gauge and show a similar result for the dipole-dipole interaction between one plasmonic mode and one matter excitation in a molecule or any other quantum emitter: this interaction can be described 
 by the SpC model (coupling terms proportional to the oscillation amplitudes $x_\text{cav}$ and $x_\text{mat}$) or with alternative equations that contain coupling terms proportional to the time derivatives $\dot{x}_\text{cav}$ and $\dot{x}_\text{mat}$, together with dressed frequencies.

We consider the same system analyzed in Sec. \ref{subsec_molecule_nanoparticle} of the main article, namely, a molecule (or another quantum emitter) placed close to a metallic nanoparticle and coupled to it through the Coulomb interaction. This system is described by the Lagrangian of Eq. \eqref{lagrangian_dipole_dipole} (here we omit laser excitation, i.e. $\mathcal{A}_\text{inc} = 0$), which leads to the SpC model in Eq. \eqref{equations_motion_eulerlagrange_SpC}, as discussed in Sec. \ref{Appendix_equationsofmotion}. To obtain the alternative model, we follow the procedure of the previous subsections and first obtain  from Eq. \eqref{lagrangian_dipole_dipole} the classical Hamiltonian of the system $H^\text{dip-dip} = \dot{d}_\text{cav} p_\text{cav} + \dot{d}_\text{mat} p_\text{mat} - L^\text{dip-dip}_\text{Cou}$, which is
\begin{align}
&H^\text{dip-dip} =  \frac{1}{2} f_{\text{cav}} p_\text{cav}^2 + \frac{1}{2} \frac{\omega^2_\text{cav}}{f_\text{cav}}  d_\text{cav}^2  + \frac{1}{2} f_{\text{mat}} p_\text{mat}^2 + \frac{1}{2} \frac{\omega^2_\text{mat}}{f_\text{mat}}  d_\text{mat}^2   \nonumber \\
& + d_\text{cav} d_\text{mat} \frac{\mathbf{n}_{\mathbf{d}\text{cav}} \cdot \mathbf{n}_{\mathbf{d}\text{mat}} - 3(\mathbf{n}_{\mathbf{d}\text{cav}} \cdot \mathbf{n}_{\mathbf{r}\text{rel}})(\mathbf{n}_{\mathbf{d}\text{mat}} \cdot \mathbf{n}_{\mathbf{r}\text{rel}})}{4\pi \varepsilon_0 |\mathbf{r}_\text{cav} - \mathbf{r}_\text{mat}|^3},
\label{hamiltonian_dipole_dipole}
\end{align}
with the canonical momenta $p_\text{cav} = \frac{\dot{d}_\text{cav}}{f_\text{cav}}$ and $p_\text{mat} = \frac{\dot{d}_\text{mat}}{f_\text{mat}}$. The Hamiltonian of Eq. \eqref{hamiltonian_dipole_dipole} has been obtained from the Coulomb gauge, but the dipole gauge leads to the same Hamiltonian for this specific system because this change of gauge affects the treatment of the electromagnetic degrees of freedom $\mathcal{A}_\alpha$ associated with the transverse fields. These degrees of freedom are not present when the interaction occurs through Coulomb coupling. 

By calculating the equations of motion for the oscillator variables $x_\text{cav} = \frac{d_\text{cav}}{\sqrt{f_\text{cav}}}$ and $x_\text{mat} = \frac{d_\text{mat}}{\sqrt{f_\text{mat}}}$ as in previous subsections, we recover the equations of the SpC model (Eq. \eqref{equations_motion_classical_position_D0} in the main text). However, we can again make another choice for the variables to obtain an alternative model of harmonic oscillators. Using the oscillator $x_\text{cav} = \frac{d_\text{cav}}{\sqrt{f_\text{cav}}}$ as before and the new oscillator $x'_\text{mat} = \frac{\sqrt{f_\text{mat}}}{\omega_\text{mat}} p_\text{mat}$, the equations of motion are
\begin{subequations}
\begin{align}
&\ddot{x}_\text{cav} + (\omega_\text{cav}^2 - 4g_{\scriptscriptstyle{\text{SpC}}}^{\prime 2}) x_\text{cav} - 2 g'_{\scriptscriptstyle{\text{SpC}}} \dot{x}'_\text{mat} = 0, \\
&\ddot{x}'_\text{mat} + \omega_\text{mat}^2 x'_\text{mat} + 2g'_{\scriptscriptstyle{\text{SpC}}} \dot{x}_\text{cav} = 0,
\end{align}
\end{subequations}
with the coupling strength $g'_{\scriptscriptstyle{\text{SpC}}} = g_{\scriptscriptstyle{\text{SpC}}}\sqrt{\frac{\omega_\text{cav}}{\omega_\text{mat}}}$, slightly modified compared to the SpC value $g_{\scriptscriptstyle{\text{SpC}}}$ used in  Eq. \eqref{g_dipole_dipole} of the main text. We have thus shown that the results of the SpC model can also be obtained with a model where the coupling terms are proportional to the time derivatives $\dot{x}_\text{cav}$ and $\dot{x}'_\text{mat}$. In this case the cavity frequency has been renormalized from $\omega_\text{cav}$ to $\sqrt{\omega_\text{cav}^2 - 4g_{\scriptscriptstyle{\text{SpC}}}^{\prime 2}}$.

\section{Comparison between cavity-QED Hamiltonians of different systems and gauges}
\label{Appendix_quantumHamiltonians}

In the previous Supplementary Sections, the SpC, \MC{}, and alternative coupled harmonic oscillator models are derived from a fully classical description based on Lagrangian and Hamiltonian mechanics. We next quantize the classical Hamiltonians to obtain the cavity-QED Hamiltonians describing the system, including those in the main text. This procedure shows that the cavity-QED Hamiltonians and the corresponding coupled-harmonic oscillator models are directly related.

The coupling between a molecular emitter (or another quantum emitter) and the transverse electromagnetic modes of a dielectric cavity is described by the minimal-coupling Hamiltonian, which for the Coulomb gauge has the classical form of Eq. \eqref{hamiltonian_coulomb_gauge} and for the dipole gauge it is given by Eq. \eqref{hamiltonian_dipole_gauge}. We quantize these classical Hamiltonians following the standard rules of quantization (Eqs. \eqref{quantization_A}-\eqref{quantization_pmat} in the main article) and obtain
\begin{equation}
\hat{H}_\text{Cou}^\text{min-c} = \hbar \omega_\text{cav} \left( \hat{a}^\dagger \hat{a} +\frac{1}{2} \right) + \hbar \omega_\text{mat} \left( \hat{b}^\dagger \hat{b} + \frac{1}{2} \right) + i \hbar g_{\scriptscriptstyle{\text{\MC{}}}} \sqrt{\frac{\omega_\text{mat}}{\omega_\text{cav}}}(\hat{a} + \hat{a}^\dagger)(\hat{b} - \hat{b}^\dagger) + \hbar \frac{g_{\scriptscriptstyle{\text{\MC{}}}}^2}{\omega_\text{cav}} (\hat{a} + \hat{a}^\dagger)^2. 
\label{hamiltonian_QED_coulomb_gauge}
\end{equation}
\begin{equation}
\hat{H}_\text{Dip}^\text{min-c} = \hbar \omega_\text{cav} \left( \hat{a}^\dagger \hat{a} +\frac{1}{2} \right) + \hbar \omega_\text{mat} \left( \hat{b}^\dagger \hat{b} + \frac{1}{2} \right) - i \hbar g_{\scriptscriptstyle{\text{\MC{}}}} \sqrt{\frac{\omega_\text{cav}}{\omega_\text{mat}}} (\hat{a} - \hat{a}^\dagger)(\hat{b} + \hat{b}^\dagger) + \hbar \frac{g_{\scriptscriptstyle{\text{\MC{}}}}^2}{\omega_\text{mat}} (\hat{b} + \hat{b}^\dagger)^2. 
\label{hamiltonian_QED_dipole_gauge}
\end{equation}
for the Coulomb and dipole gauges, respectively. In these Hamiltonians, $\hat{a}$ and $\hat{a}^\dagger$ are the annihilation and creation operators of the cavity mode, while $\hat{b}$ and $\hat{b}^\dagger$ are the corresponding operators for the  molecular excitations.
The main difference between Eqs. \eqref{hamiltonian_QED_coulomb_gauge} and \eqref{hamiltonian_QED_dipole_gauge} is the last quadratic term, which is originated from the vector potential of the electromagnetic mode in the Coulomb gauge, and from the induced dipole moment of the molecule in the dipole gauge, respectively. We further note that the relation between the quantum coupling strength $g_{\scriptscriptstyle{\text{QED}}}$ (i.e., the proportionality factor if we write the third term of the Hamiltonians as $\pm i\hbar g_{\scriptscriptstyle{\text{QED}}}(\hat{a} - \hat{a}^\dagger)(\hat{b} + \hat{b}^\dagger)$ ) and the classical coupling strength $g_{\scriptscriptstyle{\text{\MC{}}}}$ is different for each gauge, with $g_{\scriptscriptstyle{\text{QED}}} = g_{\scriptscriptstyle{\text{\MC{}}}} \sqrt{\frac{\omega_\text{mat}}{\omega_\text{cav}}}$ for the Coulomb gauge and $g_{\scriptscriptstyle{\text{QED}}} = g_{\scriptscriptstyle{\text{\MC{}}}} \sqrt{\frac{\omega_\text{cav}}{\omega_\text{mat}}}$ for the dipole gauge. We emphasize, however, that the eigenvalues of the two Hamiltonians are identical (given by Eq. \eqref{eigenfrequencies_MC} in the main text). Further, these two Hamiltonians also lead to identical results for any physical magnitude, once we consider that the operators $\hat{a}$, $\hat{a}^\dagger$, $\hat{b}$ and $\hat{b}^\dagger$ are not equivalent in the two Hamiltonians and are related to a different set of canonical momenta (and thus to different physical magnitudes) in each of them: $\Pi_\text{Cou}$ and $p_\text{Cou}$ given by Eq. \eqref{eq_canonical_momenta_coulomb_gauge} for the Hamiltonian of Eq. \eqref{hamiltonian_QED_coulomb_gauge}, or $\Pi_\text{Dip}$ and $p_\text{Dip}$ given by Eq. \eqref{eq_canonical_momenta_dipole_gauge} for the Hamiltonian of Eq. \eqref{hamiltonian_QED_dipole_gauge}.

On the other hand,  dipole-dipole interactions (for example, between a metallic nanoparticle and a molecular emitter) are modeled with the following cavity-QED Hamiltonian both in the Coulomb and dipole gauges (obtained by applying the quantization rules to Eq. \eqref{hamiltonian_dipole_dipole}):
\begin{equation}
\hat{H}^\text{dip-dip} = \hbar \omega_\text{cav} \left( \hat{a}^\dagger \hat{a} +\frac{1}{2} \right) + \hbar \omega_\text{mat} \left( \hat{b}^\dagger \hat{b} + \frac{1}{2} \right) +  \hbar g_{\scriptscriptstyle{\text{QED}}} (\hat{a} + \hat{a}^\dagger)(\hat{b} + \hat{b}^\dagger),
\label{hamiltonian_QED_dipole_dipole}
\end{equation}
with $g_{\scriptscriptstyle{\text{QED}}}=g_{\scriptscriptstyle{\text{SpC}}}$. This Hamiltonian does not have any diamagnetic term. Thus it gives different results than the minimal-coupling Hamiltonians of Eqs. \eqref{hamiltonian_QED_coulomb_gauge} and \eqref{hamiltonian_QED_dipole_gauge}. Further, the operators 
 $\hat{a}$, $\hat{a}^\dagger$, $\hat{b}$ and $\hat{b}^\dagger$ in Eq. \eqref{hamiltonian_QED_dipole_dipole} are related to the induced dipole moments of the nanoparticle and the molecule according to Eq. \eqref{quantization_d} of the main article.

The analysis of this and the previous sections establishes that the classical coupled harmonic oscillator models and the cavity-QED Hamiltonians can be derived from the same starting point of the Lagrangian in Eq. \eqref{general_EM_lagrangian}, and can thus be used to obtain equivalent physical results.

\section{Summary of classical models and their connection with cavity-QED Hamiltonians}
\label{Appendix_Comparison}

\begin{figure}
\centering
\includegraphics[scale=0.35]{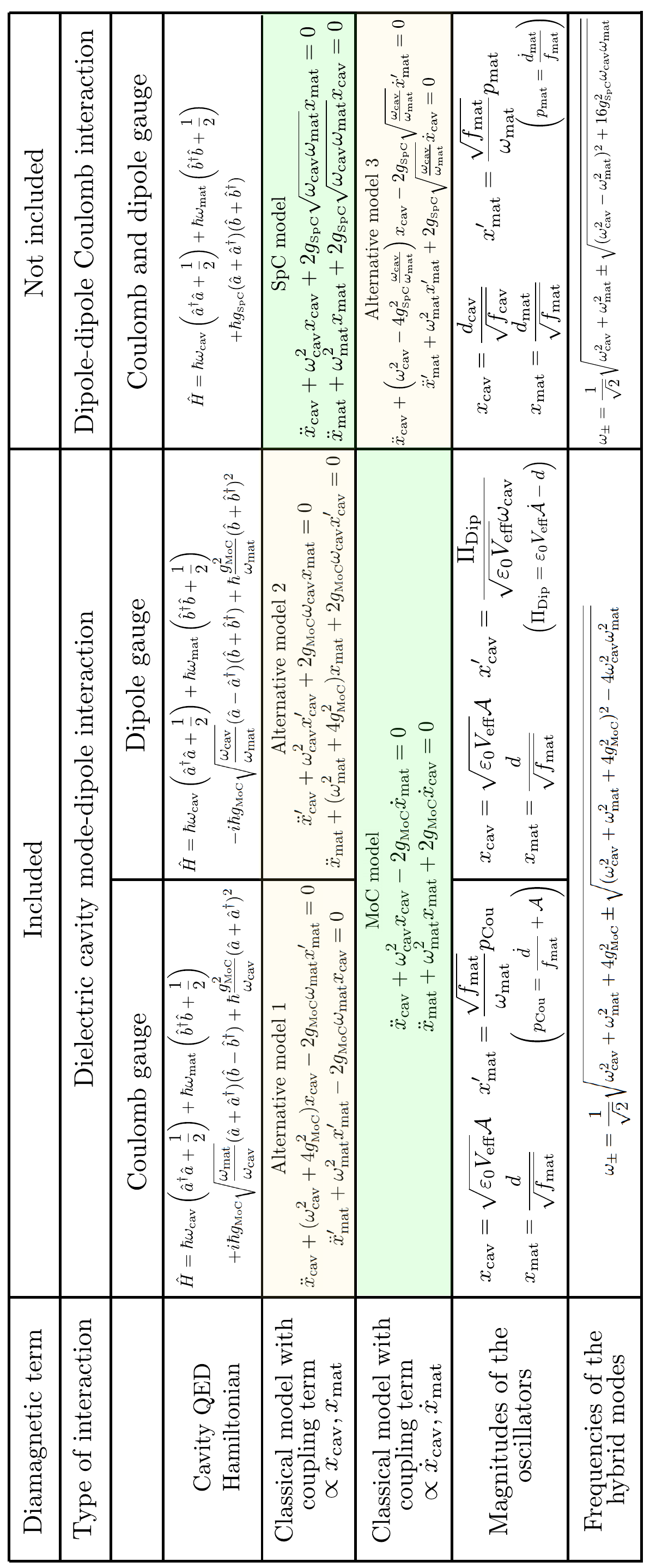}
\caption*{Table S1: Summary of the correspondences between the classical coupled harmonic oscillator models and the cavity-QED Hamiltonians. We consider the coupling between a dipole (representing, e.g., a molecular excitation) and a dielectric cavity (with transverse electromagnetic modes) or a plasmonic nanocavity  (dipole-dipole coupling via Coulomb interactions). The coupling with transverse modes is described in the Coulomb (second column) and dipole (third column) gauges, while the dipole-dipole coupling is described in the same way in both gauges as indicated in the fourth column. The fourth row shows the cavity-QED Hamiltonians that describe each of these situations. The fifth and the sixth rows indicate the corresponding classical harmonic oscillations models: the fifth row corresponds to the models associated with coupling terms proportional to oscillation amplitudes, and the sixth row to models with coupling terms proportional to their time derivatives (with coupling strengths $g_{\scriptscriptstyle{\text{\MC{}}}}$ given by Eq. \eqref{eq_coupling_strength_MC} and $g_{\scriptscriptstyle{\text{SpC}}}$ given by Eq. \eqref{g_dipole_dipole_SI}). We highlight in green the SpC and \MC{} models, which are the focus of the main text and for which the bare frequencies $\omega_\text{cav}$ and $\omega_\text{mat}$ are considered. With the yellow background, we indicate the alternative models where we use dressed frequencies, which also change the coupling term. The seventh row shows the association between the amplitudes of the oscillators and the physical magnitudes of the system, which differ for each model in the fifth and sixth rows. The last row provides the frequencies of the two hybrid modes for the two different types of interaction. To ease comparison, we write both the cavity-QED Hamiltonians and coupled harmonic oscillator models in terms of $g_{\scriptscriptstyle{\text{SpC}}}$ and $g_{\scriptscriptstyle{\text{\MC{}}}}$. }
 \label{figure_table_supp}
\end{figure} 

Table S1 summarizes all the classical models discussed in Supplementary Secs. \ref{Appendix_equationsofmotion} and \ref{Appendix_alternative_models}, as well as the cavity-QED Hamiltonians discussed in Supplementary Sec. \ref{Appendix_quantumHamiltonians}. These sections focus on two types of interactions: the coupling between a molecular excitation and transverse electromagnetic modes in a dielectric (Fabry-Pérot) cavity, and the dipole-dipole coupling due to the Coulomb interaction. The models describing the first type of interaction (second and third columns) depend on the chosen gauge (Coulomb or dipole) \cite{andrews18}, but all of them result in identical eigenfrequencies and other physical magnitudes (the latter require to take into account the specific connection of the classical oscillation amplitudes and quantum operators with, e.g., the electric field depends on the model). All the models in the fourth column describing dipole-dipole Coulomb coupling are also equivalent to each other. On the other hand, the models in the fourth column are not equivalent to those in the second and third columns. 

Table S1 shows that if the classical equations depend directly on the bare (non-dressed) frequencies of the uncoupled oscillators $\omega_\text{cav}$ and $\omega_\text{mat}$, the description of the interaction between transverse cavity modes and matter excitations requires a coupling term proportional to the time derivatives of the oscillation amplitudes (\MC{} model, equivalent to cavity-QED Hamiltonians with diamagnetic term). In contrast, the coupling term associated with dipole-dipole interactions is proportional to the oscillation amplitudes where frequencies are not dressed (SpC model, equivalent to cavity-QED Hamiltonians without diamagnetic term). These classical models are analyzed in the main text and Supplementary Sec. \ref{Appendix_equationsofmotion} and highlighted in green.
On the other hand, as discussed in Supplementary Sec. \ref{Appendix_alternative_models}, each type of interaction can also be modeled with alternative models where the type of coupling term is modified from proportional to the oscillation amplitudes to proportional to their time derivatives, or vice versa.  We highlight these models in Table S1 by the yellow squares. In these alternative models, dressing or renormalization of one of the oscillator frequencies is needed to maintain their equivalence with their corresponding cavity-QED Hamiltonians. Further, the alternative classical models also require the modification of the physical magnitudes that each oscillator represents. However, if the transformation of oscillation amplitudes and frequencies is done appropriately, all models describing the coupling between transverse fields and dipoles (second and third columns) yield identical results, and the same happens for dipole-dipole interactions (fourth column).

\section{Linearized Model}
\label{Appendix_linearized}

\begin{figure}
\centering
\includegraphics[scale=1.1]{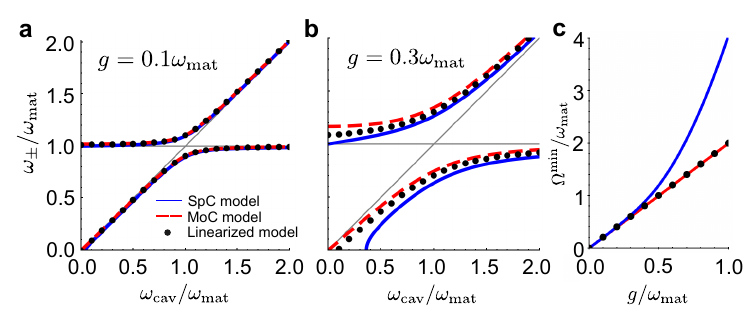}
\caption{Comparison of the Spring Coupling (SpC), Momentum Coupling (\MC{}), and linearized models. a) Eigenfrequencies $\omega_\pm$ of the hybrid states calculated from the bare values $\omega_\text{cav}$ and  $\omega_\text{mat}$, with $\omega_\text{mat}$ fixed and $\omega_\text{cav}/\omega_\text{mat}$ changing. $\omega_\pm$  obtained from the SpC model (blue solid line, corresponding to Eq. \eqref{eigenfrequencies_SpC} in the main text), \MC{} model (red dashed line, Eq. \eqref{eigenfrequencies_MC} in the main text) and the approximate linearized model (black dots, Eq. \eqref{eigenfrequencies_linear}), for coupling strength $g=g_{\scriptscriptstyle{\text{SpC}}}=g_{\scriptscriptstyle{\text{\MC{}}}}=g_{\scriptscriptstyle{\text{lin}}} = 0.1 \, \omega_\text{mat}$. The thin gray lines correspond to the bare cavity frequency $\omega_\text{cav}$ and the bare frequency of the matter excitation, $\omega_\text{mat}$.    b) Same as panel (a), for coupling strength $g=g_{\scriptscriptstyle{\text{SpC}}}=g_{\scriptscriptstyle{\text{\MC{}}}}=g_{\scriptscriptstyle{\text{lin}}} =0.3 \, \omega_\text{mat}$. 
c) Minimum splitting between the hybrid modes $\Omega^\text{min} = \omega_+ - \omega_-$, as a function of the coupling strength $g$ for the SpC model (blue solid line), the \MC{} model (red solid line) and the linearized model (black dots). All frequencies are normalized with respect to the fixed frequency of the matter excitation $\omega_\text{mat}$, so that the results do not depend on the particular value of $\omega_\text{mat}$, only on the
$\omega_\text{cav}/\omega_\text{mat}$ ratio or $g/\omega_\text{cav}$ ratio . The \MC{} and SpC results are the same as in Fig. \ref{figure_models} of the main text.} \label{figure_models_SI}
\end{figure}

We show in this section that, for $g < 0.1 \omega_\text{mat}$ (i.e., before the onset of ultrastrong coupling according to the standard definition of this regime), it is possible to reduce both the \MC{} and the SpC model to the same simplified linearized model by considering that the eigenfrequencies $\omega_\pm$ do not differ too strongly from the bare frequencies $\omega_\alpha$ ($\alpha$ = 'cav' or $\alpha$ = 'mat').

Using the approximation $\omega_\alpha+\omega \approx 2\omega\approx 2\omega_\alpha$, the frequency-domain equations of both the SpC and \MC{} models become linear in $\omega$:
\begin{subequations}
\begin{equation}
(\omega_\text{cav} - \omega) x_\text{cav} + g_{\scriptscriptstyle{\text{lin}}} x_\text{mat} =0 \\
\end{equation}
\begin{equation}
(\omega_\text{mat} - \omega) x_\text{mat} + g_{\scriptscriptstyle{\text{lin}}}^* x_\text{cav} =0,
\end{equation}
\label{equations_motion_linear}
\end{subequations}
with $g_{\scriptscriptstyle{\text{lin}}} = g_{\scriptscriptstyle{\text{SpC}}}=ig_{\scriptscriptstyle{\text{\MC{}}}}$. The resulting eigenfrequencies are 
 \begin{equation}
\omega_{\pm,\text{lin}} = \frac{\omega_\text{cav} + \omega_\text{mat}  \pm \sqrt{(\omega_\text{cav}- \omega_\text{mat})^2 + 4|g_{\scriptscriptstyle{\text{lin}}}|^2}}{2}. \label{eigenfrequencies_linear}
\end{equation}

We compare in Fig. \ref{figure_models_SI} the results of this model (black dots) to those obtained with the \MC{} (red dashed lines) and SpC (blue solid lines) models. The results are obtained with Eq. \eqref{eigenfrequencies_linear}, and Eqs. \eqref{eigenfrequencies_MC}  and \eqref{eigenfrequencies_SpC} of the main text, respectively. Fig. \ref{figure_models_SI}a shows that, for  $g = 0.1 \omega_\text{mat}$ (as in Sec. \ref{sec:comparison} of the main text, we use $g$ to refer to $g_{\scriptscriptstyle{\text{SpC}}}$, $g_{\scriptscriptstyle{\text{\MC{}}}}$ and/or $g_{\scriptscriptstyle{\text{lin}}}$ in discussions that are valid for more than one model), the three models indeed result in very similar eigenvalues for all values of $\omega_\text{cav}$/$\omega_\text{mat}$. However, this is not the case for $g = 0.3 \omega_\text{mat}$ (Fig. \ref{figure_models_SI}b), where the eigenfrequencies of the linearized model are typically in between those of the SpC and \MC{} models. Notably, the linearized model does not present any mode in a forbidden energy band that is half as wide as in the \MC{} model (while the SpC model did not present such a forbidden band). Similarly, the linearized model does not present any lower-energy mode, (i.e., negative $\omega_{-,\scriptscriptstyle{\text{lin}}}$,   for   $\frac{\omega_\text{cav}}{\omega_\text{mat}}< \left(\frac{g_{\scriptscriptstyle{\text{lin}}}}{\omega_\text{mat}} \right)^2$); in contrast,  the corresponding condition for the SpC model is $\frac{\omega_\text{cav}}{\omega_\text{mat}}< \left(\frac{2g_{\scriptscriptstyle{\text{SpC}}}}{\omega_\text{mat}}\right)^2$ (where $\omega_{-,\scriptscriptstyle{\text{SpC}}}$ is imaginary) and the \MC{} model always presents a lower-energy mode. On the other hand, the splitting at zero detuning is equal to $\Omega = 2g$ in both the linearized and \MC{} models  (but with different values of $\omega_\pm$ for each of them), which is the minimum splitting in these two models. In contrast, the minimum splitting scales non-linearly with $g$ for the SpC model, and is larger than for the \MC{} or linearized modes. These results are illustrated in Fig. \ref{figure_models_SI}(c), which shows the dependence of the normalized minimum splitting $\Omega^\text{min}/\omega_\text{mat}$ on the normalized coupling strength $g/\omega_\text{mat}$.  

\section{Evolution of the eigenvalues for a different choice of coupling strength}
\label{Appendix_eigenvalues-nonconstantg}

\begin{figure}
\centering
\includegraphics[scale=1.4]{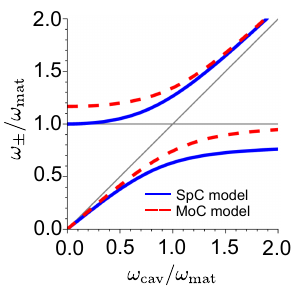}
\caption{Comparison of the Spring Coupling (SpC) and Momentum Coupling (\MC{}) models for a different choice of the coupling strength than in the main text. Eigenfrequencies $\omega_\pm$ obtained as a function of $\omega_\text{cav}$, obtained from the SpC model (blue solid line, corresponding to Eq. \eqref{eigenfrequencies_SpC} in the main text) and \MC{} model (red dashed line, Eq. \eqref{eigenfrequencies_MC} in the main text) for coupling strength $g_{\scriptscriptstyle{\text{SpC}}}= 0.3 \sqrt{\omega_\text{cav}\omega_\text{mat}}$ and $g_{\scriptscriptstyle{\text{\MC{}}}} = 0.3 \, \omega_\text{mat}$, respectively. The thin gray lines correspond to the bare cavity frequency $\omega_\text{cav}$ and the bare frequency of the matter excitation, $\omega_\text{mat}$.    All frequencies are normalized with respect to the fixed frequency of the matter excitation $\omega_\text{mat}$ ($\hbar\omega_\text{mat}=0.1$ eV), and the \MC{}  results are the same as in Fig. \ref{figure_models} of the main text. }\label{figure_models_nonconstantg}
\end{figure} 

In Sec. \ref{sec:comparison} of the main text (as well as in Sec. \ref{Appendix_linearized}), we consider that the coupling strengths $g_{\scriptscriptstyle{\text{\MC{}}}}$ and $g_{\scriptscriptstyle{\text{SpC}}}$ do not depend on the resonant frequency of the cavity $\omega_\text{cav}$. This choice is consistent with the results obtained in Secs. \ref{subsec_molecule_dielectric} and \ref{subsec_bulk_dielectric} of the main text, which used the \MC{} model to describe the coupling of a molecular emitter or an ensemble of molecular emitters with the transverse fields of an electromagnetic mode of a dielectric (Fabry-Pérot) cavity. On the other hand, when describing the Coulomb coupling within the SpC model in Sec. \ref{subsec_molecule_nanoparticle}, the dependence of the coupling strength $g_{\scriptscriptstyle{\text{SpC}}}$ on $\omega_\text{cav}$ can vary with the details of each particular configuration.  To exemplify the consequences of such details, we consider in this section a different dependence of $g_{\scriptscriptstyle{\text{SpC}}}$ on $\omega_\text{cav}$ than in the main text. 

We analyze again the Coulomb coupling between metallic spherical particles of radius $R_\text{cav}$ and a quantum emitter, described with the SpC model as in Sec. \ref{subsec_molecule_nanoparticle} of the main text. However, we now change the plasma frequency of the metal, which modifies the dipole moment
$f_\text{cav} = 4\pi\varepsilon_0 R_\text{cav}^3 \omega_\text{cav}^2$  so that, according to Eq. \eqref{g_dipole_dipole} in the main text, the coupling strength scales as $g_{\scriptscriptstyle{\text{SpC}}}\propto \sqrt{\omega_\text{cav}}$  (assuming a constant $f_\text{mat}$).

We then plot in Fig. \ref{figure_models_nonconstantg} the results obtained within the \MC{} model for $\hbar\omega_\text{mat}=0.1$ eV, $g_{\scriptscriptstyle{\text{\MC{}}}} = 0.3 \, \omega_\text{mat}$ (red dashed line, same as in the main text) and within the SpC model for coupling strength $g_{\scriptscriptstyle{\text{SpC}}}=$~$0.3 \omega_\text{mat} \sqrt{\omega_\text{cav}/\omega_\text{mat}}= 0.3 \sqrt{\omega_\text{cav}\omega_\text{mat}}$  (blue solid line), as $\omega_\text{cav}$ is changed. With this choice, $g_{\scriptscriptstyle{\text{\MC{}}}} = g_{\scriptscriptstyle{\text{SpC}}}$ under resonant conditions ($\omega_\text{cav}=\omega_\text{mat}$). We find that, for this scaling of  $g_{\scriptscriptstyle{\text{SpC}}}$, the SpC model results in two (real valued) eigenfrequencies for all values of $\omega_\text{cav}$, as well as in the opening of a  Reststrahlen band. Interestingly, however, this band appears for energies smaller than $\omega_\text{mat}$, contrary to the result for the \MC{} model.

\section{Transformation from individual to collective oscillators in the description of homogeneous materials in Fabry-Pérot cavities}
\label{Appendix_collectivetransformation}
In Sec. \ref{subsec_bulk_dielectric} of the main article, we analyze how classical models of harmonic oscillators describe an ensemble of $N_\text{dip}$ molecules (or a homogeneous material) inside a Fabry-Pérot cavity.
Each molecular emitter couples with all the other molecular emitters and also with the transverse modes of the cavity, and all these interactions can be modeled through Eq. \eqref{equations_motion_bulk} in the main text. In this supplementary section, we show in more detail how to describe this system by considering the coupling of each Fabry-Pérot mode with a single collective mode of matter oscillators. Specifically, here we demonstrate how to transform Eq. \eqref{equations_motion_bulk} in Sec. \ref{subsec_bulk_dielectric} of the main text,  written in terms of harmonic oscillators of individual molecular excitations, into Eq. \eqref{equations_motion_bulk_transformed}, which considers collective modes. This derivation can be generalized to other cavities by following the same procedure but using the spatial distribution of the transverse electric field of the corresponding cavity modes.

We assume that the Fabry-Pérot cavity contains perfect mirrors in the planes $z=0$ and $z=L_\text{cav}$ ($L_\text{cav}$ is the thickness of the cavity), so that the cavity has transverse electric (TE) modes with field distribution\footnote{To simplify the discussion, here we show explicitly the transformation under the field distribution of TE modes. Fabry-Pérot cavities also have transverse magnetic (TM) modes, and all the transformations are equivalent after substituting the field distribution of these modes into Eq. \eqref{eq_fabry_perot_field_distribution}, but additional care needs to be taken to account for the position dependence of the polarization direction of the cavity fields.}
\begin{equation}
\Xi_{n\mathbf{k}_\parallel}(\mathbf{r}) = \sin\left( \frac{n\pi z}{L_\text{cav}}   \right)  e^{i \mathbf{k}_\parallel \cdot \mathbf{r}_{\parallel}  }.
\label{eq_fabry_perot_field_distribution}
\end{equation}

The integer $ n$ indexes all modes of the cavity and the wavevector in the parallel direction $\mathbf{k}_\parallel$ is any two-dimensional vector (we consider a discrete set of $\mathbf{k}_\parallel$ by assuming that the cavity has long but finite size in the lateral dimensions and using Born-von Karman periodic boundary conditions for Eq. \eqref{eq_fabry_perot_field_distribution}). We further assume that the direction of the transition dipole moments of the molecules is the same as that of the electric field of the mode (parallel to the mirror planes). As a consequence, the coupling strength between each molecular emitter placed in the position $\mathbf{r}_i = (\mathbf{r}_{\parallel,i},z_i)$ and the $n\mathbf{k}_\parallel$ Fabry-Pérot mode is calculated with the expression $g_{\scriptscriptstyle{\text{\MC{}}}}^{(n\mathbf{k}_\parallel, i)} = \frac{1}{2} \sqrt{\frac{f_\text{dip}}{\varepsilon_0 V_\text{eff}}}  \Xi_{n\mathbf{k}_\parallel}(\mathbf{r}_i)$ (see discussion of Supplementary Sec. \ref{Appendix_equationsofmotion} and Eq. \eqref{eq_coupling_strength_MC}). By introducing the field distribution of Eq. \eqref{eq_fabry_perot_field_distribution} in the expression of the coupling strength explicitly, the equations of motion of the system (Eq. \eqref{equations_motion_bulk} in the main text) become
\begin{subequations}
\begin{equation}
\ddot{x}_{\text{cav},n\mathbf{k}_\parallel}  + \omega^2_{\text{cav},n\mathbf{k}_\parallel} x_{\text{cav},n\mathbf{k}_\parallel}  - \sum_i \sqrt{\frac{f_\text{dip}}{\varepsilon_0 V_\text{eff}}}  \sin\left( \frac{n\pi z_i}{L_\text{cav}}\right) e^{-i \mathbf{k}_\parallel \cdot \mathbf{r}_{\parallel i}  }   \dot{x}_{\text{dip},i} = 0,
\label{eq_supplementary_bulk_1}
\end{equation}
\begin{equation}
\ddot{x}_{\text{dip},i} + \omega^2_\text{dip} x_{\text{dip},i} + \sum_{n',\mathbf{k'}_\parallel}  \sqrt{\frac{f_\text{dip}}{\varepsilon_0 V_\text{eff}}}   \sin\left( \frac{n'\pi z_i}{L_\text{cav}}   \right) e^{i \mathbf{k'}_\parallel \cdot \mathbf{r}_{\parallel i}  } \dot{x}_{\text{cav},n'\mathbf{k'}_\parallel}  + \sum_{j \neq i} 2 \omega_\text{dip} g_{\scriptscriptstyle{\text{SpC}}}^{(i,j)} x_{\text{dip},j} = 0.
\label{eq_supplementary_bulk_2}
\end{equation}
\end{subequations}

 In Eq. \eqref{eq_supplementary_bulk_1}, we already observe that the oscillator $x_{\text{cav},n\mathbf{k}_\parallel}$ of the $n\mathbf{k}_\parallel$ cavity mode is coupled to a collective matter operator. By defining the collective  oscillator of the $n\mathbf{k}_\parallel$ matter mode as
\begin{equation}
x_{\text{mat},n\mathbf{k}_\parallel} = \frac{1}{\sqrt{N_\text{eff}}} \sum_i e^{-i \mathbf{k}_\parallel \cdot \mathbf{r}_{\parallel i}  }  \sin\left( \frac{n\pi z_i}{L_\text{cav}}   \right) x_{\text{dip},i},
\label{eq_collective_matter_oscillator}
\end{equation}
Equation \eqref{eq_supplementary_bulk_1} becomes
\begin{equation}
\ddot{x}_{\text{cav},n\mathbf{k}_\parallel}  + \omega^2_{\text{cav},n\mathbf{k}_\parallel} x_{\text{cav},n\mathbf{k}_\parallel}  - 2\sqrt{N_\text{eff}} g_{\scriptscriptstyle{\text{\MC{}}}}^\text{max} \dot{x}_{\text{mat},n\mathbf{k}_\parallel} = 0.
\label{eq_transformation_finished_1}
\end{equation}
where $g_{\scriptscriptstyle{\text{\MC{}}}}^\text{max} = \frac{1}{2} \sqrt{\frac{f_\text{dip}}{\varepsilon_0 V_\text{eff}}}$ is the maximum achievable coupling strength between a single molecular emitter and a cavity mode in this system, found for molecules placed in the antinodes of the mode.  $N_\text{eff}=\sum_{i} \left|\Xi_{n\mathbf{k_\parallel}}(\mathbf{r}_i)\right|^2$ is the effective number of molecular emitters that couple with the cavity mode, whose exact relation with the total number of molecular emitters $N_\text{dip}$ depends on the system and the spatial distribution of the modes. By performing the sum $N_\text{eff}=\sum_{i} \left|\Xi_{n\mathbf{k_\parallel}}(\mathbf{r}_i)\right|^2=\sum_i \left|\sin\left( \frac{n\pi z_i}{L_\text{cav}}   \right)\right|^2$ with the specific field distribution of Fabry-Pérot cavity modes, we obtain $N_\text{eff}=N_\text{dip}/2$ for this cavity. We observe in Eq. \eqref{eq_transformation_finished_1} that the coupling strength between the cavity mode and the collective oscillator mode increases as $g_{\scriptscriptstyle{\text{\MC{}}}}^\text{max} \sqrt{N_\text{eff}}$. This scaling of the coupling strength (together with the scaling as $1/\sqrt{N_\text{eff}}$ of the collective oscillator in Eq. \eqref{eq_collective_matter_oscillator}) is the same as in the quantum Dicke model \cite{dicke54rep}, further confirming that the classical oscillator models are consistent with cavity-QED descriptions.

The next step is to transform Eq. \eqref{eq_supplementary_bulk_2}, which requires considering $N_\text{dip}$ equations simultaneously, one per molecular emitter at position $\mathbf{r}_i$.  To do the transformation, we multiply  Eq. \eqref{eq_supplementary_bulk_2}  by $\frac{1}{\sqrt{N_\text{eff}}} \sin \left( \frac{n\pi z_i}{L_\text{cav}} \right) e^{-i \mathbf{k}_\parallel \cdot \mathbf{r}_{\parallel i}  }$ for each $i$ molecular emitter and sum  the $N_\text{dip}$ resulting terms. With this procedure, and using Eq. \eqref{eq_collective_matter_oscillator}, the transformation of the first two terms is straightforward as
\begin{equation}
\frac{1}{\sqrt{N_\text{eff}}}  \sum_i \sin\left( \frac{n\pi z_i}{L_\text{cav}}   \right) e^{-i \mathbf{k}_\parallel \cdot \mathbf{r}_{\parallel i}  } (\ddot{x}_{\text{dip},i} + \omega^2_\text{dip} x_{\text{dip},i})  =  \ddot{x}_{\text{mat},n\mathbf{k}_\parallel} + \omega_\text{dip}^2 x_{\text{mat},n\mathbf{k}_\parallel}.
\label{eq_first_second_terms_transformed}
\end{equation}

Repeating the procedure with the third term of Eq. \eqref{eq_supplementary_bulk_2}, we obtain
\begin{align}
&\frac{2}{\sqrt{N_\text{eff}}} g_{\scriptscriptstyle{\text{\MC{}}}}^\text{max} \sum_{n',\mathbf{k'}_\parallel} \dot{x}_{\text{cav},n'\mathbf{k'}_\parallel} \sum_i \sin\left( \frac{n'\pi z_i}{L_\text{cav}}   \right) \sin\left( \frac{n\pi z_i}{L_\text{cav}}   \right) e^{i(\mathbf{k}_\parallel-\mathbf{k'}_\parallel)\cdot\mathbf{r_\parallel}} \nonumber \\
&\qquad  = \frac{2}{\sqrt{N_\text{eff}}} g_{\scriptscriptstyle{\text{\MC{}}}}^\text{max} \sum_{n',\mathbf{k'}_\parallel} \dot{x}_{\text{cav},n\mathbf{k}_\parallel}  N_\text{eff} \delta_{n,n'} \delta_{\mathbf{k}_\parallel, \mathbf{k'}_\parallel} = 2 g_{\scriptscriptstyle{\text{\MC{}}}}^\text{max} \sqrt{N_\text{eff}}  \dot{x}_{\text{cav},n\mathbf{k}_\parallel}.
\label{eq_third_term_transformed}
\end{align}
Equation \eqref{eq_third_term_transformed} shows that, although each molecular emitter couples with all Fabry-Pérot modes of different $\mathbf{k}_\parallel$,  the collective matter oscillator of amplitude $x_{\text{mat},n\mathbf{k}_\parallel}$, described by the indexes $n$ and $\mathbf{k_\parallel}$, only couples with the cavity mode of same indexes due to the orthogonality of all these modes.

Last, we transform the fourth term of Eq. \eqref{eq_supplementary_bulk_2}, which involves molecule-molecule interactions. To perform this transformation, we consider the SpC coupling strength between molecular emitters as given by Eq. \eqref{g_dipole_dipole_SI} explicitly, which leads to
\begin{align}
& \frac{1}{\sqrt{N_\text{eff}}} \sum_i \sum_{j\neq i} 2\omega_\text{dip}  g_{\scriptscriptstyle{\text{SpC}}}^{(i,j)} \sin \left( \frac{n\pi z_i}{L_\text{cav}} \right) e^{-i \mathbf{k}_\parallel \cdot \mathbf{r}_{\parallel i}} x_{\text{mat},j} \nonumber \\
&= \frac{1}{\sqrt{N_\text{eff}}} \sum_j 2\omega_\text{dip} x_{\text{mat},j} e^{-i \mathbf{k}_\parallel \cdot \mathbf{r}_{\parallel j}}  \sum_{i\neq j}    g_{\scriptscriptstyle{\text{SpC}}}^{(i,j)} \sin\left( \frac{n\pi z_i}{L_\text{cav}}   \right) e^{-i \mathbf{k}_\parallel \cdot (\mathbf{r}_{\parallel i} - \mathbf{r}_{\parallel j}) } \nonumber \\
&= \frac{1}{\sqrt{N_\text{eff}}} \sum_j  2\omega_\text{dip} x_{\text{mat},j} e^{-i \mathbf{k}_\parallel \cdot \mathbf{r}_{\parallel j}} \sum_{i \neq j} \frac{1}{2} \frac{f_\text{dip} e^{-i \mathbf{k}_\parallel \cdot (\mathbf{r}_{\parallel i} - \mathbf{r}_{\parallel j}) }}{4\pi\varepsilon_0 |\mathbf{r}_i-\mathbf{r}_j|^3 \omega_\text{dip}} [1-3(\mathbf{n}_\mathbf{d}\cdot \mathbf{n}_{\mathbf{r}ij})] \sin \left( \frac{n\pi z_i}{L_\text{cav}} \right) \nonumber \\
&\approx \frac{1}{\sqrt{N_\text{eff}}} \sum_j  2\omega_\text{dip} x_{\text{mat},j} e^{-i \mathbf{k}_\parallel \cdot \mathbf{r}_{\parallel j}} \sin \left( \frac{n\pi z_j}{L_\text{cav}} \right) \underbrace{\sum_{i \neq j} \frac{1}{2} \frac{f_\text{dip} e^{-i \mathbf{k}_\parallel \cdot (\mathbf{r}_{\parallel i} - \mathbf{r}_{\parallel j}) }}{4\pi\varepsilon_0 |\mathbf{r}_i-\mathbf{r}_j|^3 \omega_\text{dip}} [1-3(\mathbf{n}_\mathbf{d}\cdot \mathbf{n}_{\mathbf{r}ij})]}_{g_{\text{shift}}^{(n\mathbf{k}_\parallel)}} \nonumber \\
& \qquad =   2\omega_\text{dip} g_{\text{shift}}^{(n\mathbf{k}_\parallel)} x_{\text{mat},n\mathbf{k}_\parallel}. \label{eq_fourth_term_transformed}
\end{align}
In the fourth line in Eq. \eqref{eq_fourth_term_transformed}, we have considered that the dipole-dipole coupling strength between different molecular emitters, which depends on their distance as $|\mathbf{r}_i-\mathbf{r}_j|^{-3}$, decays much faster over $z$ than the term $\sin(n\pi z/L_\text{cav})$ changes (unless $n$ is so large that it has very fast oscillations, which we do not consider here). Due to this fast decay, we have checked numerically that the term $\sin(n\pi z_i/L_\text{cav})$ can be taken outside the sum over the molecular emitters $i$ as a constant of value $\sin(n\pi z_j/L_\text{cav})$, i.e. where only the emitter $j$ is involved. The sum over the variable $i$ in Eq. \eqref{eq_fourth_term_transformed} can be then performed numerically to obtain the collective molecule-molecule coupling strength $g_{\text{shift}}^{(n\mathbf{k}_\parallel)}$.

Therefore, by gathering all transformed terms in Eqs. \eqref{eq_first_second_terms_transformed}, \eqref{eq_third_term_transformed} and \eqref{eq_fourth_term_transformed}, and using Eq. \eqref{eq_collective_matter_oscillator}, Eq. \eqref{eq_supplementary_bulk_2} becomes
\begin{equation}
\ddot{x}_{\text{mat},n\mathbf{k}_\parallel} + (\omega_\text{dip}^2 + 2\omega_\text{dip} g_{\text{shift}}^{(n\mathbf{k}_\parallel)})  x_{\text{mat},n\mathbf{k}_\parallel} + 2 g_{\scriptscriptstyle{\text{\MC{}}}}^\text{max} \sqrt{N_\text{eff}} \dot{x}_{\text{cav},n\mathbf{k}_\parallel} = 0.
\label{eq_transformation_finished_2}
\end{equation}
Equations \eqref{eq_transformation_finished_1} and \eqref{eq_transformation_finished_2} correspond to Eqs. \eqref{equations_motion_bulk_transformed_1} and \eqref{equations_motion_bulk_transformed_2} in the main article.
Importantly, the derivation carried out in this section shows two important features of light-matter coupling in this system: i) although each $n\mathbf{k}_\parallel$ cavity mode is coupled to all individual molecular emitters,  it is only coupled to the $n\mathbf{k}_\parallel$  collective mode due to the orthogonality of the modes, and ii) the only consequence of the molecule-molecule coupling for the interaction between the $n\mathbf{k}_\parallel$ cavity and collective matter modes is to shift the bare frequency of the matter oscillator from $\omega_\text{dip}$ to $\sqrt{\omega_\text{dip}^2 + 2\omega_\text{dip} g_{\text{shift}}^{(n\mathbf{k}_\parallel)}}$ \cite{ribeiro23rep}. 

\section{Reststrahlen band}\label{Appendix_reststrahlen}

In the main text, we have shown that, if we impose that the resonant cavity mode and matter excitation frequencies in the coupled harmonic oscillator model are the bare ones without any dressing, then the Reststrahlen band is only correctly recovered when we use the \MC{} model, i.e., the coupling term is proportional to the time derivative of the amplitude of the oscillators.  However, according to Sec. \ref{Appendix_alternative_models} and Table S1, if we relax this condition and dress the cavity mode or matter excitation, we find alternative classical harmonic oscillator models that are equivalent to the \MC{} model but that use a coupling term proportional to the oscillation amplitude and the appropriate dressing. Here, we apply this finding to demonstrate how to reproduce the Reststrahlen band with a coupling term proportional to the oscillator amplitudes. The results are equivalent to those obtained using the Hopfield Hamiltonian\cite{hopfield58rep}. To obtain the Reststrahlen  band, we first follow the approach in Ref. \cite{yoo21rep} to obtain the bulk dispersion of a phononic material directly from the response of an infinite material  (an equivalent demonstration could  be performed by connecting the dispersion of a Fabry-Pérot cavity with the bulk dispersion as in Sec. \ref{subsec_bulk_dielectric})

In the previous work, the authors considered a phononic material with a permittivity given by Eq. \eqref{eq_permittivity_polar_materials} in the main text and derived the system of equations 
\begin{eqnarray}
\left(\begin{matrix}
\omega^2-\omega^2_\text{TO} & \omega\omega_\text{p}\\
 \omega\omega_\text{p} & \omega^2-\omega^2_k
\end{matrix}\right)
\left(\begin{matrix}
ij_\text{latt}/(q_\text{i})\\
 \sqrt{\varepsilon_0\varepsilon_\infty}|{\bf E}(k)_\text{inc}|
\end{matrix}\right)=0
\label{eq:couplingphotonicbulk}
\end{eqnarray}
where $|{\bf E}(k)_\text{inc}|$ is the amplitude of the incident electric field, $q_\text{i}$ the (positive) charge of the lattice ions, $\varepsilon_\infty$ the high-frequency permittivity of the material, $\omega$ the frequency of the bulk dispersion modes,  $\omega_k=\frac{c k_0}{\sqrt{\varepsilon_\infty}} $ the frequency of a free photon of vacuum wavevector $k_0$ propagating in a medium of permittivity $\varepsilon_\infty$ and $k$ the wavevector in this material. $\omega_\text{p}=\sqrt{\omega^2_\text{LO}-\omega^2_\text{TO}}$ is the parameter that controls the coupling strength in this model and $\omega_\text{LO}$ and  $\omega_\text{TO}$ the frequency of the longitudinal and transverse optical phonon modes, respectively. According to the discussion in Sec. \ref{subsec_bulk_dielectric} of the main text, $\omega_\text{TO}$ is here the bare frequency of the  material.  As a difference to the previous work, we neglect losses, and we have written the equations as a function of the normalized microscopic current $j_\text{latt}$, which depends on the normalized relative displacement $x_\text{latt}$ of the atoms in the atomic lattice\cite{yoo21rep} through $j_\text{latt}=-i q_i \omega x_\text{latt}$ ($x_\text{latt}$ correspond to the normalized displacement of the atoms that are positively charged from the atoms that are mostly negatively charged). We can then identify $x_k =\sqrt{\varepsilon_0\varepsilon_\infty}|{\bf E}(k)_\text{inc}|$ (the amplitude of the oscillator associated with the field of wavevector $k$) and $x_\text{mat}=-j_\text{latt}/(q_\text{i})$ (associated with the matter excitation) and rewrite Eq. \eqref{eq:couplingphotonicbulk} as

\begin{subequations}
\begin{equation}
-i\omega^2x_\text{mat} + i\omega^2_\text{TO} x_\text{mat} + \omega\omega_\text{p} 
{x}_k =0  
\end{equation}
\begin{equation}
\omega^2x_k - \omega^2_k x_k - i\omega\omega_\text{p} 
{x}_\text{mat} =0,  
\end{equation}
\end{subequations}
which can be rewritten in the time domain to obtain
\begin{subequations}
\begin{equation}
i\ddot{x}_\text{mat} + i\omega^2_\text{TO} x_\text{mat} + i\omega_\text{p} \dot
{x}_k =0  
\end{equation}
\begin{equation}
-\ddot{x}_k - \omega^2_k x_k + \omega_\text{p} 
\dot{x}_\text{mat} =0,  
\end{equation}
\end{subequations}
or, equivalently,
\begin{subequations}\begin{equation}
\ddot{x}_k + \omega_k^2 x_k -2G_{\scriptscriptstyle{\text{\MC{}}}}   \dot{x}_\text{mat}  =0, 
\end{equation}
\begin{equation}
\ddot{x}_\text{mat} + \omega_\text{mat}^2 x_\text{mat} + 2G_{\scriptscriptstyle{\text{\MC{}}}} 
\dot{x}_k =0.  
\end{equation}\label{eq:MCbulk}
\end{subequations}
where $\omega_\text{mat}=\omega_\text{TO}$ and $G_{\scriptscriptstyle{\text{\MC{}}}}=\frac{\omega_\text{p}}{2}=\frac{\sqrt{\omega^2_\text{LO}-\omega^2_\text{TO}}}{2}$, so that we recover  Eq. \eqref{equations_motion_bulk_transformed_changedterminology} in the main text. This equivalence indicates that the result obtained here for an infinite phononic material coincides with those obtained in Sec. \ref{subsec_bulk_dielectric} of the main text, where we focused on the eigenmodes of the Fabry-Pérot cavity. Following the discussion in that section, this further confirms that Eq. \eqref{eq:MCbulk} describes the bulk dispersion within the \MC{} model.

\begin{figure}
\centering
\includegraphics[scale=1.1]{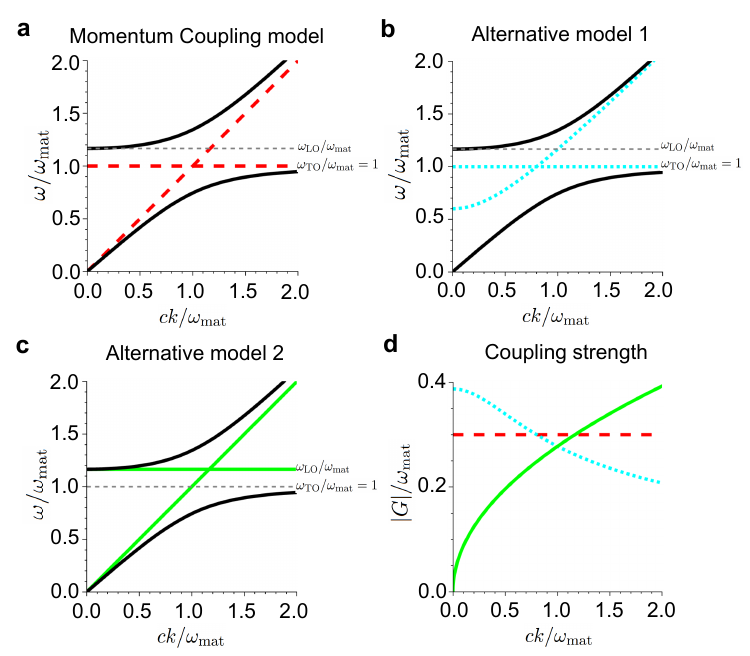}
\caption{Bulk dispersion and opening of the Reststrahlen band according to different models. (a) 
Bulk dispersion (black line) and uncoupled frequencies (bare photon frequency $\omega_{k}$, diagonal red dashed line; transverse optical frequency $\omega_{\text{TO}}$, horizontal red dashed line) according to the Momentum Coupling (\MC{}) model. The horizontal dashed line corresponds to the longitudinal optical phonon frequency.  (b) Bulk dispersion (black line) and uncoupled frequencies (dressed photon frequency $\omega^{A1}_{k}$, diagonal-like cyan short-dashed line; bare transverse optical phonon frequency $\omega_{\text{TO}}$, horizontal cyan short-dashed line) according to the alternative model 1. The horizontal dashed line corresponds to the longitudinal optical phonon frequency.  (c) 
Bulk dispersion (black line) and uncoupled frequencies (bare photon frequency $\omega_{k}$, diagonal green solid line; longitudinal optical phonon frequency $\omega_{\text{LO}}$, horizontal green solid line) according to alternative model 2. The horizontal dashed line corresponds to the transverse optical phonon frequency.  (d) Coupling strength as a function of wavevector according to the \MC{} model ($G_{\scriptscriptstyle{\text{\MC{}}}}$, red dashed line), the alternative model 1 ($|G^{\text{A1}}|$, cyan short-dashed line) and the alternative model 2 ($G^{\text{A2}}$, green solid line).      In all panels, the bulk dispersion is the same (black lines, corresponding to the one obtained for  $G_{\scriptscriptstyle{\text{\MC{}}}}=0.3\omega_{\text{mat}}$),
all frequencies and the coupling strength are normalized by $\omega_{\text{mat}}=\omega_{\text{TO}}$ and the results are plotted as a function of the normalized wavevector $ck/\omega_{\text{mat}}$.} \label{fig:comparingmodels}
\end{figure} 

On the other hand, Table S1 indicates that an equivalent dispersion  can be obtained with the following coupled harmonic oscillator model (alternative model 1):

\begin{subequations}
\begin{align}
&\ddot{x}_k + (\omega_k^2 + 4G_{\scriptscriptstyle{\text{\MC{}}}}^2) x_k - 2 G_{\scriptscriptstyle{\text{\MC{}}}} \omega_\text{mat} x'_\text{mat} = 0, \\
&\ddot{x}'_\text{mat} + \omega_\text{mat}^2 x'_\text{mat} - 2 G_{\scriptscriptstyle{\text{\MC{}}}} \omega_\text{mat} x_k = 0,
\end{align}
\end{subequations}
where the coupling term is proportional to the amplitude of the oscillators. We can rewrite this equation as

\begin{subequations}
\begin{align}
&\ddot{x}_k + \left(\omega^{\text{A1}}_{k}\right)^2  x_k + 2 G^{\scriptscriptstyle{\text{A1}}} \sqrt{\omega_\text{mat}\omega^{\text{A1}}_{k}} x'_\text{mat} = 0, \\
&\ddot{x}'_\text{mat} + \omega_\text{mat}^2 x'_\text{mat} + 2 G^{\scriptscriptstyle{\text{A1}}} \sqrt{\omega_\text{mat}\omega^{\text{A1}}_{k}} x_k = 0, \\
&\omega^{\text{A1}}_{k}=\sqrt{\omega_k^2 + 4G^2_{\scriptscriptstyle{\text{\MC{}}}}}, \\
& G^{\scriptscriptstyle{\text{A1}}} =-G_{\scriptscriptstyle{\text{\MC{}}}}\frac{\omega_\text{mat}}{\sqrt{\omega_\text{mat}\omega^{\text{A1}}_{k}}}=-G_{\scriptscriptstyle{\text{\MC{}}}}\sqrt{\frac{\omega_\text{mat}}{\sqrt{\omega_k^2 + 4G^2_{\scriptscriptstyle{\text{\MC{}}}}}}}.
\end{align}\label{eq:alternativedispersion1}
\end{subequations}
Crucially, the first two equations are formally equivalent to the SpC model, except that in this case, $\omega^{\text{A1}}_{k}$ is a dressed frequency and the coupling strengths $G^{\scriptscriptstyle{\text{A1}}}$ has been changed, as given by the last two equations (the superscript 'A1' refers to alternative model 1).

Proceeding in the same manner but for the alternative model 2 in Table 1, we obtain a second set of coupled harmonic oscillations that also give the same dispersion. 
\begin{subequations}
\begin{align}
&\ddot{x}'_k + \omega_{k}^2  x'_k + 2 G^{\text{A2}} \sqrt{\omega^{\text{A2}}_{\text{mat}}\omega_{k}} x_\text{mat} = 0, \\
&\ddot{x}_\text{mat} + \left(\omega^{\text{A2}}_{{\text{mat}}}\right)^2 x_\text{mat} + 2 G^{\text{A2}} \sqrt{\omega^{\text{A2}}_{\text{mat}}\omega_{k}} x'_k = 0, \\
&\omega^{\text{A2}}_{\text{mat}}=\sqrt{\omega_\text{mat}^2 + 4G^2_{\scriptscriptstyle{\text{\MC{}}}}}=\sqrt{\omega_\text{TO}^2 + 4G^2_{\scriptscriptstyle{\text{\MC{}}}}}=\omega_\text{LO}, \\
& G^{\text{A2}} =G_{\scriptscriptstyle{\text{\MC{}}}}\frac{\omega_k}{\sqrt{\omega^{\text{A2}}_\text{mat}\omega_{k}}}=G{\scriptscriptstyle{\text{\MC{}}}}\sqrt{\frac{\omega_k}{\sqrt{\omega_\text{mat}^2 + 4G^2_{\scriptscriptstyle{\text{\MC{}}}}}}},
\end{align}\label{eq:alternativedispersion2}
\end{subequations}
where, in this case, the dressed frequency is that of the matter excitation $\omega^{\text{A2}}_{\text{mat}}$.

 The different coupled harmonic oscillator models are compared in Fig. \ref{fig:comparingmodels}. The bulk dispersion obtained for $G_{\scriptscriptstyle{\text{\MC{}}}}=0.3\omega_{\text{mat}}=0.3\omega_{\text{TO}}$ (and corresponding values of $G^{\text{A1}}$ and $G^{\text{A2}}$) is shown in panels (a-c) by the black lines. As expected, the three models give identical results. The red dashed lines in Fig. \ref{fig:comparingmodels}(a) show the frequencies of the uncoupled modes (i.e. the frequencies that are obtained if the coupling is ignored) of the \MC{} model given by Eq. \eqref{eq:MCbulk}, corresponding to the frequency of the transverse optical mode, $\omega_{\text{TO}}$, and of the free photons in the material of permittivity $\varepsilon_\infty$, $\omega_{k}$.  Figure \ref{fig:comparingmodels}(b) shows the corresponding result for the alternative model 1, with the uncoupled modes (cyan short-dashed line) being in this is case the TO photon at frequency $\omega_{\text{TO}}$ and the dressed photon at frequency  $\omega^{A1}_{k}$. Last, the uncoupled frequencies of the alternative model 2 are indicated by the solid green line in Fig. \ref{fig:comparingmodels}(c) and correspond to the LO phonon frequency $\omega_{\text{LO}}$ and of the free photons $\omega_{k}$. The coupling strength that need to be used in each of this models to reproduce the same bulk dispersion is shown in Fig. \ref{fig:comparingmodels}(d) (red dashed line corresponds to the coupling strength in the \MC{} model, $G_{\scriptscriptstyle{\text{\MC{}}}}=0.3\omega_{\text{mat}}$; the cyan short-dashed line corresponds to the coupling strength in the alternative model 1, $G^{\text{A1}}$; the  green solid line to the coupling strength in alternative model 2, $G^{\text{A2}}$). These results thus stress that the same bulk dispersion can be obtained using different classical coupled harmonic oscillator models. 

Last, we emphasize that the possibility of obtaining the same dispersion with both the \MC{} model (Eq. \eqref{eq:MCbulk}) and the second alternative model (Eq. \eqref{eq:alternativedispersion2}) indicates that the bulk dispersion can be obtained with classical coupled harmonic oscillator models that use a coupling term that can be proportional to either the oscillator amplitude (alternative model 2) and to its derivative (\MC{} model). These two models 
offer a very different picture of the opening of the Reststrahlen band (we do not discuss here the first alternative model because it does not have a simple physical interpretation):

\begin{itemize}
\item According to the second alternative model (Eq. \eqref{eq:alternativedispersion2}),  the dressed matter excitation in the coupled equations corresponds to the longitudinal optical phonon frequency, $\omega^{\text{A2}}_{\text{mat}}=\sqrt{\omega_\text{TO}^2 + 4G^2_{\scriptscriptstyle{\text{\MC{}}}}}= \omega_\text{LO}$, the renormalized coupling strength $G^{\text{A2}}$ becomes zero for photons of energy (or momentum) tending to zero, and $\left(G^{\text{A2}}\right)^2$ scales linearly with photon energy. Thus, in this picture, i) the square of the coupling term is proportional to the energy of the photons,  ii) the longitudinal optical phonon appears as the resonant matter excitation in the harmonic oscillator equations, and can be interpreted as the dressed matter excitation of the 'bare' transverse optical phonon, iii) the (dressed) matter excitation and the photons do not couple at low energies and iv) at large energy/momentum, the coupling becomes infinite. The arbitrarily large coupling strength at large momenta explains why, in the two limits of large detuning ($\omega_k\rightarrow 0$ and $\omega_k\rightarrow \infty$), two different asymptotic frequencies ($\omega_\text{TO}$ and $\omega_\text{LO}$ respectively) are obtained, i.e., it explains the opening of the Reststrahlen band. 
 \item In contrast, in the \MC{} model, (i) the coupling constant is independent of the photon energy, and (ii) the transverse optical phonon coincides with the bare matter excitation . In this case, the Reststrahlen band opens because the coupling is proportional to the time derivative of the oscillation amplitudes. 
\end{itemize}

\end{document}